\documentclass[11pt]{article}
\usepackage{amsmath}
\usepackage{graphicx,psfrag,epsf}
\usepackage{enumerate}
\usepackage{enumitem}
\usepackage{adjustbox}
\usepackage{natbib}
\usepackage{url} 

\DeclareMathOperator*{\argmin}{arg\,min}
\usepackage{amsfonts}
\usepackage{verbatim}
\usepackage{xcolor}
\usepackage{amssymb}
\usepackage{epstopdf}%
\usepackage{caption,subcaption}
\usepackage{algorithm,algorithmicx}
\usepackage[noend]{algpseudocode}
\usepackage[detect-all]{siunitx}
\usepackage{tabu}
\usepackage{arydshln}
\usepackage{eufrak}
\usepackage{mathtools}
\usepackage{hanging}
\DeclarePairedDelimiter{\ceil}{\lceil}{\rceil}
\newcolumntype{"}{@{\hskip\tabcolsep\vrule width 1pt\hskip\tabcolsep}}
\makeatother
\providecommand{\U}[1]{\protect\rule{.1in}{.1in}}

\newtheorem{theorem}{Theorem}

\newtheorem{prop}{Proposition}

\newcommand{\bbet}{\boldsymbol{\beta}}
\newcommand{\tbbet}{\boldsymbol{\tilde{\beta}}}
\newcommand{\bbbet}{\boldsymbol{\bar{\beta}}}
\newcommand{\hbbet}{\boldsymbol{\hat{\beta}}}
\newcommand{\hbeta}{\hat{\beta}}
\newcommand{\bx}{\mathbf{x}}
\newcommand{\bX}{\mathbf{X}}
\newcommand{\by}{\mathbf{y}}
\newcommand{\bz}{\mathbf{z}}

\algrenewcommand\algorithmicrequire{\hspace*{\algorithmicindent}\textbf{Input:}}
\algrenewcommand{\algorithmicensure}{\hspace*{\algorithmicindent}\textbf{Initialize:}}
\algdef{SE}[DOWHILE]{Do}{doWhile}{\algorithmicdo}[1]{\algorithmicwhile\ #1}%

\usepackage{placeins}
\usepackage{booktabs}
\usepackage{multirow}
\usepackage{rotating}
\usepackage{afterpage}
\usepackage{setspace}
\usepackage{mathtools}
\usepackage{hyperref}
\hypersetup{
	colorlinks=true,
	linkcolor=blue,
	citecolor=blue,
	filecolor=magenta,
	linktocpage=true,      
	urlcolor=blue,
}

\begin{document}
	
	\onehalfspacing

	\def\spacingset#1{\renewcommand{\baselinestretch}%
		{#1}\small\normalsize} \spacingset{1}

	
	\title{\textbf{Data-Driven Logistic Regression Ensembles With Applications in Genomics}}
	
	\author{
		\begin{tabular}{*{2}{>{\centering}p{.5\textwidth}}}
			\large Anthony-Alexander Christidis  & \large Stefan Van Aelst \tabularnewline
			Department of Statistics & Department of Mathematics \tabularnewline
			University of British Columbia & KU Leuven \tabularnewline
			2207 Main Mall & Celestijnenlaan 200B \tabularnewline
			Vancouver, BC V6T 1Z4 &  3001 Leuven, Belgium \tabularnewline
			\url{anthony.christidis@stat.ubc.ca} & \url{stefan.vanaelst@kuleuven.be}
		\end{tabular}
		\\ \\
		\begin{tabular}{*{3}{>{\centering}p{.5\textwidth}}}
			\large Ruben Zamar  \tabularnewline
			Department of Statistics \tabularnewline
			University of British Columbia  \tabularnewline
			2207 Main Mall \tabularnewline
			Vancouver, BC V6T 1Z4  \tabularnewline
			\url{ruben@stat.ubc.ca}
		\end{tabular}
	}
	
	\date{}
	
	\maketitle
	
	\begin{abstract}
		Advances in data collecting technologies in genomics have significantly increased the need for tools designed to  study the genetic basis of many diseases.
		Effective statistical methods should excel in both prediction accuracy and biomarker identification. 
		We introduce a novel approach to high-dimensional binary classification that integrates regularization with ensembling techniques.  The method constructs compact ensembles of interpretable models derived by optimizing a global objective function.
		In medical genomics applications, the proposed approach identifies critical biomarkers overlooked by competing methods. We develop a variable importance ranking system to help researchers prioritize promising genes. The method's asymptotic properties are established, and an efficient computational algorithm is provided.
		Through extensive simulations across complex scenarios and analysis of cancer genomics datasets, we demonstrate strong predictive performance. Based on the numerical experiments, we offer practical guidelines for determining optimal ensemble size.
	\end{abstract}
	
	\noindent%
	{\it Keywords:} Consistent ensemble; Diversity penalty; High-dimensional classification; Logistic regression models; Split modeling.
	\vfill

	\section{Introduction}
	
	The surge of genomic data collection through ever-evolving technologies has necessitated the development of sophisticated statistical methods to analyze high-dimensional gene expression data, unlocking its potential for breakthroughs in healthcare and scientific discovery.
	In many medical genomics applications, the goal is to train a classifier that can accurately predict the presence of a disease or particular subtypes of a disease based on the genetic profile of a patient.
	Furthermore, in the medical sciences and other fields predictions often have important consequences in decision-making processes, so there is a high demand for interpretable learning methods \citep[see e.g.][]{murdoch2019definitions, rudin2019stop, rudin2022interpretable}. It should thus be clear how a classifier arrives at its predictions so that decisions can be fully explained. In addition, medical researchers aim to unravel the relation between the genetic profile and occurrence of a disease, so an appropriate classification method should also be able to identify patterns between the expression of  key biomarkers and the presence of a disease.
	
	The vast richness and availability of medical genomics data is evident from large publicly accessible databases such as the Gene Expression Omnibus (GEO) database \citep{edgar2002gene, barrett2012ncbi}. In many of these datasets, a large number of gene expression measurements are collected for a relatively small number of (cell tissue) samples. The status of the sample is  typically given as well, specifying e.g. whether the sample collected was affected by a disease or not.
	Classification methods can then be used to discover patterns between the expression level of certain genes and the presence of the disease.  The two types of methods that are generally deployed in such applications are sparse and ensemble methods, respectively. 
	
	On the one hand, sparse methods yield interpretable models that only use a subset of the genes to make a decision ~\citep[see e.g.][for a modern treatment of sparse methods]{hastie2015statistical}. However, because the number of predictor genes is much larger than the number of samples, there may be several models comprised of  different subsets of genes that are  equally accurate in predicting the presence of the disease \citep[a phenomenon coined ``the multiplicity of good models" by][]{mccullagh1989monographs}. Thus, several potentially important genes may be erroneously discarded from the decision-making process when using a single sparse model.  On the other hand, ensemble methods combine multiple diverse models and generally achieve superior prediction performance if the members of the ensemble are sufficiently diverse. Diversity  is often 
	achieved  using randomization~\citep[see e.g.][]{ho1998random,RF,rglm} or sequentially fitting the residuals of the previous fit~\citep[see e.g.][]{GBM,buhlmann2003boosting,boosting,yu2020submito}.  Ensemble methods have been particularly successful in high-dimensional prediction tasks with genomic data \citep[see e.g.][]{dorani2018ensemble,zahoor2020classification}.  However, while ad hoc methods have been developed to assess variable importance in some ensemble methods, such as the variable importance measure of~ \citet{RF}, interpretation of the resulting prediction rules and  identification of important predictor genes (i.e. key biomarkers) is less straightforward.    
	
	We propose a new approach to learn a diverse ensemble of sparse classification models that is especially well suited for high-dimensional medical genomics applications. Specifically, we extend the regression ensemble method proposed by \cite{christidis2020} to the binary classification setting. We demonstrate the advantage of the proposed ensemble method in terms of both prediction and identification of key biomarkers using cancer genomics datasets from the GEO database, with a broader analysis including datasets for multiple sclerosis and psoriasis provided in the supplementary material. In particular, we show that the prediction accuracy of the sparse models in the ensembles matches the prediction accuracy of standard single-model sparse methods. Since the models in the ensembles are learned simultaneously and directly from the data (free of randomization or other heuristics) by optimizing a global objective function, they each provide an alternative explanation for the relationship between the genes and disease. The objective function contains a penalty that promotes diversity in the obtained models so that individual models may be driven by potentially different biological mechanisms. With respect to the identification of important predictor genes, the examples demonstrate that the proposed ensemble method can identify key biomarkers that are discarded by state-of-the-art sparse methods and ensemble variable ranking methods. At the same time, we show that the proposed method generally includes key biomarkers selected by sparse methods or flagged as important by ensemble variable ranking methods.
	
	Each of the models in the proposed ensemble method is a penalized logistic regression with a sparsity inducing penalty such as the Lasso~\citep{donoho1994ideal,Lasso} or the elastic net~\citep{EN}. Rather than resorting to randomization or other indirect methods to generate different models, we jointly learn the models in the ensemble on the training data and incorporate  a diversity penalty~\citep{christidis2020} in the objective function with the aim to diversify the models. The degree to which these models are sparse and diverse is driven directly by the data. In this way, the method efficiently exploits the so-called accuracy-diversity trade-off  between the models, and generates an ensemble with high predictive performance that often matches or even outperforms popular state-of-the-art ensemble methods.
	Moreover, by ensembling the models at the level of their linear predictors we retain interpretability for the logistic regression coefficients in the ensembled model. 
	We use several measures of diversity \citep{kuncheva2003measures}
	to study the accuracy-diversity trade-off in the ensemble method. This trade-off provides insight in the effect of the number of ensembled models on the performance of the resulting ensemble.  This insight makes it possible to make some  recommendations for the choice of the number of ensembled models.
	
	Following the methodological development framework of \cite{heinze2024phases}, this work encompasses phases I and II, introducing a novel method with initial validation through comprehensive simulations across various scenarios (sample sizes, sparsity, correlation structures, class imbalance, non-linearity, and interactions). While the results of the numerical experiments and medical genomics data applications demonstrate the promising performance of the proposed classification ensembles across a variety of data scenarios, we present this work as a method development study with strong proof-of-concept validation, rather than a claim of universal superiority. The success observed across these carefully chosen testing scenarios suggests that phases III and IV, independent validation and practical implementation, may naturally follow as the method gains adoption in the research community.
	
	The remainder of the paper is organized as follows. Section \ref{sec:hd_class_methods} reviews the literature on sparse and ensemble methods for high-dimensional classification. Section \ref{sec:split_logistic_models} introduces the proposed data-driven ensemble of sparse logistic regression models and establishes its consistency under mild regularity assumptions. In Section \ref{sec:computing}, an efficient block coordinate descent algorithm for solving the multi-convex optimization problem is provided. Section \ref{sec:simulation} presents the results of an extensive simulation study that systematically evaluates the proposed method against a large number of state-of-the-art competitors. In Section \ref{section:gene_expression}, we demonstrate the method's strong performance and practical benefits through a comprehensive analysis of medical genomics datasets. This section presents in-depth motivating examples using lung and thyroid cancer datasets to highlight the method's ability to uncover multiple biological mechanisms and rank genes by importance. A broader benchmark study validating its predictive accuracy across ten datasets is detailed in the Supplementary Material. To offer practical guidance on choosing the ensemble size, Section \ref{sec:div_measures} investigates the accuracy-diversity trade-off and computational cost. Finally, Section \ref{sec:discussion} concludes the paper and outlines potential directions for future research. potential directions for future research. \FloatBarrier
	
	\section{High-Dimensional Classification Methods} \label{sec:hd_class_methods}
	
	Training classification models on high-dimensional data, such as gene expression profiles where the number of measurements ($p$) exceeds the number of tissue samples ($n$), presents significant statistical and computational challenges. This $p > n$ scenario is common in bioinformatics and other fields. To address these challenges, two primary families of methods have emerged as dominant strategies in the literature: sparse modeling, which aims for feature selection and interpretability, and ensemble methods, which typically focus on maximizing predictive accuracy and stability by combining multiple models. In this section, we briefly review representative state-of-the-art techniques from both families, discussing their strengths and limitations to provide context for the proposed method introduced in Section \ref{sec:split_logistic_models}.
	
	\subsection{Sparse Methods}
	
	Let $\mathcal{D}=\left( \mathbf{y}, \mathbf{X}\right)$ denote the training data, where $\mathbf{y}\in \mathbb{R}^{n}$ is the vector of class labels and $\mathbf{X}\in \mathbb{R}^{n\times p}$ is the design matrix which consists of $n$ measurements $\bx_i$ on $p$ predictors. We consider the binary classification problem where the classes are labeled by $y_i \in \{0, 1\}$. The predictor variables have been standardized, i.e., $\sum\limits_{i=1}^{n}x_{ij}/n=0$ and $\sum\limits_{i=1}^{n}x_{ij}^{2}/n=1$ for $1\leq j \leq p$. 
	Logistic regression models the class-conditional probabilities through a non-linear function of the predictor variables,
	\begin{equation} \label{eq:logistic_model}
		p_i = \mathbb{P}(y_i=1|\bx_i) = S(\beta_0 + \bx_i^T \bbet), \quad 1 \leq i \leq n,
	\end{equation}
	where $ \beta_0 $ and $ \bbet \in \mathbb{R}^p $ are the intercept and vector of regression coefficients, respectively. The function $ S(t) = e^t/(1+e^t) $ is the well-known logistic function.  With $f(\bx_i)=\beta_0 + \bx_i^T \bbet$, the negative log-likelihood for a single observation $(y_i, \bx_i)$, often referred to as the logistic loss or log loss, is given by
	\begin{equation} \label{eq:loss_function}
		\mathcal{L}\left(f(\bx_i), y_i\right)=\mathcal{L}(\beta_0,\bbet \mid y_i, \bx_i) = - y_i f(\bx_i) + \log(1+e^{f(\bx_i)}).
	\end{equation} 
	This loss function arises directly from the Bernoulli likelihood where $\mathbb{P}(y_i=1|\bx_i)=p_i=S(f(\bx_i))$ and $\mathbb{P}(y_i=0|\bx_i)=1-p_i = 1-S(f(\bx_i)) = S(-f(\bx_i))$. Minimizing the corresponding empirical loss (average negative log-likelihood)
	\begin{equation}
		\mathcal{V}_n(f) = \frac{1}{n} \sum_{i=1}^{n}\mathcal{L}\left(f(\bx_i), y_i\right)
		\label{empirical_loss}
	\end{equation}
	via standard maximum likelihood estimation requires $p < n$. When $p \ge n$, the estimation problem is typically ill-posed. Even when $p < n$, if $p$ is large relative to $n$, minimizing \eqref{empirical_loss} often leads to overfitting of the training data and poor out-of-sample prediction accuracy.
	
	By restricting model complexity, sparse methods aim to find a single (sparse) model that achieves good prediction accuracy using only a small subset of the predictors. They have proven to be highly successful approaches for high-dimensional classification problems in the genomic sciences \citep[see e.g.][]{zuo2017incorporating, rejchel2020rank}. Sparse regularization methods typically solve an optimization problem of the form
	\begin{align}
		\min_{\beta_0 \in \mathbb{R}, \, \bbet \in \mathbb{R}^p} \mathcal{V}_n(f) + \lambda P_s\left(\bbet\right),
	\end{align}
	where $ P_s(\bbet) $ is a penalty function that induces sparsity in the coefficient vector $ \bbet $. The tuning parameter $\lambda > 0 $ is usually determined in a data-driven way, typically by cross-validation (CV). A common and effective choice for the penalty term $P_s$ is the elastic net penalty \citep{EN}. It combines $\ell_1$ and $\ell_2$ regularization components:
	\begin{align} 
		P_{s}(\boldsymbol{\beta }) = \frac{1-\alpha}{2}\Vert \boldsymbol{\beta }\Vert _{2}^{2} + \alpha \Vert \boldsymbol{\beta}\Vert_{1}, \quad \alpha \in [0,1].
		\label{eq:sparsity_penalty}
	\end{align}
	The mixing parameter $\alpha$ balances the sparsity-inducing $\ell_1$ norm $\Vert \boldsymbol{\beta}\Vert_{1}$ and the squared $\ell_2$ norm $\Vert \boldsymbol{\beta}\Vert_{2}^{2}$. This penalty includes Lasso \citep{Lasso} ($\alpha=1$) and Ridge \citep{Ridge} ($\alpha=0$) as special cases.
	
	Next to Lasso and elastic net, some of the more popular sparse regularization methods are the adaptive Lasso \citep{zou2006adaptive}, the relaxed Lasso \citep{meinshausen2007relaxed}, the smoothly clipped absolute deviation (SCAD) estimator \citep{SCAD} and the minimum concave penalized (MCP) estimator \citep{MCP}. Additional approaches include Sure Independence Screening (SIS) \citep{SIS}, which first reduces dimensionality before applying a penalty like SCAD, and RuleFit \citep{friedman2008predictive}, which combines sparse linear terms with rule-based terms derived from decision trees. A vast amount of asymptotic theory has been developed for a large class of regularized estimators, see e.g. \cite{buhlmann-book} for an extensive treatment. In summary, sparse regularization methods yield a single interpretable model with good prediction accuracy and extensive asymptotic theory that describes their behavior.
	
	\subsection{Ensemble Methods}
	
	Ensemble methods have proven to be very successful in high-dimensional classification tasks, often yielding higher prediction accuracy than their sparse single-model competitors. To better understand the behavior of ensemble models, \cite{ueda1996generalization} first developed a decomposition of its generalization error for the regression case. \cite{brown2005managing} provided an in-depth analysis of the bias-variance-covariance trade-off in regression ensembles. In particular, if the ensemble of a collection of estimators $\hat{f}_1,\dots, \hat{f}_G$ is their average $\bar{f}=\sum_{g=1}^G \hat{f}_g/G$, then its mean squared prediction error (MSPE) can be decomposed as	
	$ \text{MSPE}[\bar{f}]=\text{Bias}[\bar{f}]^2+\text{Var}[\bar{f}] + \sigma^2$, where $ \sigma^2 $ is the irreducible variance of the errors. The bias and variance of the regression ensemble can be decomposed further as
	\begin{align} 
		\text{Bias}\left[\bar{f}\right] &= \overline{\text{Bias}}_G, \\
		\label{eq:split_var}
		\text{Var}\left[\bar{f}\right] &= \frac{1}{G} \ \overline{\text{Var}}_G + \frac{G-1}{G} \ \overline{\text{Cov}}_G,
	\end{align}
	where $ \overline{\text{Bias}}_G $, $ \overline{\text{Var}}_G $, and $ \overline{\text{Cov}}_G $ are the average of the biases, variances, and pairwise covariances of the $ G $ estimators in the ensemble, respectively. From~(\ref{eq:split_var}) it becomes clear that as the number of estimators increases, their correlations play a much more critical role than their average variability in obtaining a good ensemble estimator. A similar principle was derived for classifier ensembles by \cite{tumer1996error} and later refined by \cite{fumera2003linear}.
	
	The importance of diversity among the constituent models within an ensemble is a well-established principle, exemplified by popular methods like random forests \citep{RF}. In random forests, diversity is achieved by constructing individual decision trees using random subsets of candidate features at each split point. Other techniques also leverage randomization to foster diversity. For instance, the random generalized linear models (RGLM) method of \cite{rglm} combines bagging \citep{breiman1996bagging} with the random predictor subspace method \citep{ho1998random}. Sequential ensemble methods employ a different strategy, such as (extreme) gradient boosting \citep{chen2016xgboost}, where diverse trees are generated iteratively to correct the errors of the preceding ensemble. A third approach is stacking \citep{breiman-stacked}, where predictions from different algorithms are combined through a meta-learner. The specific relationship between various measures of diversity in classifier ensembles and their resulting prediction accuracy has been formally investigated by \citet{kuncheva2003measures}.
	
	An alternative to these heuristic approaches is to build diversity directly into the model fitting process. The split regularized regression framework proposed by \citet{christidis2020} introduces a competitive ensemble approach where multiple sparse regression models are learned simultaneously by optimizing a single global objective function. This objective function includes not only a standard sparsity penalty for each model but also a novel {diversity penalty} that explicitly discourages different models from selecting the same variables. By directly penalizing coefficient overlap, this method produces a small, data-driven ensemble of diverse and individually interpretable models. This deterministic, optimization-based approach to generating diversity stands in contrast to the randomization or sequential fitting strategies used by other ensemble methods.
	
	In summary, many conventional ensemble methods face challenges regarding interpretability, as they often aggregate predictions from numerous relatively simple or ``weak" base models. While ad hoc measures like variable importance scores exist, understanding the combined structure can be difficult. Although these methods often rely on heuristics, theoretical understanding has advanced, with consistency proofs available for methods like Random Forests \citep{biau2008consistency}. \FloatBarrier
	
	\section{Split Logistic Regression Models} \label{sec:split_logistic_models}
	
	Building upon the split regularized regression framework of \citet{christidis2020}, we now introduce Split Logistic Regression, a novel hybrid approach specifically designed for high-dimensional binary classification. This method extends the original framework, which was developed for linear regression using a squared error loss, to the classification setting by employing the logistic loss function. It combines the stability and interpretability of sparse methods with the high accuracy of ensemble learning by simultaneously fitting multiple sparse logistic regression models that are encouraged to be diverse.
	
	Let $\left( \mathbf{y}, \mathbf{X}\right)$ denote the training data, as before. Using the logistic loss function \eqref{eq:loss_function} for $\mathcal{L}(\beta_0^g,\bbet^g \mid y_i, \bx_i)$ in each model with $f_g(\bx_i)=\beta_0^g + \bx_i^T \bbet^g$, the split logistic regression objective function to simultaneously fit $G$ models is given by
	{\small
		\begin{equation} \label{eq:split_objective}
			\mathcal{J}\left(\beta_0^1,\bbet^1,\dots,\beta_0^G,\bbet^G \right) = \sum_{g=1}^{G} \left[ \frac{1}{n}\sum_{i=1}^n \mathcal{L}\left(\beta_0^g,\bbet^g | y_i, \bx_i\right)+ \lambda_{s} P_s(\bbet^g) \right] + \frac{\lambda_{d}}{2} \sum_{h\neq g} P_d(\bbet^h,\bbet^g),
		\end{equation}
	}
	\noindent
	which needs to be minimized with respect to all regression coefficients. In the regression ensemble setting considered in \cite{christidis2020}, the loss function $\mathcal{L}(\beta_0^g,\bbet^g | y_i, \bx_i)$ used for each model is the squared error loss. The sparsity penalty function $P_s$ regularizes each of the $G$ individual models, while the goal of the diversity penalty $P_d$ is to discourage the same variable from appearing in multiple models, thereby encouraging the models to complement each other.
	
	Note that the diversity penalty $P_d$ needs to have two desirable properties. First, it should encourage the selection of uncorrelated models. Secondly, it should be computationally tractable so that the objective function \eqref{eq:split_objective} can be minimized in a stable and timely manner. Finally, for moderate values of $G$ the ensemble tends to be sparse in the sense that the set of predictors that appear in at least one of the models will be much smaller than the complete set of candidate predictors. 
	
	For the sparsity penalty we use the elastic net penalty in \eqref{eq:sparsity_penalty}, but other penalties such as SCAD or MCP could be used as well. For the diversity penalty, which encourages the individual models to be sufficiently different, we use the proposal of \cite{christidis2020}, 
	\begin{equation}
		P_d(\bbet^h, \bbet^g) = \sum_{j=1}^{p} |\beta_j^g| |\beta_j^h|.
		\label{eq:diversity_penalty}
	\end{equation}
	This penalty directly measures the overlap between coefficient vectors of different models. When two models both include the same variable $j$ with non-zero coefficients, the penalty adds a cost proportional to the product $|\beta_j^g||\beta_j^h|$. Larger coefficient magnitudes in shared variables incur greater penalties, creating a strong incentive for models to utilize different sets of predictors.
	
	This formulation encourages diversity by forcing models to focus on different aspects of the data: when one model captures a particular signal using certain variables, other models are directed to find alternative representations using different variables. We selected this specific penalty form because it maintains the multi-convex structure of the objective function enabling efficient optimization, operates directly on model coefficients rather than predictions preserving interpretability, and allows for a continuous trade-off between diversity and accuracy through the parameter $\lambda_d$. In high-dimensional settings with correlated predictors, this leads to models that capture complementary information, approaching the prediction problem from different perspectives while sharing variables only when their predictive value sufficiently outweighs the diversity penalty cost.
	
	The tuning constants $ \lambda_s, \lambda_d \geq 0 $ control the amount of shrinkage and diversity between the models, respectively. Letting $\lambda_{d} \to  \infty$, enforces that the diversity penalty $ P_d\left(\bbet^h, \bbet^g\right) \to 0 $ for all $ g \neq h $ so that the active variables in each of the individual models are distinct. On the other hand, it can be seen that for $\lambda_{d}=0$, the solution for all $G$ individual models is the same. In this case all models are equal to the logistic elastic net solution with penalty parameter $\lambda_{s}$, which is then also the split logistic regression ensemble solution. Hence, split logistic regression is a generalization of regularized logistic regression and allows for the selection of $ G > 1$ accurate and potentially diverse models. Note that since both $\lambda_{s}$ and $\lambda_{d}$ are chosen by CV, the degree of sparsity and diversity is driven by the data.
	
	Minimizing the split logistic regression objective function~(\ref{eq:split_objective}) yields solutions $ \hat{f}_g(\bx) = \hbeta_0^g + \bx^T \hbbet^g$ for $g=1,\dots,G$ which are well-suited for creating an ensemble. We use the ensembling function 
	\begin{equation} \label{eq:ensembling_function}
		\hat{f}(\bx) = S\left(\frac{1}{G} \sum_{g=1}^G \hat{f}_g(\bx)\right)=
		S\left(\frac{1}{G}\sum_{g=1}^G \hbeta_0^g + \bx^T \left(\frac{1}{G}\sum_{g=1}^G {\hbbet}^g \right)\right).
	\end{equation}
	The advantage of this ensembling function is that the ensemble also becomes a logistic transformation of a linear function. For any predictor $j$, we denote its ensemble coefficient as $\bar{\beta}_j = \frac{1}{G}\sum_{g=1}^G \hat{\beta}_j^g$, which represents the average effect across all models in the ensemble. Let $\mathcal{M}_g = \{j: \hat{\beta}_j^g \neq 0\}$ denote the set of variables selected for model $g$. To investigate the importance of variables, we can then consider the sets:
	\begin{align} \label{eq:importance_sets}
		\mathcal{A}_k = \left\{j:  \sum_{g=1}^{G} \mathbb{I}\left(j \in \mathcal{M}_g\right) \geq k \right\}, \quad 1 \leq k \leq G,
	\end{align}
	where $\mathcal{A}_G \subseteq \mathcal{A}_{G-1} \subseteq \cdots \subseteq \mathcal{A}_1$. These sets identify variables in order of their importance, as those appearing in multiple models must contribute substantially to the loss function reduction to overcome the diversity penalty.
	
	While the resulting model maintains the structure of a logistic regression, the interpretation of the ensemble coefficients requires careful consideration. As for penalized regression coefficients in general, they exhibit bias due to shrinkage effects. However, they do provide valuable information about the direction and relative magnitude of predictor effects. The combined information—average effect ($\bar{\beta}_j$) and selection frequency through the sets $\mathcal{A}_k$—enhance interpretability compared to black-box ensemble approaches while maintaining prediction accuracy.
	
	For the ensembling function~(\ref{eq:ensembling_function}), consistency of split logistic regression is established in Theorem \ref{thm:consistency} below. The proof of Theorem \ref{thm:consistency} is provided in the supplementary material where we also study its prediction error in the more general case of model misspecification.
	
	\begin{theorem} \label{thm:consistency}
		Assume the data $\left(y_i, \mathbf{x}_i\right)$, $ 1 \leq i \leq n $, follow a logistic regression model for some $ \bbet^* \in \mathbb{R}^p$,  with $ \lVert \bbet^* \rVert_1 $ and $ \lVert \bbet^* \rVert_2 $ of order smaller than $\sqrt{n/\log(p)}$ and $ \log(p)/n  \to 0$.
		Let  $ \hat{f}_1,\dots,\hat{f}_G$ be the solution of (\ref{eq:split_objective}). If $ \lambda_s $ and $\lambda_{d}$ are of order $\sqrt{\log(p)/n}$, then the ensemble prediction $\hat{f}$ given in (\ref{eq:ensembling_function}) is consistent.
	\end{theorem}
	
	The proposed split logistic regression ensemble approach offers several potential empirical benefits compared to existing sparse regression and ensemble methods. By jointly estimating multiple sparse models with an explicit diversity penalty, the method can potentially discover different underlying mechanisms while maintaining the interpretability of each component model. By ensembling at the linear predictor level rather than the prediction level, we preserve coefficient interpretability while potentially improving predictive performance through the accuracy-diversity trade-off. This approach may also identify a broader set of relevant predictors across the ensemble compared to single sparse models. In the subsequent sections, we illustrate these benefits using simulations and cancer genomics datasets, where we demonstrate that the proposed approach achieves strong predictive performance from a small number of models, each individually accurate and interpretable, while potentially revealing multiple biological pathways associated with cancer outcomes, a feature particularly valuable in high-dimensional biomedical applications. \FloatBarrier

	\section{Algorithm} \label{sec:computing}
	
	The difficulty of obtaining a global minimizer of the objective function (\ref{eq:split_objective}) is primarily due to the non-convexity of the diversity penalty $ P_d $.
	Note that a global minimum of the nonnegative objective function (\ref{eq:split_objective}) exists for any $ \lambda_s >0 $ because $\mathcal{J}\left(\beta_0^1,\bbet^1,\dots,\beta_0^G,\bbet^G \right) \to \infty $ if $\Vert \bbet^g \Vert \to \infty $ for any $ 1 \leq g \leq G $.

	To construct an efficient algorithm, we observe that the objective function is multi-convex. That is, the parameters of the objective function can be partitioned in such a way that the problem is convex on each set when the others are kept fixed. A modern  rigorous treatment of multi-convex programming can be found in \cite{shen2017disciplined}. In the case of split logistic regression, the optimization problem (\ref{eq:split_objective}) for the parameters $(\beta_0^g,\bbet^g)$ of a particular model reduces to a penalized logistic regression problem with a weighted elastic net penalty. 
	Indeed, ignoring constant terms, the objective function for $(\beta_0^g,\bbet^g)$ reduces to
	\begin{equation*}
		\mathcal{J}\left(\beta_0^g,\bbet^g\right) = \frac{1}{n}\sum_{i=1}^n \mathcal{L}(\beta_0^g,\bbet^g | y_i, \bx_i )+\lambda_{s}\frac{(1-\alpha)}{2}\Vert\boldsymbol{\beta}^{g}\Vert_{2}%
		^{2}+\sum\limits_{j=1}^{p}|\beta_{j}^{g}|u_{j,g},
		\label{eq:split_objective_fixed}
	\end{equation*}
	where the weights $ u_{j,g} $ in the $\ell_1 $ penalty term are given by $ u_{j,g} = \alpha \lambda_{s}+\lambda_{d}/2
	\sum_{h\neq g}|\beta_{j}^{h}|. $
	For each model the problem thus reduces to a weighted elastic net optimization where the weights in the Lasso penalty depend on the value of the coefficients in the other models. We exploit the multi-convex structure of the objective function to develop a block coordinate descent algorithm~\citep{xu2013block}.

	Recent work in non-convex optimization using block coordinate descent algorithms
	for applications in statistics and machine learning has been very promising,
	see \cite{yang2019inexact} for examples. 
	The key idea is to sequentially update the current estimate for each model using a quadratic approximation $\mathcal{L}_Q$ for the logistic loss function in the objective function. To update the coefficients $(\beta_0^g,\bbet^g)$ of a particular model $g$ we thus need to solve
	\begin{equation}
		\min_{\beta_0^g\in \mathbb{R}, \, \bbet^g \in \mathbb{R}^{p}} \left\{ \frac{1}{n} \sum_{i=1}^n \mathcal{L}_Q\left(\beta_0^g,\bbet^g\mid y_i,\bx_i\right)+ \lambda_{s} P_s(\bbet^g) + \frac{\lambda_{d}}{2} \sum_{\substack{h=1\\ h\neq g}}^{G}P_d\left(\bbet^h,\bbet^g\right)\right\}.
		\label{blockmin}
	\end{equation}
	Using the quadratic approximation $\mathcal{L}_Q$ for the logistic loss function (\ref{eq:loss_function}) derived in the supplementary material, the update for each model is given in the proposition below.
	\begin{prop} \label{prop:coordinate_descent}
		Let $(\tilde{\beta}_0^1, \tilde{\bbet}^1), \dots, (\tilde{\beta}_0^G, \tilde{\bbet}^G)$ denote the current estimates.
		The coordinate descent updates for $\tilde{\beta}_0^g$ and $\tilde{\bbet}^g=(\tilde{\beta}_1^g,\dots,\tilde{\beta}_p^g)^T$  are given by
		\begin{align*}
			\hat{\beta}_0^{g} &= \tilde{\beta}_0^{g} + \frac{\langle \bz - \mathbf{\tilde{p}}^g, \boldsymbol{1}_n \rangle}{\langle \mathbf{\tilde{w}}^g, \boldsymbol{1}_n \rangle},\\
			\hat{\beta}_{j}^{g} &= \frac{\text{Soft}\left(\frac{1}{n}\left({\tilde{r}_j^g} + \tilde{\beta}_j^{g}\langle \bx_j^2, \mathbf{\tilde{w}}^g \rangle\right), \, \alpha   \lambda_{s}+ \frac{\lambda_{d}}{2} \sum_{h\neq g} \vert 
				\tilde{\beta}^{h}_{j}\vert \right)}{\frac{1}{n}\langle \bx_j^2, {\mathbf{\tilde{w}}}^g \rangle +  (1-\alpha) \lambda_s} \quad j=1,\dots,p,
		\end{align*}
		where $ \text{Soft}(\mu, \gamma) = \text{sign}(\mu) \times \text{max}(\lvert \mu \rvert - \gamma)$, $\boldsymbol{1}_n = (1,\dots,1)^T \in \mathbb{R}^n$ and $ \tilde{r}_j^g= \left\langle \bx_j, {\bz} \right\rangle - \langle \bx_j, \mathbf{\tilde{p}}^g \rangle  $. The elements of the $n$-dimensional vectors $\bz$,
		$\mathbf{\tilde{p}}^g$ and $\mathbf{\tilde{w}}^g$ are given by $ z_i= (y_i + 1)/2 $, 
		$\tilde{p}_i^g = S(\tilde{\beta_0}^g + \bx_i^T \boldsymbol{\tilde{\beta}}^g)$ and $\tilde{w}_i^g = \tilde{p}_i^g (1-\tilde{p}_i^g), 1 \leq i \leq n$, respectively.
	\end{prop}
	The algorithm cycles through the components of $ \left(\beta_0^1, \bbet^1\right) $ by applying a single coordinate descent update to each parameter, then through those of $ \left(\beta_0^2, \bbet^2\right) $, and so on until we reach $ \left(\beta_0^G, \bbet^G\right) $. Then, we check for convergence.  
	Convergence is declared when successive estimates of the coefficients in the ensemble model show little difference, i.e. 
	$\max\limits_{0 \leq j \leq p} \vert \tilde{\beta}_j - \hat{\beta}_j  \vert^2 < \delta,$
	for some small tolerance level  $ \delta > 0$, with $\tilde{\beta}_j=\sum_{g=1}^G \tilde{\beta}_{j}^{g}/G$ and $ \hat{\beta}_j=\sum_{g=1}^G \hat{\beta}_{j}^{g}/G$ the  respective estimates for the ensemble model. 
	The algorithm converges to a coordinatewise minimizer of (\ref{eq:split_objective}) by Theorem 4.1 of \citet{tseng2001convergence}. 
	More details of the algorithm are given in the supplementary material.

	To select the tuning parameters we alternate between a grid search for the sparsity penalty and a grid search for the diversity penalty, such that the cross-validated loss of the ensemble classifier is minimized. The  details are available in the supplementary material. By default, $ 10 $-fold CV is used. Note that the value $\lambda_{d}=0$ is included in the grid search for the diversity penalty, such that the (single model)  elastic net is a possible solution of split logistic regression. 
	The warm-start and active-set cycling strategies proposed by \cite{friedman2010regularization} are well suited for the computing algorithm, and have been incorporated to speed up the algorithm.  
	The choice for the ensembling function (\ref{eq:ensembling_function}) also allows the construction of coefficient solution paths for the ensembled model which is illustrated in the supplementary material.  \FloatBarrier
	
	\section{Simulation Study} \label{sec:simulation}
	
	Following the ADEMP framework \cite{morris2019using}, we structure the simulation study to systematically evaluate the proposed method against established alternatives.
	
	\subsection{Aims}
	
	The primary aims of this simulation study are to:
	\begin{itemize}
		\item Assess the predictive performance of split logistic regression compared to established sparse and ensemble methods under varying conditions.
		\item Evaluate variable selection accuracy of the proposed approach.
		\item Investigate how different correlation structures among predictors affect method performance.
		\item Examine the impact of sample size, event probability, sparsity level, non-linear effects, and interaction effects on performance outcomes.
	\end{itemize}

	\subsection{Data-Generating Mechanisms}
	
	We investigate five simulation scenarios based on the logistic regression framework as given in Equation \eqref{eq:logistic_model}. Across scenarios, we systematically vary several aspects of the data-generating process: correlation structure among predictors, correlation levels, sparsity level, sample size, class imbalance, and functional form (linear, interaction effects, and non-linear effects).

	\vspace{0.3cm}
	\noindent
	\textbf{Base Configuration for All Scenarios:}
	\begin{itemize}
		\item Active coefficient values are randomly generated as $(-1)^z u$ where $z \sim \text{Bernoulli}(0.3)$ and $u \sim \text{Uniform}(0, 1/2)$
		\item Predictors follow multivariate normal distribution with mean zero and unit variance
		\item Dimension $p = $ 1,000 with sparsity levels $\zeta \in \{0.1, 0.2, 0.4\}$ (proportion of active variables)
		\item Training sample sizes $n \in \{50, 100\}$ with event probability $\mathbb{P}(Y=1) \in \{0.2, 0.3, 0.4\}$
		\item Test sample size $m = $ 5,000 for performance evaluation
		\item $N = 50$ Monte Carlo replications per configuration
	\end{itemize}
	
	To systematically assess method performance, we consider a range of data-generating mechanisms that span common scenarios encountered in high-dimensional biomedical data. Scenarios 1--3 are designed to explore the effects of different predictor covariance structures—namely, exchangeable correlation (Scenario 1), distinct correlation levels between active and inactive predictors (Scenario 2), and a block-wise (modular) correlation structure often observed in gene expression or genomics data (Scenario 3). To further challenge the methods, Scenario 4 adopts this realistic block-wise structure and incorporates within-block interaction effects, while Scenario 5 similarly uses the block correlation structure but introduces nonlinear (quadratic) effects for a subset of predictors.
	
	\vspace{0.3cm}
	\noindent
	\textbf{Scenario 1}: All predictors have equal pairwise correlation. Data are generated according to:
	\begin{equation*}
		\log\left(\frac{p_i}{1-p_i}\right) = \beta_0 + \bx_{A,i}^{T} \, \bbet_A, \quad 1 \leq i \leq n, 
	\end{equation*}
	where $\bx_{A,i}^{T}$ are the active predictors and $\bbet_A$ the corresponding regression coefficients. All predictors are correlated with each other, with correlation $\rho \in \{0.2, 0.5, 0.8\}$.
	
	\vspace{0.3cm}
	\noindent  
	\textbf{Scenario 2}: Differential correlation between active and inactive predictors. Data follow the same logistic model with correlation $\rho_1$ between active and inactive predictors and $\rho_2$ for all other correlations, where $\rho_1 \in \{0, 0.2, 0.5\}$ and $\rho_2 \in \{0.2, 0.5, 0.8\}$ such that $\rho_1 < \rho_2$.
	
	\vspace{0.3cm}
	\noindent
	\textbf{Scenario 3}: Block structure among active predictors. Data are generated according to:
	\begin{equation*}
		\log\left(\frac{p_i}{1-p_i}\right) = \beta_0 + \sum_{b=1}^{B} \bx_{b,i}^{T} \, \bbet_b, \quad 1 \leq i \leq n,
	\end{equation*}
	where $\bx_{b,i}$ are the predictor variables for block $b$ and $\bbet_b$ are the corresponding regression coefficients. Each block contains 25 predictors, with $B=\zeta p/25$ blocks. Correlation between predictors in different blocks is $\rho_1 \in \{0.2, 0.5\}$, while correlation between predictors in the same block is $\rho_2 \in \{0.5, 0.8\}$, such that $\rho_1 < \rho_2$.
	
	\vspace{0.3cm}
	\noindent
	\textbf{Scenario 4}: Block structure with interaction effects. Data are generated from:
	\begin{equation*}
		\log\left(\frac{p_i}{1-p_i}\right) = \beta_0 + \sum_{b=1}^{B} \bx_{b,i}^{T} \, \bbet_b + \sum_{j,k \in \mathcal{I}} \gamma_{jk} x_{ij} x_{ik}, \quad 1 \leq i \leq n,
	\end{equation*}
	where $\mathcal{I}$ represents a subset of pairs of active predictors selected for interactions. Specifically, we randomly select $\lfloor p\zeta/10 \rfloor$ pairs of active predictors to form interactions. The interaction coefficients $\gamma_{jk}$ are generated using the same mechanism as the main effects: $\gamma_{jk} = (-1)^z u$ where $z \sim \text{Bernoulli}(0.3)$ and $u \sim \text{Uniform}(0, 1/4)$. The correlation structure remains identical to Scenario 3, with blocks of 25 predictors, between-block correlation $\rho_1 \in \{0.2, 0.5\}$, and within-block correlation $\rho_2 \in \{0.5, 0.8\}$.
	
	\vspace{0.3cm}
	\noindent
	\textbf{Scenario 5}: Block structure with non-linear effects. Data are generated from:
	\begin{equation*}
		\log\left(\frac{p_i}{1-p_i}\right) = \beta_0 + \sum_{b=1}^{B} \bx_{b,i}^{T} \, \bbet_b + \sum_{j \in \mathcal{N}} \delta_{j} x_{ij}^2, \quad 1 \leq i \leq n,
	\end{equation*}
	where $\mathcal{N}$ represents a subset of active predictors selected to have additional non-linear effects. We select $\lfloor p\zeta/5 \rfloor$ active predictors to have non-linear components. The non-linear coefficients $\delta_{j}$ are generated as $\delta_{j} = (-1)^z u$ where $z \sim \text{Bernoulli}(0.3)$ and $u \sim \text{Uniform}(0, 1/4)$. The correlation structure remains identical to Scenario 3, with the same blocking and correlation parameters.
	
	\subsection{Estimands}
	
	The primary estimands of interest are:
	\begin{itemize}
		\item The true coefficient vector $\bbet$ (for variable selection performance)
		\item The true conditional class probabilities $p_i = P(Y_i=1|X_i=x_i)$ (for prediction performance)
	\end{itemize}
	
	\subsection{Methods} \label{sec:sim_methods_perf}
	
	We compare the performance of the proposed split logistic regression to a suite of established methods, as detailed in Table \ref{tab:methods}. These competitors are chosen to represent the main families of techniques used for high-dimensional classification.
	
	\begin{table}[htbp]
		\centering
		\caption{Methods and Software Implementations}
		\label{tab:methods}
		\resizebox{\textwidth}{!}{
			\begin{tabular}{lllll}
				\hline
				\textbf{ID} & \textbf{Method} & \textbf{Abbreviation} & \textbf{CRAN Package} & \textbf{CRAN  Reference} \\
				\hline
				1 & Split-Lasso  & Split-Lasso & \texttt{SplitGLM} & \cite{splitglm_package} \\
				2 & Split-Elastic Net  & Split-EN & \texttt{SplitGLM} & \cite{splitglm_package} \\
				3 & Lasso  & Lasso & \texttt{glmnet} & \cite{friedman2010regularization} \\
				4 & Elastic Net  & EN & \texttt{glmnet} & \cite{friedman2010regularization} \\
				5 & Adaptive Lasso  & Adaptive & \texttt{gcdnet} & \cite{gcdnet_package} \\
				6 & Relaxed Lasso  & Relaxed & \texttt{glmnet} & \cite{friedman2010regularization} \\
				7 & Minimum Concave Penalized  & MCP & \texttt{ncvreg} & \cite{ncvreg_package} \\
				8 & Sure Independence Screening & SIS-SCAD & \texttt{SIS} & \cite{SIS_package} \\
				9 & RuleFit & RuleFit & \texttt{xrf}  & \cite{xrf_package} \\
				10 & Random Lasso & RE-Lasso & $-$  & $-$ \\
				11 & Random Elastic Net & RE-EN &  $-$ & $-$ \\
				12 & Random GLM & RGLM & \texttt{RGLM} & \cite{randomGLM_package} \\
				13 & Random Forest & RF & \texttt{ranger} & \cite{ranger_package} \\
				14 & Extreme Gradient Boosting & XGB & \texttt{xgboost} & \cite{xgboost_package} \\
				\hline
			\end{tabular}
		}
	\end{table}
	
	The competitors include standard sparse regularization methods (Lasso, Elastic Net, Adaptive Lasso, MCP, SIS-SCAD, Relaxed Lasso, and RuleFit) and several widely-used ensemble techniques. Methods 10-11 (RE-Lasso and RE-EN) represent a standard randomization-based ensemble approach. This method generates diversity by applying Lasso or Elastic Net to multiple bootstrap samples of the data while also considering only a random subset of features for each base model, providing a direct heuristic-based comparison to our optimization-based approach.
	
	For the simulation study, we use a fixed number of models for the ensembles to ensure a fair comparison of their core performance. For the proposed Split-Lasso and Split-EN, we use $G=10$ models. For RGLM, RE-Lasso, and RE-EN, we use $G=100$ models, and for Random Forest (RF), we use its package default of $G=500$ models, consistent with common practice. For Random Forest variable importance, we implement the unbiased Gini importance measure of \cite{nembrini2018revival} to address known biases when predictors are correlated. All other tuning parameters for all methods are chosen using the default procedures in their respective \texttt{R} packages.
	
	\subsection{Performance measures}
	
	We evaluate methods using the following performance metrics:
	
	\paragraph{Prediction performance metrics:}
	\begin{itemize}
		\item Accuracy (ACC): Proportion of correctly classified observations
		\item Area Under the ROC Curve (AUC): Measure of discriminative ability across thresholds
		\item Sensitivity (SNS): Proportion of true positive cases correctly classified
		\item Specificity (SPC): Proportion of true negative cases correctly classified
		\item Test-sample loss (TSL): Average negative log-likelihood (\ref{eq:loss_function}) on test data
	\end{itemize}
	
	For all classification methods, we use a probability threshold of 0.5 to determine class assignment: observations with predicted probabilities above 0.5 are classified as class 1, and those below as class 0. While this standard threshold enables fair comparison across methods, we note that in practical applications, especially in medical contexts, this threshold could be optimized, e.g. based on the relative costs of false positives versus false negatives. The AUC metric provides a threshold-independent assessment of discriminative performance.
	
	\paragraph{Variable selection performance metrics:}
	\begin{itemize}
		\item Recall (RCL): Proportion of truly active variables identified by the model
		$$\text{RCL} = \frac{\sum_{j=1}^p\mathbb{I}(\beta_j\neq 0, \hat{\beta}_j\neq0)}{\sum_{j=1}^p\mathbb{I}(\beta_j\neq 0)}$$
		\item Precision (PRC): Proportion of selected variables that are truly active
		$$\text{PRC} = \frac{\sum_{j=1}^p\mathbb{I}(\beta_j\neq 0, \hat{\beta}_j\neq0)}{\sum_{j=1}^p\mathbb{I}(\hat{\beta}_j\neq0)}$$
	\end{itemize}
	
	Since RF with its default number of models tends to use all predictors, and XGB builds sequential models on residuals making feature selection interpretation challenging, we do not compute RCL and PRC for these methods.

	\subsection{Results}
	
	Table \ref{tab:prediction_results} presents the average ranks of the prediction performance metrics across all simulation scenarios. The proposed split-ensemble approaches, particularly Split-Elastic Net (Split-EN-10) and Split-Lasso-10, demonstrate remarkable predictive performance across all scenarios. Despite using only 10 base models, these methods consistently outperform ensemble approaches that employ 100 models or more (RE-Lasso-100, RE-EN-100, RGLM-100, and RF-500). The Split-Elastic Net ensemble achieves the best overall performance, with the top average rank in accuracy (ACC), sensitivity (SNS), area under the curve (AUC), and test sample loss (TSL) metrics across nearly all scenarios, making it particularly valuable for class-imbalanced datasets where both overall prediction accuracy and detection of the minority class cases are critical.
	
	\begin{table}[htbp]
		\centering
		\caption{\label{tab:prediction_results} Average ranks of prediction performance metrics across all simulation configurations. Metrics shown are prediction accuracy (ACC), sensitivity (SNS), specificity (SPC), area under the ROC curve (AUC), and test sample loss (TSL). The three best results for each criterion are highlighted in bold.}
		\resizebox{\textwidth}{!}{
			\extrarowsep=2pt
			\begin{tabu}{l lllll lllll lllll}
				\toprule
				& \multicolumn{5}{c}{\textbf{Main Effects}} & \multicolumn{5}{c}{\textbf{Interactions}} & \multicolumn{5}{c}{\textbf{Non-Linear}} \\
				\cmidrule(lr){2-6} \cmidrule(lr){7-11} \cmidrule(lr){12-16}
				\textbf{Method} & \textbf{ACC} & \textbf{SNS} & \textbf{SPC} & \textbf{AUC} & \textbf{TSL} & \textbf{ACC} & \textbf{SNS} & \textbf{SPC} & \textbf{AUC} & \textbf{TSL} & \textbf{ACC} & \textbf{SNS} & \textbf{SPC} & \textbf{AUC} & \textbf{TSL} \\ 
				\cmidrule(lr){1-1} \cmidrule(lr){2-6} \cmidrule(lr){7-11} \cmidrule(lr){12-16}
				\addlinespace[0.25cm]
				Split-Lasso-10 & \textbf{2.68} & \textbf{2.37} & 5.65 & \textbf{3.11} & \textbf{2.00} & \textbf{2.02} & \textbf{2.09} & 5.44 & \textbf{2.19} & \textbf{1.94} & \textbf{2.37} & \textbf{2.33} & 5.75 & \textbf{2.48} & \textbf{2.46} \\ 
				Split-EN-10 & \textbf{2.06} & \textbf{2.14} & 4.84 & \textbf{2.06} & \textbf{1.60} & \textbf{1.72} & \textbf{2.00} & 5.26 & \textbf{1.59} & \textbf{1.67} & \textbf{1.81} & \textbf{1.96} & 5.10 & \textbf{1.43} & \textbf{1.72} \\ 
				\addlinespace[0.25cm]
				Lasso & 7.68 & 8.04 & 9.07 & 8.14 & 6.94 & 7.39 & 7.74 & 9.23 & 7.96 & 6.91 & 7.28 & 7.55 & 9.35 & 8.06 & 7.06 \\ 
				EN & 6.32 & 5.91 & 7.53 & 6.99 & 5.52 & 5.93 & 5.52 & 7.76 & 6.89 & 5.48 & 5.87 & 5.54 & 7.89 & 7.00 & 5.52 \\ 
				Adaptive & 10.91 & 12.47 & \textbf{3.56} & 9.88 & 9.76 & 11.52 & 13.09 & \textbf{3.26} & 10.52 & 10.07 & 11.41 & 13.20 & \textbf{2.81} & 11.15 & 9.98 \\ 
				Relaxed & 8.92 & 7.21 & 11.14 & 9.48 & 11.18 & 8.57 & 6.81 & 11.13 & 9.04 & 10.44 & 8.35 & 6.57 & 11.56 & 9.00 & 10.74 \\ 
				MCP & 12.48 & 11.76 & 12.23 & 12.58 & 11.24 & 11.85 & 11.15 & 11.78 & 12.11 & 11.09 & 11.98 & 11.20 & 11.37 & 12.31 & 11.26 \\ 
				SIS-SCAD & 13.28 & 12.81 & 11.21 & 12.51 & 12.06 & 13.37 & 12.35 & 11.76 & 12.76 & 12.46 & 13.65 & 12.44 & 11.22 & 12.54 & 12.50 \\ 
				RuleFit & 11.54 & 10.74 & 12.09 & 12.90 & 13.48 & 11.72 & 10.43 & 12.13 & 12.78 & 13.69 & 11.85 & 9.94 & 12.48 & 12.81 & 13.85 \\ 
				\addlinespace[0.25cm]
				RE-Lasso-100 & 4.28 & 4.40 & 4.66 & 4.77 & 3.81 & 4.16 & 4.30 & 4.67 & 4.67 & 3.87 & 3.72 & 3.83 & 5.22 & 4.50 & 3.52 \\ 
				RE-EN-100 & \textbf{3.35} & \textbf{3.46} & 4.94 & 3.81 & \textbf{3.41} & \textbf{3.38} & \textbf{3.20} & 5.63 & 3.91 & \textbf{3.56} & \textbf{2.89} & \textbf{2.91} & 6.26 & \textbf{3.44} & \textbf{3.02} \\ 
				RGLM-100 & \textbf{3.67} & 4.35 & \textbf{3.69} & 4.15 & 4.99 & 3.98 & 5.13 & \textbf{3.73} & 4.31 & 4.61 & 4.54 & 6.05 & \textbf{3.20} & 4.72 & 4.89 \\ 
				RF-500 & 7.09 & 9.67 & \textbf{2.67} & 3.15 & 9.04 & 8.35 & 11.54 & \textbf{1.26} & 4.70 & 8.98 & 8.50 & 11.63 & \textbf{1.19} & 4.43 & 8.44 \\ 
				XGB & 10.76 & 9.66 & 11.72 & 11.47 & 9.96 & 11.04 & 9.65 & 11.96 & 11.57 & 10.22 & 10.78 & 9.83 & 11.59 & 11.13 & 10.04 \\ 
				\addlinespace[0.25cm]
				\bottomrule
		\end{tabu}}
	\end{table}
	
	A particularly valuable finding is the consistency of the split-ensemble methods across different data structures. Whether dealing with simple main effects, complex interactions, or non-linear relationships, both Split-EN-10 and Split-Lasso-10 maintain their superior performance. This demonstrates that these methods adapt effectively to various data complexities without requiring scenario-specific adjustments. Importantly, this excellent performance is achieved with just 10 models and without relying on randomization techniques employed by RE-Lasso, RE-EN, RGLM, and RF methods. Instead, split-ensemble methods use a well-formulated objective function that balances model fit with variable selection and diversity in a mathematically sound framework. While RF achieved the highest specificity across all scenarios, this came at a significant cost in terms of sensitivity. In contrast, the split-ensemble methods maintained competitive specificity rankings while excelling in sensitivity measures, offering a more balanced classification approach.
	
\begin{figure}[htbp]
		\centering
		\includegraphics[width=15.5cm]{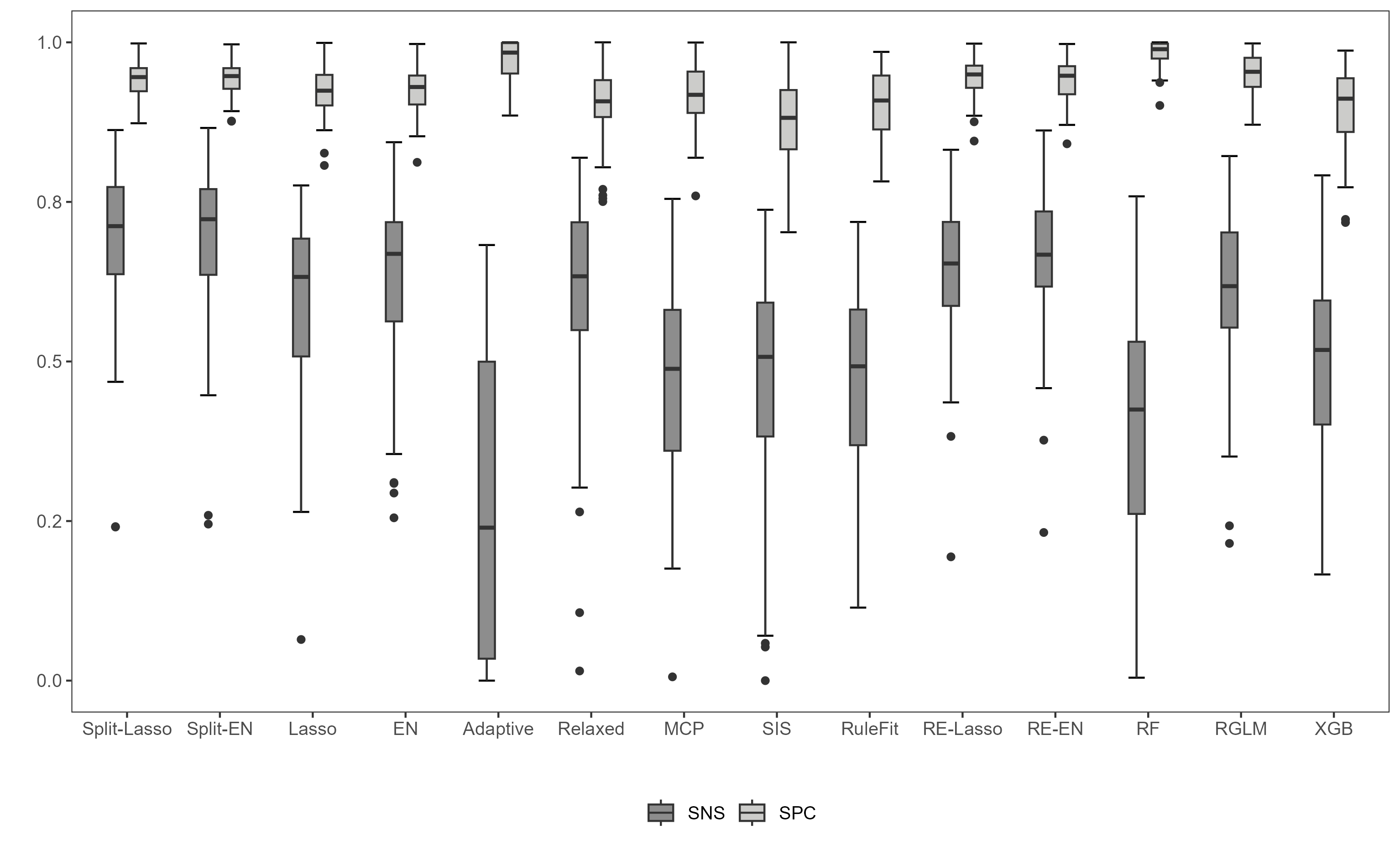}
		\caption{Sensitivity (SNS) and specificity (SPC) of sparse and ensemble classification methods over $ N=50 $ random training sets under Scenario 3 with $ \rho_1=0.2 $, $ \rho_2=0.8 $, $ p=\text{1,000} $, $ n=50 $, $ \mathbb{P}(Y=1)=0.3 $ and $\zeta=0.2$.}
		\label{fig:SplitGLM_Simulation}
	\end{figure}

	Figure \ref{fig:SplitGLM_Simulation} illustrates SNS and SPC performance across 50 random training sets under challenging high-dimensional conditions with class imbalance. The split-ensemble methods demonstrate an impressive balance of both metrics, maintaining high sensitivity while preserving competitive specificity. Despite using only 10 base models, Split-Lasso-10 and Split-EN-10 consistently outperform more complex methods. Notably, the closest competitors, RE-Lasso-100 and RE-EN-100 with 100 base models each, show lower sensitivity distributions with their 75th percentiles roughly corresponding to the split methods' median performance. While RF and Adaptive Lasso achieve high specificity, they sacrifice sensitivity considerably. The proposed split-ensemble approaches offer the best overall classification balance in this challenging scenario.

	Table \ref{tab:var_selection} presents the average ranks of the variable selection performance metrics across all simulation scenarios. The results reveal a clear trade-off between RCL and PRC among the methods. While RE-EN-100 and RE-Lasso-100 achieve the highest recall, they do so at a significant expense of precision, ranking near the bottom (10-11th place) for PRC across all scenarios. This suggests that these methods tend toward excessive variable inclusion when using 100 base models. In contrast, Split-EN-10 demonstrates an impressive balance, ranking among the top three methods for recall while maintaining mid-range precision performance, all with just 10 models. SIS-SCAD, MCP, and RuleFit excel in precision but perform poorly in recall, indicating overly conservative variable selection. Split-Lasso-10 shows competitive recall comparable to traditional sparse methods while offering better precision than randomization-based ensemble methods that use 10 times more base models. This indicates that Split-Lasso-10 is an effective approach for both predictive and interpretable modeling across various data complexity scenarios.
	
	\begin{table}[htbp]
		\centering
		\caption{\label{tab:var_selection} Average ranks of variable selection performance metrics across all simulation configurations. Metrics shown are recall (RCL) and precision (PRC). The three best results for each criterion are highlighted in bold.}
		\extrarowsep=2pt
		\begin{tabu}{l ll ll ll}
			\toprule
			& \multicolumn{2}{c}{\textbf{Main Effects}} & \multicolumn{2}{c}{\textbf{Interactions}} & \multicolumn{2}{c}{\textbf{Non-Linear}} \\
			\cmidrule(lr){2-3} \cmidrule(lr){4-5} \cmidrule(lr){6-7}
			\textbf{Method} & \textbf{RCL} & \textbf{PRC} & \textbf{RCL} & \textbf{PRC} & \textbf{RCL} & \textbf{PRC} \\ 
			\cmidrule(lr){1-1} \cmidrule(lr){2-3} \cmidrule(lr){4-5} \cmidrule(lr){6-7}
			\addlinespace[0.25cm]
			Split-Lasso-10 & 4.19 & 6.36 & 4.26 & 6.48 & 4.11 & 5.85 \\ 
			Split-EN-10 & \textbf{2.67} & 7.23 & \textbf{2.74} & 7.50 & \textbf{2.74} & 7.11 \\ 
			\addlinespace[0.25cm]
			Lasso & 8.11 & 6.64 & 8.11 & 7.26 & 8.11 & 7.72 \\ 
			EN & 6.69 & 6.12 & 6.93 & 5.39 & 6.93 & 5.61 \\ 
			Adaptive & 9.00 & 7.30 & 9.16 & 7.91 & 9.37 & 8.28 \\ 
			Relaxed & 9.81 & 4.90 & 9.73 & 5.35 & 9.52 & 6.06 \\ 
			MCP & 11.42 & \textbf{4.20} & 11.33 & \textbf{3.97} & 11.40 & \textbf{3.81} \\ 
			SIS-SCAD & 11.58 & \textbf{3.12} & 11.67 & \textbf{1.53} & 11.60 & \textbf{1.57} \\ 
			RuleFit & 6.39 & \textbf{3.30} & 6.07 & \textbf{3.61} & 6.07 & \textbf{2.98} \\ 
			\addlinespace[0.25cm]
			RE-Lasso-100 & \textbf{2.48} & 10.50 & \textbf{2.39} & 10.72 & \textbf{2.35} & 10.81 \\ 
			RE-EN-100 & \textbf{1.01} & 11.23 & \textbf{1.00} & 11.09 & \textbf{1.00} & 11.11 \\ 
			RGLM-100 & 4.65 & 7.09 & 4.61 & 7.19 & 4.80 & 7.07 \\ 
			RF-500 & $-$ & $-$ & $-$ & $-$ & $-$ & $-$ \\ 
			XGB & $-$ & $-$ & $-$ & $-$ & $-$ & $-$ \\ 
			\addlinespace[0.25cm]
			\bottomrule
		\end{tabu}
	\end{table}
	
\begin{figure}[htbp]
		\centering
		\includegraphics[width=15.5cm]{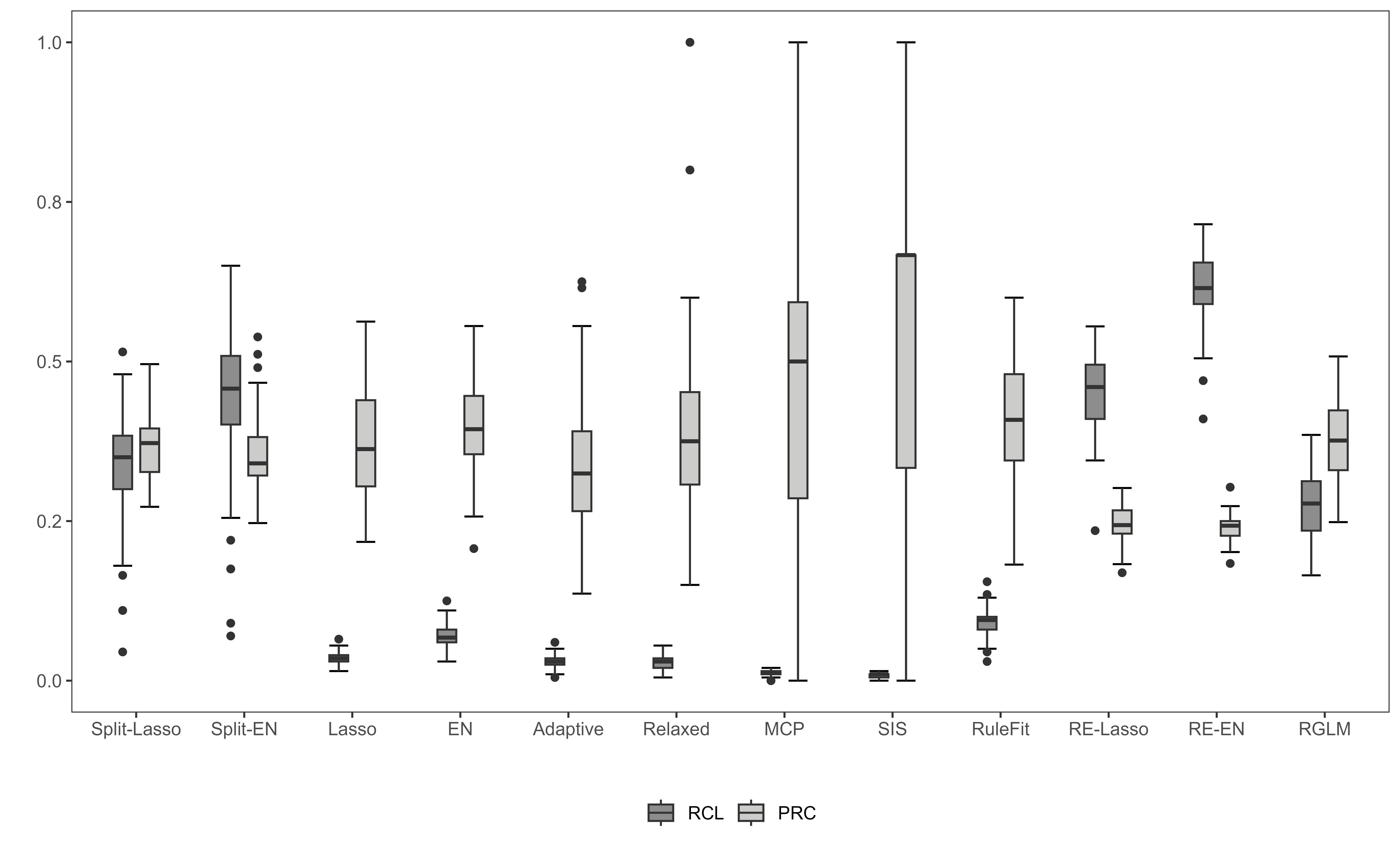}
		\caption{Recall (RCL) and precision (PRC) of sparse and split classification methods over $ N=50 $ random training sets under Scenario 3 with $ \rho_1=0.2 $, $ \rho_2=0.8 $, $ p=\text{1,000} $, $ n=50 $, $ \mathbb{P}(Y=1)=0.3 $ and $\zeta=0.2$.}
		\label{fig:SplitGLM_Simulation_RCPR}
	\end{figure}

	Figure \ref{fig:SplitGLM_Simulation_RCPR} further illustrates the variable selection performance trade-offs under challenging high-dimensional conditions. RE-Lasso-100 and RE-EN-100 achieve high recall but at a severe cost to precision, indicating they include numerous irrelevant variables. In contrast, Split-EN-10 and Split-Lasso-10 demonstrate a more balanced profile, maintaining competitive recall with substantially better precision than the randomization-based ensembles, despite using only 10 models. The split-ensemble methods' favorable position in this recall-precision trade-off space highlights their ability to balance identifying true signals while limiting false discoveries, making them particularly valuable when both accurate prediction and meaningful variable selection are required. \FloatBarrier
	
	\section{Medical Genomics Data Applications} \label{section:gene_expression}
	
	To demonstrate the practical utility and competitive performance of split logistic regression, we apply the method and its competitors to medical genomics datasets from the Gene Expression Omnibus (GEO) database. We present detailed analyses of two motivating examples to highlight the method's capabilities in both prediction and biomarker discovery. The first is an in-depth case study using an unpaired dataset (lung cancer) that fully aligns with the models' statistical assumptions. The second examines performance on a paired-sample dataset (thyroid cancer) to evaluate the method in a common and more complex practical setting.
	
	\subsection{Lung Cancer Analysis: Primary Validation on Independent Samples}
	
	Our primary motivating example uses the lung cancer dataset GSE10245, which consists of $n=58$ independent non-small cell lung cancer (NSCLC) tissue samples of subtypes adenocarcinoma (AC, $n=40$) or squamous cell carcinoma (SCC, $n=18$). The data are pre-processed using a standard four-step procedure similar to \cite{dudoit2002comparison}: (1) thresholding of expression levels, (2) filtering genes with low expression ratios, (3) filtering genes with low expression differences, and (4) a base-2 logarithmic transformation. From the remaining genes, we retain the $p=500$ with the smallest $q$-values \citep{storey2002direct} for model training. We use the same suite of methods detailed in the simulation study, including the proposed Split-Lasso and Split-Elastic Net (Split-EN) with $G=10$ models, standard sparse methods, and several ensemble methods (see Table \ref{tab:methods}). Performance is evaluated using prediction accuracy (ACC), average individual model accuracy ($\overline{\text{ACC}}$), sensitivity (SNS), and specificity (SPC), estimated over $N=50$ random splits of the dataset. For Random Forest, we use the unbiased Gini importance measure of \cite{nembrini2018revival}.
	
	The analysis provides a clear demonstration of the method's strengths. As shown in Table \ref{tab:lung_results}, Split-Lasso and Split-EN achieve excellent ensemble prediction accuracy of 0.93 with only $G=10$ models. Remarkably, the individual models within these ensembles maintain high accuracy ($\overline{\text{ACC}}$ of 0.90 and 0.92, respectively), nearly matching their single-model counterparts despite the diversity penalty deliberately distributing important genes across different models. This efficiency is highlighted by the sparsity of individual models; Split-EN models contained on average only 24 genes, compared to 38 for standard EN. The split methods show strong balance between sensitivity (0.81) and specificity (0.98-0.99), demonstrating effective classification for both majority (AC) and minority classes.
	
	\begin{table}[htbp]
		\centering
		\caption{\label{tab:lung_results} Average prediction accuracy (ACC), average individual model accuracy ($\overline{\text{ACC}}$), sensitivity (SNS), and specificity (SPC) for the lung cancer dataset (GSE10245). Performance is estimated using $N=50$ random splits into training sets with $n=29$ samples and test sets with the remaining samples. Standard errors are in parenthesis. The three best results for each criterion are highlighted in bold.}
		\extrarowsep=2pt
		\begin{tabular}{lcccc}
			\toprule
			\textbf{Method} & \textbf{ACC} & $\mathbf{\overline{ACC}}$ & \textbf{SNS} & \textbf{SPC} \\
			\midrule
			\addlinespace[0.25cm]
			Split-Lasso-10 & \textbf{0.93 (0.05)} & \textbf{0.90 (0.03)} & \textbf{0.81 (0.15)} & \textbf{0.99 (0.02)} \\
			Split-EN-10 & \textbf{0.93 (0.04)} & \textbf{0.92 (0.04)} & \textbf{0.81 (0.14)} & 0.98 (0.02) \\
			\addlinespace[0.25cm]
			Lasso & 0.91 (0.06) & $-$ & 0.76 (0.17) & 0.98 (0.03) \\
			EN & \textbf{0.92 (0.05)} & $-$ & \textbf{0.80 (0.14)} & \textbf{0.99 (0.02)} \\
			Adaptive & 0.82 (0.10) & $-$ & 0.52 (0.30) & 0.96 (0.05) \\
			Relaxed & 0.90 (0.06) & $-$ & 0.79 (0.15) & 0.96 (0.05) \\
			MCP & 0.86 (0.09) & $-$ & 0.66 (0.18) & 0.95 (0.06) \\
			SIS-SCAD & 0.85 (0.08) & $-$ & 0.62 (0.15) & 0.96 (0.06) \\
			RuleFit & 0.81 (0.09) & $-$ & 0.61 (0.21) & 0.91 (0.09) \\
			\addlinespace[0.25cm]
			RE-Lasso-10 & 0.91 (0.05) & 0.89 (0.04) & 0.76 (0.15) & \textbf{0.99 (0.02)} \\
			RE-Lasso-100 & \textbf{0.92 (0.05)} & 0.88 (0.04) & 0.77 (0.16) & \textbf{0.99 (0.02)} \\
			RE-EN-10 & \textbf{0.92 (0.05)} & \textbf{0.91 (0.04)} & \textbf{0.79 (0.16)} & \textbf{0.99 (0.02)} \\
			RE-EN-100 & \textbf{0.92 (0.05)} & \textbf{0.91 (0.04)} & \textbf{0.79 (0.15)} & \textbf{0.99 (0.02)} \\
			\addlinespace[0.25cm]
			RGLM-10 & 0.90 (0.07) & 0.82 (0.04) & 0.74 (0.18) & 0.98 (0.03) \\
			RGLM-100 & 0.91 (0.05) & 0.82 (0.03) & 0.76 (0.16) & \textbf{0.99 (0.02)} \\
			\addlinespace[0.25cm]
			RF-10 & 0.88 (0.08) & 0.76 (0.04) & 0.71 (0.19) & 0.96 (0.05) \\
			RF-500 & 0.90 (0.07) & 0.77 (0.03) & 0.72 (0.19) & \textbf{1.00 (0.01)} \\
			\addlinespace[0.25cm]
			XGB & 0.81 (0.08) & 0.76 (0.07) & 0.64 (0.21) & 0.90 (0.09) \\
			\addlinespace[0.25cm]
			\bottomrule
		\end{tabular}
	\end{table}
	
\begin{figure}[htbp]
		\centering
		\includegraphics[width=13cm]{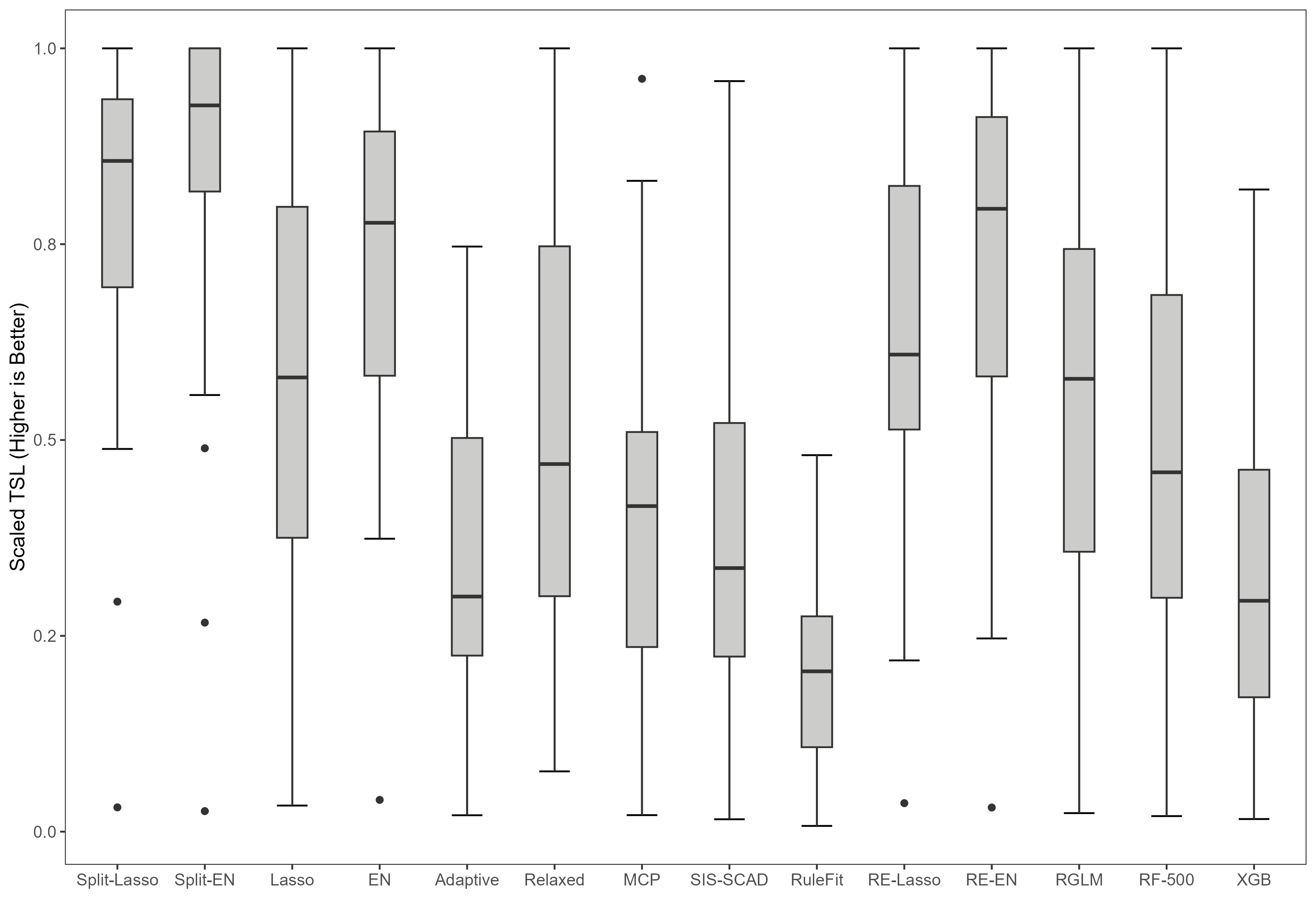}
		\caption{Comparison of model performance using scaled test sample loss (TSL), calculated as the minimum TSL on a given split divided by each method's TSL. Higher values indicate better performance, with 1.0 representing the best probability calibration on the test data. \label{fig:scaled_tsl}}
	\end{figure}

	Figure~\ref{fig:scaled_tsl} presents a comparative assessment of predictive performance using scaled test sample loss (TSL), which is essentially the deviance on test data. This dataset presents a particularly challenging case due to its class imbalance (SCC comprising only 18 samples), making accurate probability estimation crucial. For each replication, we computed the scaled TSL as:
	\begin{equation*}
		\text{Scaled TSL} = \frac{\min(\text{TSL})}{\text{TSL}}
	\end{equation*}
	where $\min(\text{TSL})$ represents the minimum test sample loss achieved by any method on that specific random training-test split. This scaling results in values between 0 and 1, where higher values represent superior performance. The Split-Lasso and Split-EN methods demonstrate remarkable efficiency, consistently achieving optimal or near-optimal performance despite leveraging only $G = 10$ models. Their distributions are tightly concentrated near 1, indicating these methods either achieved the lowest TSL or came remarkably close across replications. In contrast, the ensembles based on randomization exhibited greater variability and often produced less reliable probability estimates, as evidenced by their substantially lower scaled TSL values. Split-logistic regression's ability to maintain high-quality probability calibration with minimal ensemble complexity represents a significant advantage for implementation in clinical contexts, particularly for imbalanced datasets like this one.
	
	Beyond prediction accuracy, a key goal in genomics is identifying important biomarkers. We examine the variable selection capabilities by analyzing gene selection frequencies across the $N=50$ random data splits. Table \ref{tab:split_en_genes} presents the top 10 genes most frequently selected by Split-EN-10, along with their selection frequencies in other methods. The analysis reveals that Split-EN consistently selects clinically relevant biomarkers often overlooked by other methods. For example, PTGFRN, a gene potentially linked to lung cancer metastasis \citep{aguila2019ig, marquez2024prostaglandin}, was selected in 94\% of Split-EN models but only 38\% of standard EN models. Similarly, TMC5, a critical marker for differentiating SCC from AC subtypes \citep{xiao2017eight}, appeared in 86\% of Split-EN models but was included in only 22\% of EN models. 
	
	\begin{table}[htbp]
		\centering
		\caption{Top 10 genes most frequently selected by Split-EN-10 across $N=50$ random splits of lung cancer data, and their selection frequencies in other methods. Bold values indicate Split-EN-10 selection frequencies used to rank the top 10 genes.}
		\label{tab:split_en_genes}
		\begin{tabular}{lccccc}
			\toprule
			\textbf{Gene} & \textbf{Split-EN-10} & \textbf{EN} & \textbf{RE-EN-10} & \textbf{RE-EN-100} & \textbf{RF-500} \\
			\midrule
			CGN & \textbf{1.00} & 0.98 & 1.00 & 1.00 & 0.90 \\
			PTGFRN & \textbf{0.94} & 0.38 & 0.78 & 1.00 & 0.76 \\
			MISP & \textbf{0.90} & 0.08 & 0.52 & 0.96 & 0.32 \\
			C11orf54 & \textbf{0.88} & 0.18 & 0.54 & 0.90 & 0.48 \\
			SLC6A8 & \textbf{0.88} & 0.54 & 0.86 & 0.92 & 0.78 \\
			LACTB2 & \textbf{0.86} & 0.18 & 0.46 & 0.84 & 0.36 \\
			TMC5 & \textbf{0.86} & 0.22 & 0.68 & 0.88 & 0.52 \\
			CSRP2 & \textbf{0.86} & 0.72 & 0.84 & 0.90 & 0.66 \\
			LPCAT1 & \textbf{0.86} & 0.04 & 0.50 & 0.90 & 0.60 \\
			CCDC68 & \textbf{0.84} & 0.36 & 0.66 & 0.86 & 0.34 \\
			\bottomrule
		\end{tabular}
	\end{table}
	
	It is important to interpret the high selection frequencies of the RE-EN-100 method with caution. By constructing 100 models, this approach tends to select a very large number of unique predictors across the full ensemble. As was systematically observed in the simulation study across all scenarios (see Table \ref{tab:var_selection}), this behavior leads to very high recall at the cost of poor precision. This can inflate the appearance of good variable selection in the real-data analysis, as many genes may be included by chance in at least one model without contributing meaningfully to the overall prediction. In contrast, Split-EN-10 identifies these key biomarkers with high frequency using only a tenth of the models, indicating a more targeted and efficient selection process where each included gene carries more weight in the final decision.

	Table \ref{tab:rf_genes} shows that Split-EN-10 also maintains impressively high selection frequencies for genes identified as important by RF-500, demonstrating the method's ability to capture diverse signals in the data.
	
	\begin{table}[htbp]
		\centering
		\caption{Top 10 genes most frequently selected by RF-500 across $N=50$ random splits of lung cancer data, and their selection frequencies in other methods. Bold values indicate RF-500  selection frequencies used to rank the top 10 genes.}
		\label{tab:rf_genes}
		\begin{tabular}{lccccc}
			\toprule
			\textbf{Gene} & \textbf{Split-EN-10} & \textbf{EN} & \textbf{RE-EN-10} & \textbf{RE-EN-100} & \textbf{RF-500} \\
			\midrule
			CGN & 1.00 & 0.98 & 1.00 & 1.00 & \textbf{0.90} \\
			CD55 & 0.62 & 0.30 & 0.70 & 0.92 & \textbf{0.84} \\
			SLC6A8 & 0.88 & 0.54 & 0.86 & 0.92 & \textbf{0.78} \\
			TMEM125 & 0.74 & 0.42 & 0.76 & 0.98 & \textbf{0.78} \\
			PTGFRN & 0.94 & 0.38 & 0.78 & 1.00 & \textbf{0.76} \\
			AGR2 & 0.76 & 0.30 & 0.62 & 0.84 & \textbf{0.68} \\
			CSRP2 & 0.86 & 0.72 & 0.84 & 0.90 & \textbf{0.66} \\
			SLC16A1 & 0.66 & 0.32 & 0.74 & 0.82 & \textbf{0.66} \\
			PRSS8 & 0.42 & 0.02 & 0.30 & 0.74 & \textbf{0.64} \\
			\bottomrule
		\end{tabular}
		
	\end{table}
	
	\subsection{Thyroid Cancer Analysis: Performance on Paired-Sample Data}
	
	The second example analyzes the thyroid cancer dataset GSE5364, which contains $n=51$ tissue samples comprising 16 thyroid cancer tumors and 35 adjacent normal tissues from the same patients. This dataset features a paired-sample design, which is ubiquitous in genomics research. While methods specifically designed for paired data, such as conditional logistic regression or mixed-effects models, would be a natural choice, the goal here is to evaluate the relative performance of our proposed method and its competitors in this common practical setting where standard classification tools are often applied. This analysis provides valuable insight into the method's behavior when the independence assumption is not perfectly met.
	
	As shown in Table \ref{tab:thyroid_results}, the split-ensemble methods continue to demonstrate competitive performance in this challenging setting. Split-Lasso and Split-EN both achieve a high accuracy of 0.90, outperforming most competitors while maintaining excellent balance between sensitivity (0.81-0.82) and specificity (0.94). This suggests that even when core statistical assumptions are violated, the proposed framework remains a powerful tool for building predictive models and its relative performance remains strong.
	
	\begin{table}[htbp]
		\centering
		\caption{\label{tab:thyroid_results} Average prediction accuracy (ACC), average individual model accuracy ($\overline{\text{ACC}}$), sensitivity (SNS), and specificity (SPC) for the thyroid cancer dataset (GSE5364). Performance is estimated using $N=50$ random splits into training sets with $n=26$ samples and test sets with the remaining samples. Standard errors are in parenthesis. The three best results for each criterion are highlighted in bold.}
		\extrarowsep=2pt
		\begin{tabular}{lcccc}
			\toprule
			\textbf{Method} & \textbf{ACC} & $\mathbf{\overline{ACC}}$ & \textbf{SNS} & \textbf{SPC} \\
			\midrule
			\addlinespace[0.25cm]
			Split-Lasso-10 & \textbf{0.90 (0.05)} & \textbf{0.85 (0.05)} & \textbf{0.81 (0.11)} & \textbf{0.94 (0.08)} \\
			Split-EN-10 & \textbf{0.90 (0.05)} & \textbf{0.86 (0.05)} & \textbf{0.82 (0.10)} & \textbf{0.94 (0.08)} \\
			\addlinespace[0.25cm]
			Lasso & 0.86 (0.08) & $-$ & 0.77 (0.18) & 0.91 (0.10) \\
			EN & 0.88 (0.07) & $-$ & 0.79 (0.16) & 0.93 (0.09) \\
			Adaptive & 0.87 (0.10) & $-$ & 0.71 (0.27) & \textbf{0.94 (0.11)} \\
			Relaxed & 0.84 (0.09) & $-$ & 0.77 (0.17) & 0.88 (0.12) \\
			MCP & 0.82 (0.11) & $-$ & 0.66 (0.24) & 0.90 (0.12) \\
			SIS-SCAD & 0.82 (0.11) & $-$ & 0.67 (0.23) & 0.89 (0.11) \\
			RuleFit & 0.85 (0.09) & $-$ & 0.70 (0.22) & 0.93 (0.07) \\
			\addlinespace[0.25cm]
			RE-Lasso-10 & 0.89 (0.07) & 0.83 (0.05) & \textbf{0.81 (0.13)} & 0.93 (0.09) \\
			RE-Lasso-100 & \textbf{0.90 (0.08)} & 0.82 (0.05) & 0.79 (0.19) & \textbf{0.95 (0.07)} \\
			RE-EN-10 & 0.89 (0.06) & \textbf{0.86 (0.06)} & \textbf{0.83 (0.11)} & 0.93 (0.09) \\
			RE-EN-100 & \textbf{0.90 (0.06)} & \textbf{0.86 (0.05)} & \textbf{0.84 (0.10)} & 0.93 (0.08) \\
			\addlinespace[0.25cm]
			RGLM-10 & 0.89 (0.07) & 0.76 (0.05) & \textbf{0.82 (0.12)} & 0.92 (0.09) \\
			RGLM-100 & \textbf{0.90 (0.06)} & 0.76 (0.03) & \textbf{0.82 (0.11)} & \textbf{0.94 (0.09)} \\
			\addlinespace[0.25cm]
			RF-10 & 0.87 (0.07) & 0.73 (0.05) & 0.74 (0.17) & \textbf{0.94 (0.07)} \\
			RF-500 & \textbf{0.91 (0.07)} & 0.73 (0.03) & 0.79 (0.16) & \textbf{0.97 (0.06)} \\
			\addlinespace[0.25cm]
			XGB & 0.85 (0.09) & 0.79 (0.09) & 0.72 (0.21) & 0.91 (0.08) \\
			\addlinespace[0.25cm]
			\bottomrule
		\end{tabular}
		
	\end{table}
	
	Split logistic regression provides a natural framework for ranking genes by importance using the sets $\mathcal{A}_k$ defined in \eqref{eq:importance_sets}, where genes appearing in more individual models are considered more important. In this exploratory paired-sample analysis, we found that this ranking system provides powerful biological insights. Applying this to the thyroid cancer dataset with cross-validated values for $\lambda_s$ and $\lambda_d$ on the full dataset, the set $\mathcal{A}_4$ (genes appearing in at least 4 of 10 models) consistently contained four genes with known biological significance: TRPC1 \citep{asghar2015transient}, APOD \citep{huang2001gene}, F11R \citep{czubak2022f11}, and SPON2 \citep{tang2023comprehensive}. Remarkably, three of these were selected in fewer than 10\% of standard EN models, highlighting the unique advantage of the proposed method in uncovering multiple, potentially independent biological pathways. This built-in ranking system provides a data-driven tool to help researchers prioritize genes for further investigation.
	
	In addition to the detailed case studies, we performed a comprehensive benchmark across ten genomics datasets to assess the generalizability of our method's performance. The complete results of this large-scale comparison are presented in the Supplementary Material. The benchmark intentionally includes datasets with both independent and paired-sample designs, offering a comprehensive evaluation of relative model performance across varied experimental conditions found in practice. The findings, detailed in the supplement, show that Split-EN consistently ranks among the top methods, validating its efficiency and high accuracy across a wide range of genomic applications. \FloatBarrier

	\section{The Number of Models} \label{sec:div_measures}
	
	Constructing  accurate and diverse models for an ensemble are opposite objectives \citep{krogh1995neural}. 
	We perform  an empirical study to explore this accuracy-diversity trade-off for split logistic regression, which will drive the choice for the number of models in real medical genomics data applications. 
	
	The analysis of genomics data  via high-throughput technologies has generated the need for classification algorithms that can handle high-dimensional data containing correlated predictors (genes) within different pathways or networks, see  \cite{yousefi2011multiple} and \cite{zhang2012sources} for example. In light of this, to investigate the accuracy-diversity trade-off of split logistic regression, we use the high-dimensional block correlation setting of Scenario 3 in Section~\ref{sec:simulation} with configuration parameters $ (n, p)=(50,\, $1,000), $ (\rho_1 , \rho_2) = (0.2, 0.5)$, $ \zeta \in \{0.1,0.2,0.4\}$ and $\mathbb{P}(Y=1)= 0.2$. 
	
	To quantify diversity for ensemble classifiers, we adopt the entropy diversity measure of \cite{kuncheva2003measures}. Given an ensemble comprised of $ G $ individual classifiers, the entropy measure (EM) for a given $ \bx $ is defined as 
	\begin{align}
		\text{EM}(\bx) = \frac{1}{G - \ceil{G/2}} \, \min\left(\ell(\bx), G - \ell(\bx)\right),
	\end{align}
	where $\ell(\bx)$ denotes the numbers of individual classifiers in the ensemble that correctly classify $ \bx $. The entropy measure ranges between 0 and 1, with $ \text{EM}(\bx)=0 $ and $ \text{EM}(\bx)=1$ corresponding to no diversity   and to the highest possible diversity between the individual classifiers, respectively. The overall entropy $\mathbb{E}[\text{EM}(\bx)]$ can be estimated by averaging $\text{EM}(\bx)$ over a test set.
	
	\subsection{Results}

	Table \ref{tab:div_measures} shows the evolution of the ensemble prediction accuracy (ACC), the average prediction accuracy of the individual models ($\overline{\text{ACC}}$)
	and the entropy diversity measure (EM) averaged over the test sets, as a function of the number of models for split logistic regression. It also contains the overlap (OVP) between the individual models in the ensemble, defined as
	\begin{align*}
		\text{OVP} = \frac{\sum\limits_{j=1}^{p} o_{j} \mathbb{I}\lbrace \ o_j \neq 
			0\rbrace}{\sum\limits_{j=1}^{p} \mathbb{I}\lbrace o_j \neq 0\rbrace}, \quad o_j = \frac{1}{G}\sum_{g=1}^{G} \mathbb{I}\lbrace \hbeta_{j}^{g} \neq 0\rbrace.
	\end{align*} 
	It can be seen that in all three settings $ \overline{\text{ACC}} $ decreases and EM increases with the number of models, while  the ensemble ACC increases. Hence, as the number of models increases, the accuracy of the individual models has less impact on the ensemble ACC compared to their level of diversity. Split logistic regression manages to achieve a proper balance for this trade-off, resulting in a high ACC for the ensemble.
	\begin{table}[htbp]
		\centering
		\caption{Ensemble prediction accuracy (ACC), average individual model accuracy (${\overline{\text{ACC}}}$), entropy diversity measure (EM), and variable overlap (OVP) as a function of the number of models ($G$). Results are averaged over test sets under Scenario 3 with $ \rho_1=0.2 $, $ \rho_2=0.5 $, $ p=\text{1,000} $, $ n=50 $, $ \mathbb{P}(Y=1)=0.2 $, and sparsity $\zeta \in \{0.1, 0.2, 0.4\}$. \label{tab:div_measures}} 
		\resizebox{\textwidth}{!}{
			\extrarowsep =2pt
			\begin{tabular}{lllllllllllll}
				\toprule
				& \multicolumn{4}{c}{$\mathbf{\boldsymbol{\zeta}=0.1}$} & \multicolumn{4}{c}{$\mathbf{\boldsymbol{\zeta}=0.2}$} & \multicolumn{4}{c}{$\mathbf{\boldsymbol{\zeta}=0.4}$} \\ \cmidrule(lr){1-1} \cmidrule(lr){2-5} \cmidrule(lr){6-9} \cmidrule(lr){10-13}  
				$\mathbf{G}$ & \textbf{ACC} & $\mathbf{\overline{ACC}}$ & \textbf{EM} & \textbf{OVP} & \textbf{ACC} & $\mathbf{\overline{ACC}}$ & \textbf{EM} & \textbf{OVP} & \textbf{ACC} & $\mathbf{\overline{ACC}}$ & \textbf{EM} & \textbf{OVP} \\ \cmidrule(lr){1-1} \cmidrule(lr){2-5} \cmidrule(lr){6-9} \cmidrule(lr){10-13} 
				\addlinespace[0.25cm]
				1 & 0.86 & $-$ & $-$ & $-$ & 0.87 & $-$ & $-$ & $-$ & 0.87 & $-$ & $-$ & $-$ \\ 
				2 & 0.87 & 0.86 & 0.04 & 0.81 & 0.89 & 0.88 & 0.05 & 0.76 & 0.88 & 0.88 & 0.05 & 0.73 \\ 
				5 & 0.87 & 0.86 & 0.10 & 0.53 & 0.89 & 0.88 & 0.12 & 0.44 & 0.89 & 0.88 & 0.11 & 0.42 \\ 
				7 & 0.87 & 0.86 & 0.10 & 0.47 & 0.90 & 0.88 & 0.14 & 0.32 & 0.89 & 0.88 & 0.11 & 0.39 \\ 
				10 & 0.87 & 0.85 & 0.13 & 0.29 & 0.90 & 0.87 & 0.16 & 0.17 & 0.90 & 0.87 & 0.14 & 0.19 \\ 
				15 & 0.87 & 0.84 & 0.18 & 0.20 & 0.90 & 0.86 & 0.20 & 0.12 & 0.90 & 0.87 & 0.18 & 0.07 \\ 
				20 & 0.87 & 0.84 & 0.19 & 0.12 & 0.90 & 0.86 & 0.21 & 0.08 & 0.90 & 0.86 & 0.19 & 0.05 \\ 
				25 & 0.87 & 0.83 & 0.20 & 0.13 & 0.90 & 0.85 & 0.24 & 0.04 & 0.90 & 0.86 & 0.21 & 0.04 \\ 
				\addlinespace[0.25cm]
				\bottomrule
		\end{tabular}}
	\end{table}

	When $ G $ is small, the average  ${\overline{\text{ACC}}}$ of the individual models in Table \ref{tab:div_measures} is close to the ACC of the logistic elastic net (the case $G=1$ in Table \ref{tab:div_measures}). 
	The choice of diversity tuning parameter $\lambda_{d}$ is driven by the data based on a CV criterion. For a small number of models, a smaller value of $\lambda_{d}$ is selected such that the OVP of the models is large and they share a lot of important predictors. This results in accurate individual models but  a relatively low diversity as seen from the EM values.
	As the number of models increases, the OVP becomes smaller, resulting in individual models that have a higher average EM. Indeed, for a large number of models it becomes beneficial to increase diversity between the models to decrease the misclassification rate of the ensemble. 
	In this case the diversity penalty in split logistic regression reduces the overlap between individual models, leading to high diversity which results in high classification accuracy. 
	In summary, split logistic regression thus successfully achieves the proper balance between individual model accuracy and diversity regardless of the number of models. Alternative diversity measures are considered in the supplementary material and lead to the same conclusions.
	
	\subsection{Computational Cost}
	
	Table \ref{tab:div_measures} indicates that a larger number of models results in an ensemble with higher prediction accuracy. However, ACC stabilizes quickly, so there is a diminishing returns type of behavior in terms of prediction accuracy versus computational cost. Indeed, we also ran split logistic regression using $ G=50 $ models, but in all cases there is hardly any improvement in ACC compared to the ensemble with $ G=25 $ models shown in Table \ref{tab:div_measures}. 		  
	In fact, with $ G=25 $ models split logistic regression already achieves nearly full diversity (OV $\approx 1/G $), so little gain is expected by increasing the number of models further while computation time does grow. 
	
	Table \ref{tab:cpu} shows the average computation time (in CPU seconds) across all sparsity levels of scenario 3 as a function of the number of models. As $ G $ increases there is a price to pay in average individual  ${\overline{\text{ACC}}}$ which may not be compensated by a higher ensemble ACC.	
	The computation time seems to depend linearly on the number of models and is approximately given by $ 0.27 + 0.15 \times G $ for $ G \geq 2 $. 
	In real data applications such as the gene expression applications in the previous section it is generally a good strategy to use a data-driven choice of the number of models in the ensemble by increasing $G$ until the CV performance has stabilized.
	
	\begin{table}[htbp]
		\centering
		\caption{\label{tab:cpu}Computation time of \texttt{R} function call for split logistic regression in CPU seconds for varying number of models, using multithreading (5 threads). CPU seconds are on a 2.1 GHz Intel Xeon Platinum 8468 processor in a machine running  CentOS Linux 7.9 with 32 GB of RAM.} 
		\extrarowsep=2pt
		\begin{tabu}{lrrrrrrr}
			\toprule
			$ \mathbf{G} $    & 2    & 5     & 7     & 10    & 15    & 20    & 25    \\ \hline
			$ \textbf{Time} $ & 0.55 & 1.05 & 1.39 & 1.76 & 2.53 & 3.40 & 4.06 \\
			\bottomrule
		\end{tabu}
	\end{table}
	
	\FloatBarrier

	\section{Discussion and Future Directions} \label{sec:discussion}

	We presented a new approach to learn a diverse ensemble of sparse logistic regression models that is well suited for high-dimensional medical genomics data. The individual models for the ensemble are learned simultaneously by optimizing an objective function which balances between individual model strength and diversity between the models. 
	The sparsity penalty in the objective function controls the stability of the individual models while the diversity penalty favorably exploits the accuracy-diversity trade-off to achieve excellent performance for the resulting ensemble. In contrast to other popular ensemble methods, split logistic regression models remain logistic regression models and thus are highly interpretable.  Moreover, the individual models in the ensemble may be of interest in their own right because they each provide a relationship between the predictor genes and disease status that can provide insight to very complex biological mechanisms.

	In detailed analyses of lung and thyroid cancer datasets, split logistic regression achieved state-of-the-art prediction accuracy. The case studies demonstrated the method's ability to identify unique, clinically-relevant biomarkers that were missed by competing methods, while simultaneously capturing key markers consistently found by other approaches. A variable ranking method native to the split-modeling framework was also shown to be effective for prioritizing genes for further investigation. These strong results were further validated in a comprehensive benchmark study across ten diverse datasets, detailed in the supplementary material, where the proposed method consistently ranked among the top performers.

	While this manuscript focuses on microarray gene expression data, which has been a canonical example of the $p \gg n$ scenario in statistical genomics for decades, the proposed split logistic regression approach is equally applicable to other types of high-dimensional omics data. Modern technologies like RNA-seq, single-cell sequencing, proteomics, and metabolomics all generate data with similar dimensional characteristics. For instance, single-cell RNA sequencing data can be transformed into pseudobulk profiles representing cell populations, methylation data can be summarized at the gene or region level, and proteomics datasets frequently contain thousands of measured proteins across limited samples. The common thread across these data types is that they all present the fundamental statistical challenge of extracting meaningful signals from thousands of features measured on relatively few subjects, precisely the scenario where the proposed method demonstrates advantages over traditional approaches.
	
	Due to the diversity penalty, split logistic regression makes use of different groups of variables in the individual models to build an ensemble. Allowing interactions among predictors can be beneficial to further improve the prediction performance of classifiers. Since split logistic regression can have much higher recall than single-model methods such as Lasso and the elastic net, the proposed methodology can also be useful to detect interaction effects that would be missed by such single-model methods. This is important for example for genomics data applications where it is known that gene interaction effects are common. The diversity penalty can also be combined more generally with different sparsity penalties such as the group Lasso \citep{group} for categorical variables or the fused Lasso \citep{tibshirani2005sparsity} for data exhibiting spatial or temporal structures. Furthermore, our analysis of paired-sample data highlights a valuable direction for future research: extending the split-modeling framework to explicitly account for data dependencies, for instance by incorporating it into mixed-effects models or conditional logistic regression frameworks.

	Block coordinate descent is an effective approach to solve the multi-convex optimization problem underlying split logistic regression. Multi-convex programming is an emerging field in optimization with many applications in statistics and machine learning, see e.g. \cite{shen2017disciplined} and \cite{pardalos2017non}. In future research we will investigate whether alternative approaches can further decrease the computational cost of the method. 
	
	In split logistic regression the models are ensembled at the level of the linear predictors. This guarantees high interpretability of the ensemble model, but is not necessarily optimal from a prediction point of view. In future research it will be examined whether alternative ensembling functions can improve on the prediction accuracy of the ensemble.  
	
	Ensemble methods are very popular to analyze small sample data with a large number of predictor variables, and the proposed method provides a framework to build an optimal classification ensemble model. Similarly to logistic regression, the general split modeling framework could be applied to multi-class classification problems to obtain a powerful ensemble classifier. The split modeling framework could also be extended to generalized linear models in general. \FloatBarrier

	\section*{Acknowledgments}
	
	Part of this work was conducted while Anthony-Alexander Christidis was a UBC Doctoral Researcher at KU Leuven's Department of Mathematics under a Mitacs Globalink Research Award.
	
	\section*{Funding}
	
	Funding was provided by a Mitacs Globalink Research Award and through graduate assistantships at the University of British Columbia.
	
	\section*{Supplementary Material}
	
	The \texttt{R}/\texttt{C++} package \texttt{SplitGLM}  along with its reference manual is publicly available on CRAN at \url{https://CRAN.R-project.org/package=SplitGLM}. The data and scripts to replicate the numerical experiments are available at \url{https://doi.org/10.5281/zenodo.15588653}.
	
	\FloatBarrier
	
	\section*{Appendix}
	
	The Appendix is organized as follows:
	\begin{itemize}
		\item \textbf{Appendix \ref{app:asymptotics}: Ensemble Asymptotics.} Provides a detailed proof of the asymptotic properties of the proposed split logistic regression ensemble, establishing consistency under mild regularity conditions.
		
		\item \textbf{Appendix \ref{app:algorithms}: Details of the Algorithm.} Presents the derivation of the quadratic approximation for the logistic loss, followed by detailed pseudocode for the block coordinate descent algorithm (Algorithm \ref{alg:splitglm_algo_supp}) and the alternating grid search for tuning parameter selection (Algorithm \ref{alg:split_algo_CV_supp}).
		
		\item \textbf{Appendix \ref{app:diversity}: Alternative Diversity Measures.} Complements the analysis in the main paper by evaluating the accuracy-diversity trade-off using several alternative diversity measures, with results presented in Table \ref{tab:alternative_diversity}.
		
		\item \textbf{Appendix \ref{app:results_genomics}: Benchmark Across Ten Genomics Datasets.} Details the setup and summary results for the comprehensive benchmark study. Table \ref{tab:gene_description_supp} describes the datasets, while Tables \ref{tab:gene_ranks_supp} and \ref{tab:gene_ranks_summary_supp} summarize the performance rankings.
		
		\item \textbf{Appendix \ref{app:full_sim_results}: Full Results of Simulation Study.} Contains the complete, unabridged results for the simulation study. The tables are organized by scenario as described in the main paper:
		\begin{itemize}
			\item Scenario 1 (Main Effects, Exchangeable Correlation): Tables \ref{tab:sim_pred_1_50_0.2} -- \ref{tab:sim_sel_1_100_0.8}.
			\item Scenario 2 (Main Effects, Differential Correlation): Tables  \ref{tab:sim_pred_2_50_0.5_0.2} -- \ref{tab:sim_sel_2_100_0.8_0.5}.
			\item Scenario 3 (Main Effects, Block Correlation): Tables \ref{tab:sim_pred_3_50_0.5_0.2} -- \ref{tab:sim_sel_3_100_0.8_0.5}.
			\item Scenario 4 (Interactions, Block Correlation): Tables \ref{tab:sim_pred_4_50_0.5_0.2} -- \ref{tab:sim_sel_4_100_0.8_0.5}.
			\item Scenario 5 (Non-Linear Effects, Block Correlation): Tables \ref{tab:sim_pred_5_50_0.5_0.2} -- \ref{tab:sim_sel_5_100_0.8_0.5}.
		\end{itemize}
		
		\item \textbf{Appendix \ref{app:full_results_genomics}: Full Results for Medical Genomics Data.} Provides the detailed performance tables for each of the ten genomics datasets evaluated in the benchmark study. Full results are presented in Tables \ref{tab:results_GSE20347} -- \ref{tab:results_GSE14905}.
	\end{itemize}
	
	\FloatBarrier

	\appendix
	\section{Ensemble Asymptotics} \label{app:asymptotics}
	
	In this section, we prove a general result for the asymptotic behavior of the prediction error of the ensemble split regression method and show that it implies consistency of the prediction under the assumptions stated in Theorem 1 of the article.
	
	\subsection{Preliminaries} \label{sec:conistency_preliminaries}
	
	Consider data $ \{y_i, \bx_i\}_{i=1}^n $ where $ \bx_i \in \mathbb{R}^p $ and $y_i \in \{-1,1\}$ for $ i=1,2,\dots,n $. Without loss of generality, we assume each column of the design matrix $ \bX \in \mathbb{R}^{n \times p}$ has been scaled by its maximum value such that $ \underset{1 \leq j \leq p}{\max} \lVert \bx_{\cdot j} \rVert_{\infty} \leq 1 $ where $ \bx_{\cdot j} $ is the $ j $-th column of $ \bX $. Let $\mathcal{H}$ be a (rich) parameter space that includes the of space linear functions, and for each $ f \in \mathcal{H} $ we take the (convex) logistic loss function $\mathcal{L}: \mathcal{H} \times \mathbb{R}^p \times \{-1,1\} \mapsto \mathbb{R}$ as defined in the main article, 
	\begin{equation} \label{eq:loss_function}
		\mathcal{L}(f(\bx_i), y_i) =  \log (1+ e^{-y_i f(\bx_i)}).
	\end{equation}
	We denote the \textit{empirical risk} by
	\begin{align*}
		\mathcal{V}_n (f) = \frac{1}{n}\sum_{i=1}^n \mathcal{L}( f(\bx_i), y_i),
	\end{align*}
	such that $\mathcal{V}_n$ is the \textit{empirical measure} that puts mass $ 1/n $ for each observation $ (y_i, \bx_i) $, and the \textit{expected risk} by 
	\begin{align*}
		\mathcal{V}(f) =\frac{1}{n} \sum_{i=1}^n \mathbb{E}\left[\mathcal{L}( f(\bx_i), y_i)\right]
	\end{align*}
	We also denote the \textit{target function} 
	\begin{align*}
		f^* = \argmin_{f \in \mathcal{H}} \mathcal{V}(f)
	\end{align*}
	as minimizer of the expected risk. For any $f \in \mathcal{H}$, the \textit{excess risk} is given by 
	\begin{align*}
		\mathcal{E}(f) = \mathcal{V}(f) - \mathcal{V}(f^*),
	\end{align*}
	where by definition $ \mathcal{E}(f) \geq 0 $ for all $f \in \mathcal{H}$. In the case of model misspecification where the target function $ f^* $ is not necessarily linear, we define the linear subspace $\mathcal{H}_{\bbet} = \{f_{\bbet} : \bbet \in \mathbb{R}^{p+1} \} \subset \mathcal{H}$, where the map $ \bbet \mapsto f_{\bbet} $ is linear. As in Section 6.6 of \cite{buhlmann-book} we consider the notationally simpler case without intercept. We denote the best linear approximation of the target function $ f^* $ by
	\begin{align*}
		f_{\bbet^*} = \argmin_{f \in \mathcal{H}_{\bbet}} \mathcal{V}(f),
	\end{align*}
	We define the \textit{empirical process} for the linear subspace as
	\begin{align*}
		\left\{\mathcal{P}_n(f_{\bbet}) = \mathcal{V}_n(f_{\bbet}) - \mathcal{V}(f_{\bbet}): f_{\bbet} \in \mathcal{H}_{\bbet}\right\}.
	\end{align*}
	For a fixed (and arbitrary) $ f_{\tbbet}$, we define
	\begin{align} \label{eq:Z_M}
		Z_M = \sup_{\lVert {\bbet} - \tbbet \rVert_1 \leq M} \bigg\lvert \mathcal{P}_n({f}_{\bbet})  - \mathcal{P}_n({f}_{\tbbet})\bigg\rvert
	\end{align}
	and the set 
	\begin{align*}
		\mathcal{B} = \{Z_M \leq \lambda_0 M \},
	\end{align*}
	where
	\begin{align*}
		\lambda_0 &= 4{T}\left(n, p\right) + \frac{t}{3n} + \sqrt{\frac{2t}{n}} \sqrt{1 +8T(n, p) },\\
		T(n, p) &= \sqrt{\frac{2 \log(2p)}{n}} + \frac{\log(2p)}{3n}.
	\end{align*}
	By Lemma 14.20 and Theorem 14.5 of \cite{buhlmann-book}, we have the probability inequality
	\begin{align*}
		\mathbb{P}(\mathcal{B}) \geq 1 - \exp(-t),
	\end{align*}
	i.e. for some $ M $ sufficiently small
	\begin{align*}
		\mathbb{P}(Z_M \leq \lambda_0 M) \geq 1 - \eta
	\end{align*}
	for some $\lambda_0$ that depends on the sample size $ n $, the dimensionality of the data $ p $, and the confidence level $ 1 - \eta $.
	
	We denote the total sparsity and diversity penalties of the split logistic regression parameters for any set of $ G $ linear functions $ f_{\bbet^1}, \dots, f_{\bbet^G} $ by
	\begin{align} \label{eq:sparsity_penalty}
		P(f_{\bbet^1}, \dots, f_{\bbet^G}) = \sum_{g=1}^G P_s\left(\bbet^g\right)= \sum_{g=1}^G  \left[ \frac{1-\alpha}{2} \lVert \bbet^g \rVert_2^2 + \alpha \lVert \bbet^g\rVert_1 \right]
	\end{align}
	where $ \alpha \in [0,1] $, and
	\begin{align}\label{eq:diversity_penalty}
		Q(f_{\bbet^1}, \dots, f_{\bbet^G}) = \sum_{h\neq g} P_d\left(\bbet^g, \bbet^h\right)= \sum_{h\neq g}  \sum_{j=1}^{p} \lvert \beta_j^g\rvert \lvert \beta_j^h\rvert,
	\end{align} 
	respectively.	
	
	\subsection{Ensemble Consistency} \label{sec:conistency_statement}
	
	Let the solution to the split logistic regression objective function be the collection of functions
	{\small\begin{equation} 
			\left({f}_{\hbbet^1},\dots,{f}_{\hbbet^G}\right) = \argmin_{{f}_{\bbet^1},\dots,{f}_{\bbet^G } \in \mathcal{H}_{\bbet}} \left\{\sum_{g=1}^{G}  \left[ \frac{1}{n}\sum_{i=1}^n \mathcal{L}\left(f_{\bbet^g}(\bx_i), y_i\right)+ \lambda_{s} P_s(\bbet^g) \right] + \frac{\lambda_{d}}{2} \sum_{h\neq g} P_d(\bbet^h,\bbet^g)\right\},
			\label{eq:split_objective}
	\end{equation}}
	and let 
	{\small\begin{equation} 
			({f}_{\tbbet^1},\dots,{f}_{\tbbet^G}) = \argmin_{{f}_{\bbet^1},\dots,{f}_{\bbet^G} \in \mathcal{H}_{\bbet}}  \left\{ \sum_{g=1}^{G} \left[ \mathcal{E}(f_{\bbet^g})+ \lambda_{s} P_s(\bbet^g) \right] + \frac{\lambda_{d}}{2} \sum_{h\neq g} P_d(\bbet^h,\bbet^g)\right\},
			\label{eq:split_excess}
	\end{equation}}
	For $ Z_M $ taken as \eqref{eq:Z_M} using the solution from \eqref{eq:split_excess}, define
	\begin{align} \label{eq:M_definition}
		\widetilde{M} =  \frac{1}{G\lambda_0}\left[\sum_{g=1}^G\mathcal{E}({f}_{\tbbet^g}) + 2 \lambda_{s} P({f}_{\tbbet^1}, \dots, {f}_{\tbbet^G}) + \frac{\lambda_{d}}{2} Q({f}_{\tbbet^1}, \dots, {f}_{\tbbet^G}) \right].
	\end{align}
	Let the set
	\begin{align}
		\mathcal{\widetilde{B}} = \left\{ Z_{\widetilde{M}} \leq \lambda_0 \widetilde{M}  \right\}
	\end{align}
	where  $ \lambda_{s} \geq 4G\lambda_0/\alpha $ if $\alpha \in (0,1]$ and $ \lambda_{s} \geq 8G^3\lambda_0 /\widetilde{M} $ if $\alpha=0$. Then, we will prove below that on the set $\mathcal{\widetilde{B}}$, it holds that
	{\small
		\begin{align} \label{eq:consistency_inequality}
			\mathcal{V}\left(\frac{1}{G}\sum_{g=1}^G{f}_{\hbbet^g}\right) - \mathcal{V}(f^*) \leq  2 \left[  \mathcal{E}({f}_{\bbet^*}) 
			+ 2\alpha \lambda_s   \lVert \bbet^* \rVert_1 + \frac{1-\alpha}{2} \lambda_{s} \lVert \bbet^* \rVert_2^2 + \frac{\lambda_{d}(G-1)}{2} \lVert \bbet^* \rVert_2^2 \right].
	\end{align}}
	Hence, if the target is linear, i.e. $ f^* = f_{\bbet^*} $, then it holds that
	\begin{align} \label{eq:consistency_inequality_linear}
		\mathcal{V}\left(\frac{1}{G}\sum_{g=1}^G{f}_{\hbbet^g}\right) - \mathcal{V}(f^*) \leq  4\alpha \lambda_s   \lVert \bbet^* \rVert_1 + {(1-\alpha)} \lambda_{s} \lVert \bbet^* \rVert_2^2 + {\lambda_{d}(G-1)} \lVert \bbet^* \rVert_2^2 .
	\end{align}
	Therefore, if the data come from a logistic model it follows that if we take $\lambda_{s}$ and $\lambda_{d}$ to be or order $\sqrt{log(p)/n}$, and we assume that $ \lVert \bbet^* \rVert_1 $ and $ \lVert \bbet^* \rVert_2^2 $ are of order smaller than $\sqrt{n/log(p)}$ and $ log(p)/n \to 0 $, then the ensemble prediction $ (1/G)\sum_{g=1}^G{f}_{\hbbet^g} $ is consistent. In the more general case of model misspecification ($ f^* \neq f_{\bbet^*} $),  the prediction error converges to $2 \, \mathcal{E}(f_{\bbet^*})$.
	
	\subsection{Ensemble Consistency Proof} \label{sec:consistency_proof}

	Let ${f}_{\hbbet^1}, \dots, {f}_{\hbbet^G}$ be the solution to split logistic regression with $ G $ groups for data $ \{y_i, \bx_i\}_{i=1}^n $. Then, for any $ f_{\bbet^1}, \dots, f_{\bbet^G} \in \mathcal{H}_{\bbet}$ it holds that
	\begin{align*}
		&\sum_{g=1}^G \mathcal{V}_n( {f}_{\hbbet^g}) +  \lambda_s P({f}_{\hbbet^1}, \dots, {f}_{\hbbet^G}) + \frac{\lambda_{d}}{2} Q({f}_{\hbbet^1}, \dots, {f}_{\hbbet^G})
		\\
		\leq &\sum_{g=1}^G \mathcal{V}_n( f_{\bbet^g})+  \lambda_s P(f_{\bbet^1}, \dots, f_{\bbet^G}) + \frac{\lambda_{d}}{2} Q(f_{\bbet^1}, \dots, f_{\bbet^G}).
	\end{align*}
	Note that if $ \bbbet^g= t\hbbet^g + (1-t)\tbbet^g $  for any $ t \in [0,1] $, by a convexity argument
	{\begin{align*}
			&\sum_{g=1}^G \mathcal{V}_n(f_{\bbbet^g}) + \lambda_{s} P(f_{\bbbet^1}, \dots, f_{\bbbet^G})\\
			\leq &t \left[\sum_{g=1}^G \mathcal{V}_n({f}_{\hbbet^g}) + \lambda_{s} P({f}_{\hbbet^1}, \dots, {f}_{\hbbet^G}) \right]\\
			&+ (1-t) \left[\sum_{g=1}^G \mathcal{V}_n(f_{\tbbet^g}) + \lambda_{s} P(f_{\tbbet^1}, \dots, f_{\tbbet^G}) \right] \\
			\leq &t \left[\sum_{g=1}^G \mathcal{V}_n({f}_{\hbbet^g}) + \lambda_{s} P({f}_{\hbbet^1}, \dots, {f}_{\hbbet^G}) + \frac{\lambda_{d}}{2} Q({f}_{\hbbet^1}, \dots, {f}_{\hbbet^G})  \right]
			\\
			&+ (1-t) \left[\sum_{g=1}^G \mathcal{V}_n(f_{\tbbet^g}) + \lambda_{s} P(f_{\tbbet^1}, \dots, f_{\tbbet^G}) + \frac{\lambda_{d}}{2} Q(f_{\tbbet^1}, \dots, f_{\tbbet^G}) \right] \\
			\leq & \sum_{g=1}^G \mathcal{V}_n(f_{\tbbet^g}) + \lambda_{s} P(f_{\tbbet^1}, \dots, f_{\tbbet^G}) + \frac{\lambda_{d}}{2} Q(f_{\tbbet^1}, \dots, f_{\tbbet^G}).
	\end{align*}}
	We can write
	\begin{align*}
		&\sum_{g=1}^G\mathcal{E}(f_{\bbbet^g}) + \lambda_{s} P(f_{\bbbet^1}, \dots, f_{\bbbet^G}) \\
		= & \sum_{g=1}^G\mathcal{E}(f_{\bbbet^g}) + \lambda_{s} P(f_{\bbbet^1}, \dots, f_{\bbbet^G})\\
		& + \left[\sum_{g=1}^G \mathcal{V}_n(f_{\bbbet^g}) 
		- \sum_{g=1}^G \mathcal{V}_n(f_{\bbbet^g})\right]+ \left[\sum_{g=1}^G \mathcal{V}_n(f_{\tbbet^g}) - \sum_{g=1}^G \mathcal{V}_n(f_{\tbbet^g})\right]  \\
		& + \left[\sum_{g=1}^G \mathcal{V}(f_{\tbbet^g}) - \sum_{g=1}^G \mathcal{V}(f_{\tbbet^g}) \right] + \left[{\lambda_{s}} P(f_{\tbbet^1}, \dots, f_{\tbbet^G}) - \lambda_{s} P(f_{\tbbet^1}, \dots, f_{\tbbet^G}) \right] \\ &+\left[\frac{\lambda_{d}}{2} Q(f_{\tbbet^1}, \dots, f_{\tbbet^G}) - \frac{\lambda_{d}}{2} Q(f_{\tbbet^1}, \dots, f_{\tbbet^G}) \right] \\
		= &-\left[\sum_{g=1}^G \mathcal{P}_n(f_{\bbbet^g}) - \sum_{g=1}^G \mathcal{P}_n(f_{\tbbet^g})\right] + \sum_{g=1}^G \mathcal{E}(f_{\tbbet^g})\\
		&+ \left[\left(\sum_{g=1}^G \mathcal{V}_n(f_{\bbbet^g}) + \lambda_{s} P(f_{\bbbet^1}, \dots, f_{\bbbet^G})\right) - \left( \sum_{g=1}^G \mathcal{V}_n(f_{\tbbet^g}) + \lambda_{s} P(f_{\tbbet^1}, \dots, f_{\tbbet^G}) + \frac{\lambda_{d}}{2} Q(f_{\tbbet^1}, \dots, f_{\tbbet^G})\right)\right] \\
		&+ \lambda_{s} P(f_{\tbbet^1}, \dots, f_{\tbbet^G}) + \frac{\lambda_{d}}{2} Q(f_{\tbbet^1}, \dots, f_{\tbbet^G}).
	\end{align*}
	Thus we get the basic inequality 
	\begin{align*}
		&\sum_{g=1}^G\mathcal{E}(f_{\bbbet^g}) + \lambda_{s} P(f_{\bbbet^1}, \dots, f_{\bbbet^G}) \\
		\leq &-\left[\sum_{g=1}^G \mathcal{P}_n(f_{\bbbet^g}) - \sum_{g=1}^G \mathcal{P}_n(f_{\tbbet^g})\right] + \sum_{g=1}^G \mathcal{E}(f_{\tbbet^g})+ \lambda_{s} P(f_{\tbbet^1}, \dots, f_{\tbbet^G}) + \frac{\lambda_{d}}{2} Q(f_{\tbbet^1}, \dots, f_{\tbbet^G})\\
		= &-\sum_{g=1}^G\left[ \mathcal{P}_n(f_{\bbbet^g}) -  \mathcal{P}_n(f_{\tbbet^g})\right] + \sum_{g=1}^G \mathcal{E}(f_{\tbbet^g})+ \lambda_{s} P(f_{\tbbet^1}, \dots, f_{\tbbet^G}) + \frac{\lambda_{d}}{2} Q(f_{\tbbet^1}, \dots, f_{\tbbet^G}).
	\end{align*}
	In other words, to bound the sum of the excess risk $ \sum_{g=1}^G \mathcal{E}(f_{\bbbet^g})  $ we need to control the sum of the increments of the empirical processes  $ \mathcal{P}_n( f_{\bbbet^g})  -  \mathcal{P}_n(f_{\tbbet^g}) $,  $ 1 \leq g \leq G $.
	
	Let 
	\begin{align*}
		t^g = \frac{\widetilde{M}}{\widetilde{M} + \lVert {\hbbet^g} - \tbbet^g \rVert_1}.
	\end{align*}
	If $ {\bbbet}^{g} = t^g \hbbet^g + (1-t^g)\tbbet^g $, then $ \lVert \tbbet^g - \bbbet^g \rVert_1 \leq \widetilde{M}$. 
	Then on the set $\mathcal{\widetilde{B}}$,
	\begin{align*}
		&\sum_{g=1}^G \mathcal{E}(f_{\bbbet^g}) +  \lambda_s P(f_{\bbbet^1}, \dots, f_{\bbbet^G}) \\
		\leq &\sum_{g=1}^G Z_{\widetilde{M}}  + \sum_{g=1}^G \mathcal{E}(f_{\tbbet^g}) 
		+ \lambda_s P(f_{\tbbet^1}, \dots, f_{\tbbet^G}) + \frac{\lambda_{d}}{2} Q(f_{\tbbet^1}, \dots, f_{\tbbet^G}) \\
		= &\sum_{g=1}^G \lambda_0 \widetilde{M}  + \sum_{g=1}^G \mathcal{E}(f_{\tbbet^g}) 
		+ \lambda_s P(f_{\tbbet^1}, \dots, f_{\tbbet^G}) + \frac{\lambda_{d}}{2} Q(f_{\tbbet^1}, \dots, f_{\tbbet^G}).
	\end{align*}
	
	For the case $\alpha \in (0,1]$ we obtain
	\begin{align*}
		&\sum_{g=1}^G \mathcal{E}(f_{\bbbet^g}) +  \lambda_s P(f_{\bbbet^1}, \dots, f_{\bbbet^G}) + \lambda_s P(f_{\tbbet^1}, \dots, f_{\tbbet^G})\\
		\leq &G \lambda_0 \widetilde{M}  + \sum_{g=1}^G \mathcal{E}(f_{\tbbet^g}) 
		+ 2\lambda_s P(f_{\tbbet^1}, \dots, f_{\tbbet^G}) + \frac{\lambda_{d}}{2} Q(f_{\tbbet^1}, \dots, f_{\tbbet^G}) \\
		= & 2G \lambda_0 \widetilde{M} \leq \alpha \lambda_s  \frac{ \widetilde{M} }{2 }
	\end{align*}
	since $  \lambda_{s} \geq 4G\lambda_0/\alpha $. Notice that 
	\begin{align*}
		&\sum_{g=1}^G \mathcal{E}(f_{\bbbet^g}) +  \lambda_s P(f_{\bbbet^1}, \dots, f_{\bbbet^G}) + \lambda_s P(f_{\tbbet^1}, \dots, f_{\tbbet^G})\\
		\geq & \sum_{g=1}^G \mathcal{E}(f_{\bbbet^g}) +  \lambda_s \alpha \sum_{g=1}^G \Vert 
		\bbbet^g -\tbbet^{g}\Vert_{1} + \lambda_s \frac{1-\alpha}{2} \sum_{g=1}^G \Vert 
		\bbbet^g -\tbbet^{g}\Vert_{2}^2\\
		\geq & \sum_{g=1}^G \mathcal{E}(f_{\bbbet^g}) +  \lambda_s \alpha \sum_{g=1}^G \Vert 
		\bbbet^g -\tbbet^g\Vert_{1}.
	\end{align*}
	This implies
	\begin{align*}
		\lVert 
		\bbbet^g -\tbbet^g\rVert_{1} \leq \frac{\widetilde{M}}{2}
	\end{align*}
	for all $ 1 \leq g \leq G $, which in turn implies
	\begin{align*}
		\lVert
		\hbbet^g -\tbbet^g\rVert_{1} \leq \widetilde{M}
	\end{align*}
	for all  $ 1 \leq g \leq G $. 
	
	In the case $\alpha=0$, we obtain
	\begin{align*}
		&\sum_{g=1}^G \mathcal{E}(f_{\bbbet^g}) +  \lambda_s P(f_{\bbbet^1}, \dots, f_{\bbbet^G}) + \lambda_s P(f_{\tbbet^1}, \dots, f_{\tbbet^G})\\
		\leq &G \lambda_0 \widetilde{M}  + \sum_{g=1}^G \mathcal{E}(f_{\tbbet^g}) 
		+ 2\lambda_s P(f_{\tbbet^1}, \dots, f_{\tbbet^G}) + \frac{\lambda_{d}}{2} Q(f_{\tbbet^1}, \dots, f_{\tbbet^G}) \\
		= & 2G \lambda_0 \widetilde{M} \leq  \frac{\lambda_s}{G^2}  \frac{ \widetilde{M}^2 }{4},
	\end{align*}
	since $ \lambda_{s} \geq 8G^3\lambda_0 /\widetilde{M} $, and 
	\begin{align*}
		&\sum_{g=1}^G \mathcal{E}(f_{\bbbet^g}) +  \lambda_s P(f_{\bbbet^1}, \dots, f_{\bbbet^G}) + \lambda_s P(f_{\tbbet^1}, \dots, f_{\tbbet^G})\\
		\geq & \sum_{g=1}^G \mathcal{E}(f_{\bbbet^g}) +  \frac{\lambda_{s}}{2} \sum_{g=1}^G \Vert 
		\bbbet^g -\tbbet^{g}\Vert_{2}^2\\
		\geq & \sum_{g=1}^G \mathcal{E}(f_{\bbbet^g}) +  \frac{\lambda_{s}}{2{G}} \sum_{g=1}^G \Vert 
		\bbbet^g -\tbbet^{g}\Vert_{1}^2\\
		\geq & \sum_{g=1}^G \mathcal{E}(f_{\bbbet^g}) +  \frac{\lambda_{s}}{2{G}^2} \left(\sum_{g=1}^G \Vert 
		\bbbet^g -\tbbet^{g}\Vert_{1}\right)^2.
	\end{align*}
	This again implies that $ \lVert 
	\bbbet^g -\tbbet^g\rVert_{1} \leq {\widetilde{M}}/{2} $ and  $	\lVert
	\hbbet^g -\tbbet^g\rVert_{1} \leq \widetilde{M} $ for $ 1 \leq g \leq G $.
	
	Repeating the argument with $ \bbbet $ replaced by $ \hbbet $ yields on the set $\mathcal{\widetilde{B}}$ the inequality
	\begin{align*}
		\sum_{g=1}^G \mathcal{E}({f}_{\hbbet^g}) \leq 2 \left[ \sum_{g=1}^G \mathcal{E}(f_{\tbbet^g}) 
		+ 2\lambda_s P(f_{\tbbet^1}, \dots, f_{\tbbet^G}) + \frac{\lambda_{d}}{2} Q(f_{\tbbet^1}, \dots, f_{\tbbet^G}) \right]
	\end{align*}
	Notice that for the best linear predictor $ f_{\bbet^*} $ used for all $ G $ functions, we can rewrite \eqref{eq:sparsity_penalty} and \eqref{eq:diversity_penalty} as
	\begin{align*}
		P(f_{\bbet^*}, \dots, f_{\bbet^*}) &= \alpha  G \lVert \bbet^* \rVert_1 + \frac{1-\alpha}{2}  G \lVert \bbet^* \rVert_2^2, \, \text{and}\\
		Q(f_{\bbet^*}, \dots, f_{\bbet^*}) &= G(G-1) \lVert \bbet^* \rVert_2^2,
	\end{align*}
	respectively. Thus
	\begin{align*}
		\frac{1}{G}\sum_{g=1}^G \mathcal{E}({f}_{\hbbet^g}) \leq 2 \left[  \mathcal{E}(f_{\bbet^*}) 
		+ 2\alpha \lambda_s   \lVert \bbet^* \rVert_1 + \frac{1-\alpha}{2} \lambda_{s} \lVert \bbet^* \rVert_2^2 + \frac{\lambda_{d}(G-1)}{2} \lVert \bbet^* \rVert_2^2 \right].
	\end{align*}
	By the convexity of \eqref{eq:loss_function},
	\begin{align*}
		\frac{1}{G} \sum_{g=1}^G \mathcal{E}({f}_{\hbbet^g}) = \frac{1}{G}\sum_{g=1}^G \mathcal{V}({f}_{\hbbet^g}) - \mathcal{V}(f^*) \geq  \mathcal{V}\left(\frac{1}{G}\sum_{g=1}^G{f}_{\hbbet^g}\right) - \mathcal{V}(f^*),
	\end{align*}
	so we get the desired inequality \eqref{eq:consistency_inequality},
	\begin{align*}
		\mathcal{V}\left(\frac{1}{G}\sum_{g=1}^G{f}_{\hbbet^g}\right) - \mathcal{V}(f^*) \leq  2 \left[  \mathcal{E}(f_{\bbet^*}) 
		+ 2\alpha \lambda_s   \lVert \bbet^* \rVert_1 + \frac{1-\alpha}{2} \lambda_{s} \lVert \bbet^* \rVert_2^2 + \frac{\lambda_{d}(G-1)}{2} \lVert \bbet^* \rVert_2^2 \right].
	\end{align*}
	
	\FloatBarrier
	
	\section{Details of the Algorithm} \label{app:algorithms}
	
	In this section, we provide the derivation for the quadratic approximation of the logistic regression loss, the high-level steps of the block coordinate descent algorithm, and a detailed description of the alternating grid search for the tuning parameters.

\subsection{Quadratic Approximation}

For the binary classification problem with the classes labeled as $ Y = \{-1,1\} $, let $ \mathbf{y} \in \mathbb{R}^n $ be the vector of class labels and $ \bX \in \mathbb{R}^{n \times p} $ be the design matrix with sample size $ n $ and number of features $ p $. The logistic regression loss function is given by
\begin{align} \label{eq:logistic_loss}
	\mathcal{L}\left(f(\bx_i), y_i\right)=\mathcal{L}(\beta_0,\bbet \mid y_i, \bx_i) = \log \left(1+ e^{-y_i f(\bx_i)}\right), \quad 1 \leq i \leq n,
\end{align}
where $ f(\bx_i) = \beta_0 + \bx_i^T \bbet $ is a linear function of the predictor variables, $ \beta_0 \in \mathbb{R} $ and $ \bbet \in \mathbb{R}^p $ are the intercept and vector of regression coefficients.

We denote $ {\bX_A} \in \mathbb{R}^{n \times (p+1)} $ the augmented design matrix whose first column is a column of ones and $ \bbet_A \in \mathbb{R}^{p+1}= (\beta_0,\bbet^T)^T $ the vector with all regression parameters. The quadratic approximation for the logistic regression loss in \eqref{eq:logistic_loss} at the current estimates $ \boldsymbol{\tilde{\beta}}_A $ is given by
{
	\begin{align*}
		\frac{1}{n} \sum_{i=1}^{n}\mathcal{L}_Q(\beta_0,\bbet \mid y_i, \bx_i)  = \frac{1}{n} \sum_{i=1}^{n}\mathcal{L}(\tilde{\beta}_0,\boldsymbol{\tilde{\beta}} \mid y_i, \bx_i) +  \mathcal{V}\left(\tilde{\beta}_0, \boldsymbol{\tilde{\beta}} \mid \by, \bX_A\right) \left(\bbet_A - \boldsymbol{\tilde{{\beta}}}_A\right)  \nonumber \\
		+\frac{1}{2} \left(\bbet_A - \boldsymbol{\tilde{{\beta}}}_A\right)^T \mathcal{H}\left(\tilde{\beta}_0,\boldsymbol{\tilde{\beta}} \mid \by, \bX_A\right) \left(\bbet_A - \boldsymbol{\tilde{{\beta}}}_A\right),
\end{align*}}
where the gradient vector and hessian matrix are given by
\begin{align*}
	\mathcal{V}\left(\tilde{\beta}_0, \boldsymbol{\tilde{\beta}} \mid \by, \bX\right) &= \nabla \left( \frac{1}{n} \sum_{i=1}^{n}  \mathcal{L}({\beta}_0, \boldsymbol{{\beta}} \mid y_i, \bx_i) \right) \Bigg\vert_{\left({\beta}_0, \, \boldsymbol{{\beta}}\right) = \left(\tilde{\beta}_0, \, \boldsymbol{\tilde{\beta}}\right)} \\
	&= \frac{1}{n}\bX_A^T (\bz - \boldsymbol{\tilde{p}}),\\[5pt]
	\mathcal{H}\left(\tilde{\beta}_0,\boldsymbol{\tilde{\beta}} \mid \by, \bX\right) &= \nabla \left( \frac{1}{n}\sum_{i=1}^{n}  \mathcal{L}({\beta}_0, \boldsymbol{{\beta}} \mid y_i, \bx_i) \right)\nabla^T \Bigg\vert_{\left({\beta}_0, \, \boldsymbol{{\beta}}\right) = \left(\tilde{\beta}_0, \, \boldsymbol{\tilde{\beta}}\right)} \\
	&= - \frac{1}{n}\bX_A^T \mathbf{\tilde{W}} \bX_A.
\end{align*}
The elements of the $ n $-dimensional vectors $ \bz $, $ \mathbf{\tilde{p}} $ and $ \mathbf{\tilde{w}} $ are given by $ z_i = (y_i +1)/2 $, $ \tilde{p}_i = S(\tilde{\beta}_0 + \bx_i^T \boldsymbol{\tilde{\beta}}) $ and $ \tilde{w}_i = p_i(1-p_i) $, $ 1 \leq i \leq n $ respectively. The $ n \times n $ weight matrix at the current parameter estimates is given by $ \mathbf{\tilde{W}} = \text{diag}(\mathbf{\tilde{w}}) $. 
The quadratic approximation can subsequently be rewritten as a weighted least-squares problem
\begin{align}
	\frac{1}{n}\sum_{i=1}^{n}\mathcal{L}_Q(\beta_0,\bbet \mid y_i, \bx_i) 
	&= \frac{1}{2n}\left(\boldsymbol{\tilde{y}} - \bX_A \bbet_A\right)^T \boldsymbol{\tilde{W}}\left(\boldsymbol{\tilde{y}} - \bX_A \bbet_A\right) +  C\left(\tilde{\beta}_0,\boldsymbol{\tilde{\beta}}\right) \nonumber \\
	&= \frac{1}{2n} \sum_{i=1}^{n} \tilde{{w}}_i \left(\tilde{y}_i - f(\bx_i)\right)^2 + C\left(\tilde{\beta}_0,\boldsymbol{\tilde{\beta}}\right),
	\label{approx_loss}
\end{align}
where the elements of the $ n $-dimensional vector $ \mathbf{\tilde{y}} $ are given by $ \tilde{y}_i = \tilde{\beta}_0 + \bx_i^T \boldsymbol{\tilde{\beta}} + (z_i - \tilde{{p}}_i)/\tilde{{w}}_i $, $ 1 \leq i \leq n  $,  and $ C\left(\tilde{\beta}_0,\boldsymbol{\tilde{\beta}}\right) $ is a constant term. 

\subsection{Block Coordinate Descent Algorithm}

The objective function is multi-convex and can be written as a weighed elastic net problem for each individual model, where the $ L_1 $ penalty depends on the parameters in the other models. In particular, for a given model $ g $, the objective function is given by 
\begin{align}
	\mathcal{J}\left(\beta_0^g,\bbet^g\bigm| \by, \bX \right) = \frac{1}{n}\sum_{i=1}^n \mathcal{L}(\beta_0^g,\bbet^g\mid y_i, \bx_i )+\lambda_{s}\frac{(1-\alpha)}{2}\Vert\boldsymbol{\beta}^{g}\Vert_{2}%
	^{2}+\sum\limits_{j=1}^{p}|\beta_{j}^{g}|u_{j,g}, \quad 1 \leq g \leq G.
	\label{obj_function}
\end{align}
with weights 
$$ u_{j,g} = \alpha \lambda_{s}+\frac{\lambda_{d}}{2}%
\sum_{h\neq g}|\beta_{j}^{h}|. $$

We apply a block coordinate descent algorithm by cycling through the parameters of one model at a time and we apply the coordinate descent updates in a deterministic, cyclic order.
When updating the parameters of an individual model, a single coordinate descent update is applied for each parameter as follows. 
For notational convenience, we denote by $\mathbf{\tilde{{p}}}^g$, $\mathbf{\tilde{{w}}}^g$ and $ \mathbf{\tilde{{y}}}^g $ the $ n $-dimensional vectors with elements $\tilde{p}_i^g = S(\tilde{\beta_0}^g + \bx_i^T \boldsymbol{\tilde{\beta}}^g)$, $\tilde{w}_i^g = \tilde{p}_i^g (1-\tilde{p}_i^g)$ and $ \tilde{y}_i^g = \tilde{\beta}_0^g + \bx_i^T \boldsymbol{\tilde{\beta}}^g + (z_i - \tilde{{p}}_i^g)/\tilde{{w}}_i^g $, $ 1 \leq i \leq n $, respectively.
To obtain the coordinate descent updates we replace the logistic loss in the objective function~(\ref{obj_function}) by its quadratic approximation~(\ref{approx_loss}) at the current parameter estimates for the ensemble.
For parameter $ j $ of a particular model $ g $, $ 1 \leq j \leq p $, the coordinate descent update is then given by 
\begin{align*}
	\hat{\beta}_j^g &= \argmin_{\beta_j^g \in \mathbb{R}} \frac{1}{n}\sum_{i=1}^n \mathcal{L}_Q(\beta_0^g,\bbet^g\mid y_i, \bx_i )+\lambda_{s}\frac{(1-\alpha)}{2}\Vert\boldsymbol{\beta}^{g}\Vert_{2}%
	^{2}+\sum\limits_{j=1}^{p}|\beta_{j}^{g}|u_{j,g} \\[5pt]
	&=  \argmin_{\beta_j^g \in \mathbb{R}} \frac{1}{2n}\sum_{i=1}^n \tilde{{w}}_i \left(\tilde{y}_i^g - \beta_0^g - \sum_{i=1}^n\sum_{k\neq j}^p x_{ik} \tilde{\beta}_k^g - {\beta}_j^g \, x_{ij} \right)^2 + \lambda_{s}\frac{(1-\alpha)}{2}\left(\beta_{j}^g\right)^{2}+|\beta_{j}^{g}|u_{j,g} \\[5pt]
	&= \frac{\text{Soft}\left(\frac{1}{n}\left({\tilde{r}_j^g} + \tilde{\beta}_j^{g}\langle \bx_j^2, \mathbf{\tilde{w}}^g \rangle\right), \, \alpha   \lambda_{s}+ \frac{\lambda_{d}}{2} \sum_{h\neq g} \vert 
		\tilde{\beta}^{h}_{j}\vert \right)}{\frac{1}{n}\langle \bx_j^2, \mathbf{\tilde{w}}^g \rangle +  (1-\alpha) \lambda_s},
\end{align*}
where $\tilde{r}_j^g= \left\langle \bx_j , {\bz} \right\rangle - \langle \bx_j, \mathbf{\tilde{p}}^g \rangle $, and the last equality follows from the optimality condition for subgradients. A similar derivation can be made for the coordinate descent update of the intercept term
\begin{align*}
	\hat{\beta}_0^{g} &= \tilde{\beta}_0^{g} + \frac{\langle \bz - \mathbf{\tilde{p}}^g, \boldsymbol{1}_n \rangle}{\langle \mathbf{\tilde{w}}^g, \boldsymbol{1}_n \rangle},
\end{align*}
which yields the results in Proposition 1 of the article.
When all parameter estimates of model $ g $ have been updated, also the vectors $ \mathbf{\tilde{{p}}}^g $ and $ \mathbf{\tilde{{w}}}^g  $  are updated.
The active set cycling strategy \citep{friedman2010regularization} is also adopted and available in our software implementation.
In Algorithm \ref{alg:splitglm_algo} we provide the steps to generate solutions for split logistic regression when $\lambda_{s}$ and $\lambda_{d}$ are fixed.

\begin{algorithm}[H] 
	\caption{\label{alg:splitglm_algo}Split Logistic Regression  for Fixed $\lambda_{s}$ and $\lambda_{d}$} \label{alg:splitglm_algo_supp} 
	\begin{algorithmic}[1]
		\Require{Design matrix $\bX \in \mathbb{R}^{n \times p}$, response vector $\by \in \mathbb{R}^n$,  current solutions  $ \tbbet_{1:G} $, $ \ell_1 $-$\ell_{2}$ mixing parameter $\alpha\in[0,1]$, sparsity and diversity tuning parameters $ \lambda_{s}, \lambda_{d} \geq 0$, and convergence tolerance parameter $\delta>0$.}
		\Statex
		\State Compute the current probabilities  $\tilde{p}_i^g = S(\tilde{\beta_0}^g + \bx_i^T \boldsymbol{\tilde{\beta}}^g)$, weights $ \tilde{w}_i^g = \tilde{p}_i^g (1-\tilde{p}_i^g) $ and residuals $ \tilde{r}_j^g= \left\langle \bx_j, {\bz} \right\rangle - \langle \bx_j, \mathbf{\tilde{p}}^g \rangle $, $ 1 \leq i \leq n $, $ 1 \leq j \leq p $, $ 1 \leq g \leq G $.
		\Statex
		\State Repeat the following steps until convergence.
		\begin{enumerate}
			\item[\footnotesize 2.1:] For each model $ g $, $ 1 \leq g \leq G $:
			\begin{enumerate}
				\item[\footnotesize 2.1.1:] Perform a single (block) coordinate descent update for the intercept and each predictor $ j $, $ 1 \leq j \leq p $.
				\begin{enumerate}
					\item[\footnotesize 2.1.1.1:] Compute the new intercept in model $ g $,
					\begin{align*}
						\hat{\beta}_0^{g} = \tilde{\beta}_0^{g} + \frac{\langle \bz - \mathbf{\tilde{p}}^g, \boldsymbol{1}_n \rangle}{\langle \mathbf{\tilde{w}}^g, \boldsymbol{1}_n \rangle}.
					\end{align*}
					\item[\footnotesize 2.1.1.2:] If $\hat{\beta}_0^g \neq \tilde{\beta}_0^g$, then update the probabilities $ \mathbf{\tilde{p}}^g $, weights $ \mathbf{\tilde{w}}^g $ and residuals $ \mathbf{\tilde{r}}^g $ for model $ g $.
					\item[\footnotesize 2.1.1.3:] Update $ j $-th coefficient in model $ g $,
					\begin{align*}
						\hat{\beta}^{g}_{j} = \frac{\operatorname{Soft}\left(\frac{1}{n}\left({\tilde{r}_j^g} + \tilde{\beta}_j^{g}\langle \bx_j^2, \mathbf{\tilde{w}}^g \rangle\right), \, \alpha   \lambda_{s}+ \frac{\lambda_{d}}{2} \sum_{h\neq g} \vert 
							\tilde{\beta}^{h}_{j}\vert \right)}{\frac{1}{n}\langle \bx_j^2, {\mathbf{\tilde{w}}}^g \rangle +  (1-\alpha) \lambda_s}.
					\end{align*}
					\item[\footnotesize 2.1.1.4:] If $\hat{\beta}_j^g \neq \tilde{\beta}_j^g$, then update the probabilities $ \mathbf{\tilde{p}}^g $, weights $ \mathbf{\tilde{w}}^g $ and residuals $ \mathbf{\tilde{r}}^g $ for model $ g $.
				\end{enumerate}
			\end{enumerate}
		\end{enumerate}
		\State If successive estimates of the coefficients in the ensemble model show little difference, i.e.
		\begin{align*}
			\max_{1 \leq j \leq p}\left(\frac{1}{G}\sum_{g=1}^{G} \tilde{\beta}_j^g - \frac{1}{G} \sum_{g=1}^G \hat{\beta}_j^g \right)^2 < \delta,
		\end{align*}
		then convergence is declared.
		\Statex
		\State Return the coefficients for each model $ (\hat{\beta}_0^g, \hbbet^g)$, $ 1 \leq g \leq G $. 
	\end{algorithmic}
\end{algorithm}

\subsection{Alternating Grid Search for Tuning Parameters}

The selection of the sparsity and diversity tuning parameters, $\lambda_{s}$ and $\lambda_{d}$, is done by an alternating grid search. The first grid search is over $\lambda_{s}$ with the diversity tuning parameter  fixed  at $\lambda_{d}^{(0)}=0$, which yields a first value $ \lambda_s^{\text{opt}} $ minimizing the cross-validated loss. Keeping the sparsity parameter fixed at  value $ \lambda_s^{\text{opt}} $, we now perform a grid search over $\lambda_{d}$ which yields $\lambda_{d}^{\text{opt}}$. This process is repeated until the cross-validated loss no longer decreases.  The high-level steps of the alternating grid search are given in Algorithm \ref{alg:split_algo_CV}.

To construct a grid for $\lambda_{s}$, we estimate a value $ \lambda_s^{\max} $ that makes all models null. In the special case where $ \lambda_d = 0 $ and $ \alpha>0 $, it can easily be shown that $ \lambda_s^{\text{max}} = \frac{1}{2 \alpha} \max_{1 \leq j \leq p} |\bar{\bx}_j| $. For $ \lambda_d > 0 $, we estimate the smallest $ \lambda_s^{\text{max}} $ that makes all models null by performing an internal grid search. Based on this maximal value $ \lambda_s^{\text{max}} $ we then construct the grid for the sparsity penalty $\lambda_{s}$ similarly to the case of (single-model) penalized logistic regression. that is, we use (by default) 100 log-equispaced points between $ \epsilon \lambda_s^{\text{max}} $ and $ \lambda_s^{\text{max}} $, where $\epsilon = 10^{-4}$ if $ p < n $ and $ 10^{-2} $ otherwise. 

The smallest diversity penalty $ \lambda_d^{\text{max}} $ that makes the models fully disjoint for some fixed $ \lambda_s\geq 0 $ is similarly estimated via a grid search. We then analogously generate the diversity penalty grid using (by default) 100 log-equispaced points between $ \epsilon \lambda_d^{\text{max}} $ and $ \lambda_d^{\text{max}} $. 
For a grid search over one of the tuning parameters while keeping the other one fixed, we use warm-starts by computing solutions for a decreasing sequence of $ \lambda_s $ or $ \lambda_d $, leading to a more  stable algorithm.

\begin{algorithm}[H]
	\caption{Alternating CV Procedure  \label{alg:split_algo_CV}} \label{alg:split_algo_CV_supp}
	\begin{algorithmic}[1]
		\Require{Design matrix $\bX \in \mathbb{R}^{n \times p}$, response vector $\by \in \mathbb{R}^n$, $ \ell_1 $-$\ell_{2}$ mixing parameter $\alpha\in[0,1]$ and convergence tolerance parameter $\delta>0$.}
		\Statex
		\State Set $\lambda_{d}^{\text{opt}}=0$ and the next search is for the sparsity tuning parameter $\lambda_{s}^{\text{opt}}$. 
		\Statex
		\State Alternate between a search for $ \lambda_{s}^{\text{opt}} $ or $\lambda_{d}^{\text{opt}}$ until CV MSPE no longer decreases.
		\begin{enumerate}[label*=\footnotesize 2.\arabic*:]
			\item If the search is for $\lambda_{s}^{\text{opt}}$:
			\begin{enumerate}
				\item[\footnotesize 2.1.1:] If $\lambda_{d}^{\text{opt}}=0$, set $ \lambda_s^{\text{max}} = ({1}/{2 \alpha}) \max_{1 \leq j \leq p} |\bar{\bx}_j| $. Otherwise perform a grid search to find the smallest $\lambda_{s}^{\text{max}}$ such that each model is null.
				\item[\footnotesize 2.1.2:] Generate the log-equispaced grid between $\epsilon \lambda_{s}$ and $ \lambda_{s}^{\text{max}} $.
				\item[\footnotesize 2.1.3:] For each $\lambda_{s}$ in the log-equispaced grid compute $\hbbet_{1:G}(\lambda_s) = (\hbbet^1(\lambda_s), \dots, \hbbet^G(\lambda_s)) $ with Algorithm \ref{alg:splitglm_algo}, using the previous solution in  the grid as a warm-start. 
				\item[\footnotesize 2.1.4:] Set $\lambda_{s}^{\text{opt}}$ using the value in the grid that minimized the CV MSPE.
			\end{enumerate}
			Otherwise if the search is for $\lambda_{d}^{\text{opt}}$:
			\begin{enumerate}[label*=\footnotesize \arabic*:]
				\item[\footnotesize 2.1.1:] Perform a grid search to find the smallest $\lambda_{d}^{\text{max}}$ such that makes models fully disjoint.
				\item[\footnotesize 2.1.2:] Generate the log-equispaced grid between $\epsilon \lambda_{d}$ and $ \lambda_{d}^{\text{max}} $.
				\item[\footnotesize 2.1.3:] For each $\lambda_{d}$ in the log-equispaced grid compute $\hbbet_{1:G}(\lambda_d) = (\hbbet^1(\lambda_d), \dots, \hbbet^G(\lambda_d)) $ with Algorithm \ref{alg:splitglm_algo}, using the previous solution in  the grid as a warm-start. 
				\item[\footnotesize 2.1.4:] Set $\lambda_{d}^{\text{opt}}$ using the value in the grid that minimized the CV MSPE.
			\end{enumerate}
		\end{enumerate}
		\State For $\lambda_{d}^{\text{opt}}$ and the smallest $\lambda_{s}^{\text{max}}$ such that each model is null, generate the log-equispaced grid between  $\epsilon \lambda_{s}$ and $ \lambda_{s}^{\text{max}} $. 
		\Statex
		\State For each $\lambda_{s}$ in the log-equispaced grid compute $\hbbet_{1:G}(\lambda_s) = (\hbbet^1(\lambda_s), \dots, \hbbet^G(\lambda_s)) $ with Algorithm \ref{alg:splitglm_algo}, using the previous solution in  the grid as a warm-start. 
		\Statex
		\State Return the coefficients of the models $ \hbbet_{1:G}(\lambda_s) = (\hbbet^1(\lambda_s), \dots, \hbbet^G(\lambda_s)) $ for each $ \lambda_{s} $ in the grid. 
	\end{algorithmic}
\end{algorithm}

\FloatBarrier
	
	\section{Alternative Diversity Measures} \label{app:diversity}
	
	In this section, we investigate the accuracy-diversity trade-off using several alternative diversity measures to complement and consolidate the results obtained in Section 6 of the main article based on the entropy diversity measure.
	
	\subsection{Disagreement Measure}
	
	The disagreement (DIS) diversity measure \citep{skalak1996sources, ho1998random} of an ensemble comprised of $ G $ individual classifiers for a given input $ \bx $ is defined as 
	\begin{align*}
		\text{DIS}(\bx) = \frac{1}{G(G-1)} \sum_{g=1}^G \sum_{h\neq g}^G  \text{DIS}_{g,h}(\bx)  
	\end{align*}
	where the disagreement between between classifiers $ g $ and $ h $ is given by
	\begin{align*}
		\text{DIS}_{g,h}(\bx) = \begin{cases}
			1, & \text{if classifiers $ g $ and $ h $ disagree on the class of } \bx, \\
			0, & \text{if classifiers $ g $ and $ h $ agree on the class of } \bx.
		\end{cases}
	\end{align*}
	The disagreement measure is a pairwise diversity measure and ranges between 0 and 1, where $ \text{DIS}(\bx)=0 $ corresponds to no disagreement and increasing values of $ \text{DIS}(\bx)$ correspond to more disagreement between the individual classifiers.
	
	\subsection{Double-Fault Measure}
	
	The double-fault (DF) diversity measure \citep{giacinto2001design} of an ensemble comprised of $ G $ individual classifiers for some a given $ \bx $ is defined as 
	\begin{align*}
		\text{DF}(\bx) = \frac{1}{G(G-1)} \sum_{g=1}^G \sum_{h\neq g}^G  \text{DF}_{g,h}(\bx)  
	\end{align*}
	where the double-fault between between classifiers $ g $ and $ h $ is given by
	\begin{align*}
		\text{DF}_{g,h}(\bx) = \begin{cases}
			1, & \text{if classifiers $ g $ and $ h $ both misclassify } \bx, \\
			0, & \text{if at most one of classifiers $ g $ and $ h $ misclassify } \bx.
		\end{cases}
	\end{align*}
	The double-fault measure is a pairwise diversity measure and ranges between 0 and 1, where $ \text{DF}(\bx)=0 $ corresponds to no double-faults and increasing values of $ \text{DF}(\bx)$ correspond to more double-faults between the individual classifiers.
	
	\subsection{Kohavi-Wolpert Variance}
	
	The Kohavi-Wolpert variance (KW) diversity measure \citep{kohavi1996bias} of an ensemble comprised of $ G $ individual classifiers for a given input $ \bx $ is defined as 
	\begin{align*}
		\text{KW}(\bx) = \frac{1}{G^2} \, \ell(\bx)(G-\ell(\bx))
	\end{align*}
	where $\ell(\bx)$ denotes the numbers of individual classifiers that correctly classified input $ \bx $. The Kohavi-Wolpert variance is a non-pairwise diversity measure with $ \text{KW}(\bx)=0 $ corresponding to no diversity and increasing values of $ \text{KW}(\bx)$ corresponding to more diversity between the individual classifiers. \cite{kuncheva2003measures} have shown that
	\begin{align*}
		\text{KW}(\bx) = \frac{G-1}{2G} \, \text{DIS}(\bx).
	\end{align*}
	
	\subsection{Generalized Diversity}
	
	The generalized diversity (GD) measure \citep{partridge1997software} of an ensemble comprised of $ G $ individual classifiers for a given input $ \bx $ is defined as 
	\begin{align*}
		\text{GD}(\bx) = 1 - \frac{\sum_{g=1}^G \frac{g(g-1)}{G(G-1)} \, \mathbb{P}(\ell(\bx)=g)}{\sum_{g=1}^G \frac{g}{G} \, \mathbb{P}(\ell(\bx)=g)}
	\end{align*}
	where $\ell(\bx)$ denotes the numbers of individual classifiers from the ensemble that correctly classify input $ \bx $. The generalized diversity is a non-pairwise diversity measure and ranges between 0 and 1, where $ \text{GD}(\bx)=0 $ corresponds to no diversity and $ \text{GD}(\bx)=1$ corresponds to maximum diversity between the individual classifiers.
	
	\subsection{Results}
	
	In Table \ref{tab:alternative_diversity} we report the results of the alternative diversity measures for the same simulation settings as in Table~2 in the article. Similarly to the entropy diversity in the article, the DIS, DF, KW and GD measures are reported as a function of the number of models in split logistic regression, averaged over the test sets. It can be seen that the DIS, KW and GD diversity measures all increase with the number of models, while the DF diversity measure decreases. Hence, all measures confirm that the individual models become more diverse when the number of models increases.
	
	\begin{table}[H]
		\centering
		\caption{DIS, DF, KW and GD as a function of the number of models under Scenario 3. \label{tab:alternative_diversity}} 
		\begin{tabular}{rrrrrrrrrrrrr}
			\toprule
			& \multicolumn{4}{c}{$\mathbf{\boldsymbol{\zeta}=0.1}$} & \multicolumn{4}{c}{$\mathbf{\boldsymbol{\zeta}=0.2}$} & \multicolumn{4}{c}{$\mathbf{\boldsymbol{\zeta}=0.4}$} \\ \cmidrule(lr){1-1} \cmidrule(lr){2-5} \cmidrule(lr){6-9} \cmidrule(lr){10-13}  $\mathbf{G}$ & \textbf{DIS} & \textbf{DF} & \textbf{KW} & \textbf{GD} & \textbf{DIS} & \textbf{DF} & \textbf{KW} & \textbf{GD} & \textbf{DIS} & \textbf{DF} & \textbf{KW} & \textbf{GD} \\ \cmidrule(lr){1-1} \cmidrule(lr){2-5} \cmidrule(lr){6-9} \cmidrule(lr){10-13} 
			2  & 0.04 & 0.84 & 0.01 & 0.37 & 0.05 & 0.86 & 0.01 & 0.50 & 0.05 & 0.85 & 0.01 & 0.45 \\ 
			3  & 0.07 & 0.83 & 0.03 & 0.59 & 0.08 & 0.84 & 0.03 & 0.63 & 0.07 & 0.84 & 0.03 & 0.63 \\ 
			4  & 0.07 & 0.82 & 0.03 & 0.61 & 0.10 & 0.83 & 0.04 & 0.72 & 0.08 & 0.84 & 0.03 & 0.68 \\ 
			5  & 0.10 & 0.80 & 0.05 & 0.72 & 0.13 & 0.81 & 0.06 & 0.78 & 0.11 & 0.82 & 0.05 & 0.75 \\ 
			6  & 0.12 & 0.78 & 0.06 & 0.75 & 0.14 & 0.79 & 0.07 & 0.79 & 0.13 & 0.80 & 0.06 & 0.79 \\ 
			7  & 0.14 & 0.77 & 0.07 & 0.77 & 0.15 & 0.78 & 0.07 & 0.80 & 0.14 & 0.79 & 0.07 & 0.80 \\ 
			8  & 0.14 & 0.76 & 0.07 & 0.77 & 0.17 & 0.77 & 0.08 & 0.81 & 0.15 & 0.78 & 0.07 & 0.81 \\ 
			\bottomrule
		\end{tabular}
	\end{table}
	
	\FloatBarrier
	
	\section{Benchmark Across Ten Genomics Datasets} \label{app:results_genomics}
	
	This section presents the comprehensive benchmark study summarized in the main article. The study systematically evaluates the predictive performance of the proposed split logistic regression method against a wide range of state-of-the-art competitors on ten publicly available medical genomics datasets.
	
	\subsection{Data and Pre-processing}
	
	The ten datasets used in this benchmark cover a variety of common diseases, including several types of cancer, multiple sclerosis, and psoriasis. Key characteristics of these datasets, such as sample size, class distribution, and experimental design (paired vs. unpaired samples), are summarized in Table \ref{tab:gene_description_supp}.
	
	\begin{table}[htbp]
		\centering
		\caption{\label{tab:gene_description_supp} GEO identification (ID) codes, dataset descriptions, sample sizes ($n$), class distributions, pairing status, and the number of genes retained after matching probes to genes.}
		\resizebox{\textwidth}{!}{
			\extrarowsep=2pt
			\begin{tabu}{llrrrr}
				\toprule
				\textbf{GEO ID} & \textbf{Description} & $ \boldsymbol{n} $ & \textbf{Class Distribution} & \textbf{Paired} & \textbf{Genes} \\ 
				\cmidrule(lr){1-2} \cmidrule(lr){3-5}  \cmidrule(l){6-6}
				\addlinespace[0.25cm]
				GSE20347 & Esophageal cancerous cell tissue &      34 & 17 cancer, 17 normal & Yes & 13,515 \\
				GSE23400 & Esophageal cancerous cell tissue &   106 & 53 cancer, 53 normal & Yes & 13,515 \\
				GSE23400 & Esophageal cancerous cell tissue &  102 & 51 cancer, 51 normal & Yes & 11,271 \\
				GSE5364 & Esophageal cancerous cell tissue &           29 & 16 cancer, 13 normal & Yes & 13,515 \\
				GSE25869 & Gastric cancerous cell tissue &  64 & 32 cancer, 32 normal & Yes & 14,476 \\
				GSE5364 & Lung cancerous cell tissue & 30 & 16 cancer, 14 normal & Yes & 13,515 \\
				GSE10245 & Lung cancerous cell tissue & 58 & 40 adeno, 18 squamous & No & 13,515 \\
				GSE5364 & Thyroid cancerous cell tissue & 51 & 16 cancer, 35 normal & Yes & 13,515 \\
				GSE21942 & Multiple sclerosis cell tissue & 29 & 14 sclerosis, 15 normal & No & 23,520 \\ 
				GSE14905 & Psoriasis cell tissue & 54 & 34 psoriasis, 20 normal & No & 23,520 \\
				\addlinespace[0.25cm]
				\bottomrule
			\end{tabu}
		}
	\end{table}
	
	As noted in the main manuscript, several of these datasets feature a paired-sample design. This analysis is intended to provide a comprehensive evaluation of the \textit{relative performance} of the methods in diverse, real-world settings where such experimental designs are common, even though they violate the independence assumption of the underlying models.
	
	All datasets were pre-processed using a standard procedure: (1) thresholding expression levels, (2) filtering genes with low expression ratios, (3) filtering genes with low expression differences, and (4) a base-2 logarithmic transformation. Following this, for each dataset, we created four versions by retaining the top $p \in \{100, 250, 500, 1000\}$ genes with the smallest $q$-values.
	
	\subsection{Experimental Setup and Methods}
	
	For each version of each dataset, we performed $N=50$ random splits into a training set and a test set. The training set size was set to 50\% of the data, or 35\% if the 50\% split would result in a training set smaller than 20 observations. We compared the performance of fourteen classification methods, including the proposed Split-Lasso and Split-Elastic Net (with the number of models $G \in \{5, 10, 25\}$ chosen by cross-validation, denoted Split-Lasso-CV and Split-EN-CV), standard sparse methods (Lasso, Elastic Net, Adaptive Lasso, MCP, SIS-SCAD), and other ensemble methods (RE-Lasso-100, RE-EN-100, RGLM-100, RF-500, XGB). The performance was evaluated on the test set using two primary metrics: prediction accuracy (ACC) and test sample loss (TSL), which is the average negative log-likelihood.
	
	\subsection{Results and Discussion}
	
	Table \ref{tab:gene_ranks_supp} shows the average ranks of all methods across all ten datasets and training proportions for each level of $p$. The results reveal that Split-EN-CV consistently ranks among the top three methods across all feature dimensions. This performance is particularly impressive for high dimensions ($p \geq 500$), where it either matches or surpasses RE-EN-100 in accuracy while using only 5 to 25 interpretable sparse models compared to the 100 models used by RE-EN-100. Overall, Split-EN-CV achieves the second-best average rank for ACC (2.96), only slightly behind RE-EN-100 (2.75), despite its significantly smaller ensemble size and deterministic construction.
	
	\begin{table}[htbp]
		\centering
		\caption{\label{tab:gene_ranks_supp} Average ranks for prediction accuracy (ACC) and test sample loss (TSL) over the ten gene expression datasets for different numbers of genes ($p$) retained after pre-processing. The three best results for each column are highlighted in bold.}
		\extrarowsep=2pt
		\resizebox{\textwidth}{!}{
			\begin{tabular}{lllllllllll}
				\toprule
				& \multicolumn{2}{c}{ $\mathbf{\boldsymbol{p}=\textbf{100}}$ } & \multicolumn{2}{c}{$\mathbf{\boldsymbol{p}=\textbf{250}}$} & \multicolumn{2}{c}{$\mathbf{\boldsymbol{p}=\textbf{500}}$} & \multicolumn{2}{c}{$\mathbf{\boldsymbol{p}=\textbf{1,000}}$} & \multicolumn{2}{c}{\bf{Rank}}\\
				\cmidrule(lr){1-1} \cmidrule(lr){2-3} \cmidrule(lr){4-5} \cmidrule(lr){6-7} \cmidrule(lr){8-9} \cmidrule(lr){10-11}
				\textbf{Method} & \textbf{ACC} & \textbf{TSL} & \textbf{ACC} & \textbf{TSL} & \textbf{ACC} & \textbf{TSL} & \textbf{ACC} & \textbf{TSL}& \textbf{ACC} & \textbf{TSL} \\ 
				\cmidrule(lr){1-1} \cmidrule(lr){2-3} \cmidrule(lr){4-5} \cmidrule(lr){6-7} \cmidrule(lr){8-9} \cmidrule(lr){10-11}
				\addlinespace[0.25cm]
				Split-Lasso-CV & {3.50} & {5.30} & \textbf{2.85} & {4.30} & \textbf{2.95} & \textbf{3.50} & \textbf{3.15} & \textbf{3.50} & \textbf{3.11} & {4.15} \\
				Split-EN-CV & \textbf{3.30} & \textbf{3.60} & \textbf{2.90} & \textbf{2.90} & \textbf{2.80} & \textbf{2.85} & \textbf{2.85} & \textbf{2.65} & \textbf{2.96} & \textbf{3.00} \\
				\addlinespace[0.25cm]
				Lasso & 7.20 & 7.50 & 7.35 & 7.55 & 7.30 & 7.25 & 7.60 & 7.25 & 7.36 & 7.39 \\
				EN & {4.70} & {5.05} & 5.15 & 4.75 & 5.00 & 4.75 & 4.85 & 4.80 & 4.92 & 4.84 \\
				Adaptive & 9.10 & 7.85 & 9.30 & 8.70 & 9.70 & 9.35 & 10.65 & 9.85 & 9.69 & 8.94 \\
				Relaxed & 8.60 & 13.30 & 8.80 & 13.10 & 8.95 & 13.15 & 8.70 & 13.00 & 8.76 & 13.14 \\
				MCP & 10.55 & 9.50 & 10.85 & 9.70 & 11.30 & 10.00 & 11.35 & 9.65 & 11.01 & 9.71 \\
				SIS-SCAD & 11.40 & 10.00 & 11.30 & 10.15 & 10.90 & 9.90 & 10.70 & 9.95 & 11.07 & 10.00 \\
				RuleFit & 12.45 & 13.10 & 12.20 & 13.45 & 11.95 & 13.25 & 11.65 & 13.30 & 12.06 & 13.28 \\
				\addlinespace[0.25cm]
				RE-Lasso-100 & \textbf{3.70} & \textbf{3.30} & {4.35} & \textbf{3.30} & {4.05} & {3.75} & {3.90} & {3.90} & {4.00} & \textbf{3.56} \\
				RE-EN-100 & \textbf{2.45} & \textbf{2.25} & \textbf{2.80} & \textbf{2.40} & \textbf{2.80} & \textbf{2.15} & \textbf{2.95} & \textbf{2.15} & \textbf{2.75} & \textbf{2.24} \\
				RGLM-100 & 6.65 & 5.60 & 5.70 & 5.50 & 5.85 & 5.50 & 5.35 & 5.45 & 5.89 & 5.51 \\
				RF-500 & 9.20 & 8.65 & 9.30 & 8.90 & 9.20 & 9.10 & 9.20 & 9.10 & 9.22 & 8.94 \\
				XGB & 12.20 & 10.00 & 12.15 & 10.30 & 12.25 & 10.50 & 12.10 & 10.45 & 12.18 & 10.31 \\
				\addlinespace[0.25cm]
				\bottomrule
			\end{tabular}
		}
	\end{table}
	
	To further summarize these findings, Table \ref{tab:gene_ranks_summary_supp} counts the number of times each method achieved a top-one, top-three, or bottom-three rank across the ten datasets. Split-EN-CV demonstrates remarkable consistency, achieving the top rank for ACC in five of the ten datasets and never placing among the worst three performers. This highlights the proposed method's ability to deliver state-of-the-art accuracy with high reliability across a diverse range of genomic applications.
	
	\begin{table}[htbp]
		\centering
		\caption{\label{tab:gene_ranks_summary_supp} Number of top and lowest ranks for prediction accuracy (ACC) and test sample loss (TSL) over the ten gene expression datasets.}
		\extrarowsep=2pt
		\begin{tabular}{lllllll}
			\toprule
			& \multicolumn{2}{c}{ \bf{Top 1} } & \multicolumn{2}{c}{\bf{Top 3} } & \multicolumn{2}{c}{\bf{Low 3}}  \\
			\cmidrule(lr){1-1} \cmidrule(lr){2-3} \cmidrule(lr){4-5} \cmidrule(lr){6-7}
			\textbf{Method} & \textbf{ACC} & \textbf{TSL} & \textbf{ACC} & \textbf{TSL} & \textbf{ACC} & \textbf{TSL} \\ 
			\cmidrule(lr){1-1} \cmidrule(lr){2-3} \cmidrule(lr){4-5} \cmidrule(lr){6-7}
			\addlinespace[0.25cm]
			Split-Lasso-CV &   0 &   0 &   7 &   5 &   0 &   0 \\
			Split-EN-CV &  5 &   3 &   6 &   6 &   0 &   0 \\
			\addlinespace[0.25cm]
			Lasso &   0 &   0 &   0 &   0 &   0 &   1 \\
			EN &   1 &   1 &   2 &   2 &   0 &   0 \\
			Adaptive &  0 &   0 &   1 &   0 &   2 &   0 \\
			Relaxed &   0 &   0 &   0 &   0 &   0 &   10 \\
			MCP &   0 &   0 &   0 &   0 &   6 &   1 \\
			{SIS-SCAD } &   {0} &   {0} &   {0} &   {0} &   {4} &   {0} \\
			{RuleFit } &   {0} &   {0} &   {0} &   {0} &   {6} &   {10} \\
			\addlinespace[0.25cm]
			RE-Lasso-100 &   2 &  1 &   4 &  4 &  0 &   0 \\
			RE-EN-100 &   0 &   4 &  8 &   9 &  0 &   0 \\
			RGLM-100 &   0 &  1 &   0 &  2 &  0 &   0 \\
			RF-500 &   2 &   0 &  2 &   2 &  5 &   4 \\
			XGB &   0 &   0 &   0 &   0 &   7 &   4 \\
			\addlinespace[0.25cm]
			\bottomrule
		\end{tabular}
	\end{table}
	
	The complete, unabridged results for each individual genomic dataset, including performance metrics for all choices of $p$ and training proportions, are provided in Section \ref{app:full_results_genomics} of this supplement. These detailed tables offer a granular view of model performance and further substantiate the summary findings presented here.
	
	\FloatBarrier

\section{Full Results of Simulation Study} \label{app:full_sim_results}

This section provides the full, unabridged results for the simulation study discussed in Section 5 of the main paper. For each of the five scenarios and all combinations of the data-generating parameters (correlation, sample size, sparsity, and event probability), the following tables report the mean and standard deviation (in parentheses) for all performance metrics: prediction accuracy (ACC), sensitivity (SNS), specificity (SPC), area under the ROC curve (AUC), test-sample loss (TSL), recall (RCL), and precision (PRC).

The results are organized by scenario as described in the main paper, with tables grouped as follows:

\begin{itemize}
	\item \textbf{Scenario 1 (Main Effects, Exchangeable Correlation):} Prediction performance is detailed in Tables \ref{tab:sim_pred_1_50_0.2} through \ref{tab:sim_pred_1_100_0.8}. Variable selection performance is in Tables \ref{tab:sim_sel_1_50_0.2} through \ref{tab:sim_sel_1_100_0.8}.
	
	\item \textbf{Scenario 2 (Main Effects, Differential Correlation):} Prediction performance is detailed in Tables \ref{tab:sim_pred_2_50_0.5_0.2} through \ref{tab:sim_pred_2_100_0.8_0.5}. Variable selection performance is in Tables \ref{tab:sim_sel_2_50_0.5_0.2} through \ref{tab:sim_sel_2_100_0.8_0.5}.
	
	\item \textbf{Scenario 3 (Main Effects, Block Correlation):} Prediction performance is detailed in Tables \ref{tab:sim_pred_3_50_0.5_0.2} through \ref{tab:sim_pred_3_100_0.8_0.5}. Variable selection performance is in Tables \ref{tab:sim_sel_3_50_0.5_0.2} through \ref{tab:sim_sel_3_100_0.8_0.5}.
	
	\item \textbf{Scenario 4 (Interactions, Block Correlation):} Prediction performance is detailed in Tables \ref{tab:sim_pred_4_50_0.5_0.2} through \ref{tab:sim_pred_4_100_0.8_0.5}. Variable selection performance is in Tables \ref{tab:sim_sel_4_50_0.5_0.2} through \ref{tab:sim_sel_4_100_0.8_0.5}.
	
	\item \textbf{Scenario 5 (Non-Linear Effects, Block Correlation):} Prediction performance is detailed in Tables \ref{tab:sim_pred_5_50_0.5_0.2} through \ref{tab:sim_pred_5_100_0.8_0.5}. Variable selection performance is in Tables \ref{tab:sim_sel_5_50_0.5_0.2} through \ref{tab:sim_sel_5_100_0.8_0.5}.
\end{itemize}

\subsection{Scenario 1: Main Effects, Exchangeable Correlation}
\begin{table}[H]
\centering
\caption{Mean prediction accuracy (ACC), sensitivity (SNS) and specificity (SPC), area under ROC curve (AUC) and test-sample loss (TSL) for Scenario 1, $\rho =$ 0.2, $n =$ 50, $p =$ 1000.} \label{tab:sim_pred_1_50_0.2}
\resizebox{\textwidth}{!}{
\begin{tabular}{llrrrrrrrrrrrrrrr}
  \hline \rule{0pt}{3ex} & & \multicolumn{5}{c}{$\mathbf{\boldsymbol{\zeta}=0.1}$} & \multicolumn{5}{c}{$\mathbf{\boldsymbol{\zeta}=0.2}$} & \multicolumn{5}{c}{$\mathbf{\boldsymbol{\zeta}=0.4}$} \\ \cmidrule(lr){1-2} \cmidrule(lr){3-7} \cmidrule(lr){8-12} \cmidrule(lr){13-17}  $\mathbf{\pi_1}$ & \textbf{Method} & \textbf{ACC} & \textbf{SNS} & \textbf{SPC} & \textbf{AUC} & \textbf{TSL} & \textbf{ACC} & \textbf{SNS} & \textbf{SPC} & \textbf{AUC} & \textbf{TSL} & \textbf{ACC} & \textbf{SNS} & \textbf{SPC} & \textbf{AUC} & \textbf{TSL}  \\ \cmidrule(lr){1-2} \cmidrule(lr){3-7} \cmidrule(lr){8-12} \cmidrule(lr){13-17}  \rule{0pt}{3ex} & Split-Lasso-10 & 0.73 & 0.58 & 0.84 & 0.81 & 1.11 & 0.87 & 0.81 & 0.91 & 0.96 & 0.57 & 0.90 & 0.83 & 0.94 & 0.97 & 0.47 \\ 
   & Split-EN-10 & 0.73 & 0.58 & 0.84 & 0.82 & 1.09 & 0.88 & 0.82 & 0.91 & 0.96 & 0.56 & 0.90 & 0.84 & 0.94 & 0.97 & 0.46 \\ 
   & Lasso & 0.69 & 0.54 & 0.81 & 0.76 & 1.23 & 0.81 & 0.72 & 0.87 & 0.90 & 0.85 & 0.82 & 0.72 & 0.89 & 0.91 & 0.77 \\ 
   & Elastic Net & 0.71 & 0.55 & 0.82 & 0.78 & 1.17 & 0.83 & 0.76 & 0.89 & 0.92 & 0.74 & 0.86 & 0.77 & 0.91 & 0.94 & 0.64 \\ 
   & Adaptive & 0.60 & 0.20 & 0.92 & 0.65 & 1.32 & 0.76 & 0.55 & 0.89 & 0.86 & 1.00 & 0.78 & 0.58 & 0.91 & 0.88 & 0.95 \\ 
   & Relaxed & 0.69 & 0.54 & 0.81 & 0.76 & 1.28 & 0.80 & 0.73 & 0.85 & 0.89 & 1.11 & 0.82 & 0.72 & 0.88 & 0.90 & 0.96 \\
\hline
\end{tabular}
}
\end{table}
\begin{table}[H]
\centering
\caption{Mean prediction accuracy (ACC), sensitivity (SNS) and specificity (SPC), area under ROC curve (AUC) and test-sample loss (TSL) for Scenario 1, $\rho =$ 0.5, $n =$ 50, $p =$ 1000.}
\resizebox{\textwidth}{!}{
\begin{tabular}{llrrrrrrrrrrrrrrr}
  \hline \rule{0pt}{3ex} & & \multicolumn{5}{c}{$\mathbf{\boldsymbol{\zeta}=0.1}$} & \multicolumn{5}{c}{$\mathbf{\boldsymbol{\zeta}=0.2}$} & \multicolumn{5}{c}{$\mathbf{\boldsymbol{\zeta}=0.4}$} \\ \cmidrule(lr){1-2} \cmidrule(lr){3-7} \cmidrule(lr){8-12} \cmidrule(lr){13-17}  $\mathbf{\pi_1}$ & \textbf{Method} & \textbf{ACC} & \textbf{SNS} & \textbf{SPC} & \textbf{AUC} & \textbf{TSL} & \textbf{ACC} & \textbf{SNS} & \textbf{SPC} & \textbf{AUC} & \textbf{TSL} & \textbf{ACC} & \textbf{SNS} & \textbf{SPC} & \textbf{AUC} & \textbf{TSL}  \\ \cmidrule(lr){1-2} \cmidrule(lr){3-7} \cmidrule(lr){8-12} \cmidrule(lr){13-17}  \rule{0pt}{3ex} & Split-Lasso-10 & 0.83 & 0.76 & 0.88 & 0.92 & 0.76 & 0.92 & 0.91 & 0.93 & 0.98 & 0.35 & 0.94 & 0.91 & 0.96 & 0.99 & 0.30 \\ 
   & Split-EN-10 & 0.83 & 0.76 & 0.88 & 0.92 & 0.76 & 0.93 & 0.91 & 0.93 & 0.99 & 0.35 & 0.94 & 0.91 & 0.96 & 0.99 & 0.29 \\ 
   & Lasso & 0.81 & 0.73 & 0.87 & 0.90 & 0.85 & 0.88 & 0.86 & 0.90 & 0.96 & 0.52 & 0.90 & 0.85 & 0.93 & 0.97 & 0.47 \\ 
   & Elastic Net & 0.82 & 0.75 & 0.87 & 0.91 & 0.81 & 0.90 & 0.88 & 0.92 & 0.97 & 0.43 & 0.92 & 0.88 & 0.94 & 0.98 & 0.38 \\ 
   & Adaptive & 0.76 & 0.55 & 0.91 & 0.87 & 0.99 & 0.85 & 0.75 & 0.92 & 0.95 & 0.69 & 0.87 & 0.79 & 0.94 & 0.96 & 0.62 \\ 
   & Relaxed & 0.81 & 0.73 & 0.86 & 0.90 & 1.08 & 0.88 & 0.86 & 0.89 & 0.96 & 0.77 & 0.89 & 0.86 & 0.91 & 0.97 & 0.72 \\
\hline
\end{tabular}
}
\end{table}
\begin{table}[H]
\centering
\caption{Mean prediction accuracy (ACC), sensitivity (SNS) and specificity (SPC), area under ROC curve (AUC) and test-sample loss (TSL) for Scenario 1, $\rho =$ 0.8, $n =$ 50, $p =$ 1000.} 
\resizebox{\textwidth}{!}{
\begin{tabular}{llrrrrrrrrrrrrrrr}
  \hline \rule{0pt}{3ex} & & \multicolumn{5}{c}{$\mathbf{\boldsymbol{\zeta}=0.1}$} & \multicolumn{5}{c}{$\mathbf{\boldsymbol{\zeta}=0.2}$} & \multicolumn{5}{c}{$\mathbf{\boldsymbol{\zeta}=0.4}$} \\ \cmidrule(lr){1-2} \cmidrule(lr){3-7} \cmidrule(lr){8-12} \cmidrule(lr){13-17}  $\mathbf{\pi_1}$ & \textbf{Method} & \textbf{ACC} & \textbf{SNS} & \textbf{SPC} & \textbf{AUC} & \textbf{TSL} & \textbf{ACC} & \textbf{SNS} & \textbf{SPC} & \textbf{AUC} & \textbf{TSL} & \textbf{ACC} & \textbf{SNS} & \textbf{SPC} & \textbf{AUC} & \textbf{TSL}  \\ \cmidrule(lr){1-2} \cmidrule(lr){3-7} \cmidrule(lr){8-12} \cmidrule(lr){13-17}  \rule{0pt}{3ex} & Split-Lasso-10 & 0.88 & 0.85 & 0.90 & 0.96 & 0.55 & 0.95 & 0.94 & 0.96 & 1.00 & 0.22 & 0.96 & 0.94 & 0.98 & 1.00 & 0.19 \\ 
   & Split-EN-10 & 0.88 & 0.85 & 0.90 & 0.96 & 0.55 & 0.96 & 0.94 & 0.96 & 1.00 & 0.22 & 0.97 & 0.95 & 0.98 & 1.00 & 0.18 \\ 
   & Lasso & 0.87 & 0.84 & 0.89 & 0.95 & 0.61 & 0.93 & 0.90 & 0.95 & 0.99 & 0.32 & 0.94 & 0.91 & 0.96 & 0.99 & 0.28 \\ 
   & Elastic Net & 0.88 & 0.84 & 0.90 & 0.96 & 0.58 & 0.95 & 0.92 & 0.96 & 0.99 & 0.26 & 0.96 & 0.93 & 0.98 & 1.00 & 0.22 \\ 
   & Adaptive & 0.85 & 0.78 & 0.91 & 0.95 & 0.71 & 0.93 & 0.88 & 0.95 & 0.99 & 0.42 & 0.93 & 0.88 & 0.97 & 0.99 & 0.38 \\ 
   & Relaxed & 0.86 & 0.83 & 0.89 & 0.95 & 0.94 & 0.92 & 0.90 & 0.94 & 0.98 & 0.58 & 0.93 & 0.90 & 0.94 & 0.98 & 0.64 \\
\hline
\end{tabular}
}
\end{table}
\begin{table}[H]
\centering
\caption{Mean prediction accuracy (ACC), sensitivity (SNS) and specificity (SPC), area under ROC curve (AUC) and test-sample loss (TSL) for Scenario 1, $\rho =$ 0.2, $n =$ 100, $p =$ 1000.}
\resizebox{\textwidth}{!}{
\begin{tabular}{llrrrrrrrrrrrrrrr}
  \hline \rule{0pt}{3ex} & & \multicolumn{5}{c}{$\mathbf{\boldsymbol{\zeta}=0.1}$} & \multicolumn{5}{c}{$\mathbf{\boldsymbol{\zeta}=0.2}$} & \multicolumn{5}{c}{$\mathbf{\boldsymbol{\zeta}=0.4}$} \\ \cmidrule(lr){1-2} \cmidrule(lr){3-7} \cmidrule(lr){8-12} \cmidrule(lr){13-17}  $\mathbf{\pi_1}$ & \textbf{Method} & \textbf{ACC} & \textbf{SNS} & \textbf{SPC} & \textbf{AUC} & \textbf{TSL} & \textbf{ACC} & \textbf{SNS} & \textbf{SPC} & \textbf{AUC} & \textbf{TSL} & \textbf{ACC} & \textbf{SNS} & \textbf{SPC} & \textbf{AUC} & \textbf{TSL}  \\ \cmidrule(lr){1-2} \cmidrule(lr){3-7} \cmidrule(lr){8-12} \cmidrule(lr){13-17}  \rule{0pt}{3ex} & Split-Lasso-10 & 0.75 & 0.66 & 0.82 & 0.83 & 1.02 & 0.88 & 0.84 & 0.91 & 0.96 & 0.52 & 0.91 & 0.88 & 0.93 & 0.97 & 0.42 \\ 
   & Split-EN-10 & 0.75 & 0.66 & 0.82 & 0.83 & 1.02 & 0.88 & 0.84 & 0.91 & 0.96 & 0.52 & 0.91 & 0.88 & 0.93 & 0.97 & 0.41 \\ 
   & Lasso & 0.72 & 0.61 & 0.80 & 0.80 & 1.10 & 0.84 & 0.77 & 0.89 & 0.93 & 0.71 & 0.86 & 0.81 & 0.90 & 0.94 & 0.64 \\ 
   & Elastic Net & 0.73 & 0.62 & 0.81 & 0.80 & 1.09 & 0.85 & 0.79 & 0.89 & 0.94 & 0.65 & 0.88 & 0.83 & 0.91 & 0.95 & 0.57 \\ 
   & Adaptive & 0.67 & 0.39 & 0.88 & 0.74 & 1.21 & 0.82 & 0.69 & 0.90 & 0.91 & 0.83 & 0.84 & 0.74 & 0.91 & 0.93 & 0.76 \\ 
   & Relaxed & 0.72 & 0.61 & 0.80 & 0.80 & 1.12 & 0.84 & 0.77 & 0.88 & 0.92 & 0.76 & 0.86 & 0.81 & 0.89 & 0.94 & 0.71 \\
\hline
\end{tabular}
}
\end{table}
\begin{table}[H]
\centering
\caption{Mean prediction accuracy (ACC), sensitivity (SNS) and specificity (SPC), area under ROC curve (AUC) and test-sample loss (TSL) for Scenario 1, $\rho =$ 0.5, $n =$ 100, $p =$ 1000.}
\resizebox{\textwidth}{!}{
\begin{tabular}{llrrrrrrrrrrrrrrr}
  \hline \rule{0pt}{3ex} & & \multicolumn{5}{c}{$\mathbf{\boldsymbol{\zeta}=0.1}$} & \multicolumn{5}{c}{$\mathbf{\boldsymbol{\zeta}=0.2}$} & \multicolumn{5}{c}{$\mathbf{\boldsymbol{\zeta}=0.4}$} \\ \cmidrule(lr){1-2} \cmidrule(lr){3-7} \cmidrule(lr){8-12} \cmidrule(lr){13-17}  $\mathbf{\pi_1}$ & \textbf{Method} & \textbf{ACC} & \textbf{SNS} & \textbf{SPC} & \textbf{AUC} & \textbf{TSL} & \textbf{ACC} & \textbf{SNS} & \textbf{SPC} & \textbf{AUC} & \textbf{TSL} & \textbf{ACC} & \textbf{SNS} & \textbf{SPC} & \textbf{AUC} & \textbf{TSL}  \\ \cmidrule(lr){1-2} \cmidrule(lr){3-7} \cmidrule(lr){8-12} \cmidrule(lr){13-17}  \rule{0pt}{3ex} & Split-Lasso-10 & 0.84 & 0.80 & 0.87 & 0.92 & 0.72 & 0.93 & 0.91 & 0.94 & 0.99 & 0.31 & 0.95 & 0.93 & 0.96 & 0.99 & 0.26 \\ 
   & Split-EN-10 & 0.84 & 0.80 & 0.87 & 0.92 & 0.72 & 0.93 & 0.91 & 0.94 & 0.99 & 0.30 & 0.95 & 0.93 & 0.96 & 0.99 & 0.25 \\ 
   & Lasso & 0.82 & 0.77 & 0.86 & 0.91 & 0.78 & 0.91 & 0.88 & 0.92 & 0.97 & 0.43 & 0.92 & 0.88 & 0.94 & 0.98 & 0.39 \\ 
   & Elastic Net & 0.83 & 0.78 & 0.86 & 0.91 & 0.76 & 0.92 & 0.89 & 0.93 & 0.98 & 0.38 & 0.93 & 0.90 & 0.95 & 0.98 & 0.33 \\ 
   & Adaptive & 0.81 & 0.71 & 0.88 & 0.90 & 0.88 & 0.90 & 0.84 & 0.93 & 0.97 & 0.53 & 0.90 & 0.85 & 0.95 & 0.97 & 0.50 \\ 
   & Relaxed & 0.82 & 0.77 & 0.85 & 0.90 & 0.84 & 0.90 & 0.88 & 0.92 & 0.97 & 0.51 & 0.91 & 0.88 & 0.93 & 0.97 & 0.49 \\
\hline
\end{tabular}
}
\end{table}
\begin{table}[H]
\centering
\caption{Mean prediction accuracy (ACC), sensitivity (SNS) and specificity (SPC), area under ROC curve (AUC) and test-sample loss (TSL) for Scenario 1, $\rho =$ 0.8, $n =$ 100, $p =$ 1000.} \label{tab:sim_pred_1_100_0.8}
\resizebox{\textwidth}{!}{
\begin{tabular}{llrrrrrrrrrrrrrrr}
  \hline \rule{0pt}{3ex} & & \multicolumn{5}{c}{$\mathbf{\boldsymbol{\zeta}=0.1}$} & \multicolumn{5}{c}{$\mathbf{\boldsymbol{\zeta}=0.2}$} & \multicolumn{5}{c}{$\mathbf{\boldsymbol{\zeta}=0.4}$} \\ \cmidrule(lr){1-2} \cmidrule(lr){3-7} \cmidrule(lr){8-12} \cmidrule(lr){13-17}  $\mathbf{\pi_1}$ & \textbf{Method} & \textbf{ACC} & \textbf{SNS} & \textbf{SPC} & \textbf{AUC} & \textbf{TSL} & \textbf{ACC} & \textbf{SNS} & \textbf{SPC} & \textbf{AUC} & \textbf{TSL} & \textbf{ACC} & \textbf{SNS} & \textbf{SPC} & \textbf{AUC} & \textbf{TSL}  \\ \cmidrule(lr){1-2} \cmidrule(lr){3-7} \cmidrule(lr){8-12} \cmidrule(lr){13-17}  \rule{0pt}{3ex} & Split-Lasso-10 & 0.89 & 0.87 & 0.91 & 0.96 & 0.50 & 0.96 & 0.95 & 0.97 & 1.00 & 0.19 & 0.96 & 0.96 & 0.97 & 1.00 & 0.18 \\ 
   & Split-EN-10 & 0.89 & 0.87 & 0.91 & 0.96 & 0.50 & 0.96 & 0.95 & 0.97 & 1.00 & 0.19 & 0.97 & 0.96 & 0.97 & 1.00 & 0.18 \\ 
   & Lasso & 0.88 & 0.85 & 0.91 & 0.96 & 0.54 & 0.94 & 0.92 & 0.96 & 0.99 & 0.27 & 0.94 & 0.93 & 0.95 & 0.99 & 0.25 \\ 
   & Elastic Net & 0.89 & 0.86 & 0.91 & 0.96 & 0.52 & 0.95 & 0.93 & 0.97 & 0.99 & 0.23 & 0.96 & 0.94 & 0.96 & 0.99 & 0.22 \\ 
   & Adaptive & 0.88 & 0.81 & 0.92 & 0.96 & 0.64 & 0.94 & 0.90 & 0.96 & 0.99 & 0.34 & 0.94 & 0.92 & 0.96 & 0.99 & 0.33 \\ 
   & Relaxed & 0.88 & 0.85 & 0.91 & 0.96 & 0.55 & 0.93 & 0.92 & 0.95 & 0.99 & 0.47 & 0.94 & 0.93 & 0.94 & 0.99 & 0.39 \\
\hline
\end{tabular}
}
\end{table}
\begin{table}[H]
\centering
\caption{Mean recall (RCL) and precision (PRC) for Scenario 1, $\rho =$ 0.2, $n =$ 50, $p =$ 1000.} \label{tab:sim_sel_1_50_0.2}
\resizebox{0.6\textwidth}{!}{
\begin{tabular}{llrrrrrr}
  \hline \rule{0pt}{3ex} & & \multicolumn{2}{c}{$\mathbf{\boldsymbol{\zeta}=0.1}$} & \multicolumn{2}{c}{$\mathbf{\boldsymbol{\zeta}=0.2}$} & \multicolumn{2}{c}{$\mathbf{\boldsymbol{\zeta}=0.4}$} \\ \cmidrule(lr){1-2} \cmidrule(lr){3-4} \cmidrule(lr){5-6} \cmidrule(lr){7-8}  $\mathbf{\pi_1}$ & \textbf{Method} & \textbf{RCL} & \textbf{PRC} & \textbf{RCL} & \textbf{PRC} & \textbf{RCL} & \textbf{PRC}  \\ \cmidrule(lr){1-2} \cmidrule(lr){3-4} \cmidrule(lr){5-6} \cmidrule(lr){7-8}  \rule{0pt}{3ex} & Split-Lasso-10 & 0.18 & 0.14 & 0.29 & 0.25 & 0.28 & 0.45 \\ 
   & Split-EN-10 & 0.22 & 0.14 & 0.35 & 0.24 & 0.36 & 0.44 \\ 
   & Lasso & 0.02 & 0.15 & 0.03 & 0.24 & 0.03 & 0.43 \\ 
   & Elastic Net & 0.04 & 0.15 & 0.05 & 0.24 & 0.05 & 0.43 \\ 
   & Adaptive & 0.02 & 0.16 & 0.03 & 0.25 & 0.03 & 0.42 \\ 
   & Relaxed & 0.02 & 0.16 & 0.03 & 0.23 & 0.02 & 0.42 \\
\hline
\end{tabular}
}
\end{table}
\begin{table}[H]
\centering
\caption{Mean recall (RCL) and precision (PRC) for Scenario 1, $\rho =$ 0.5, $n =$ 50, $p =$ 1000.}
\resizebox{0.6\textwidth}{!}{
\begin{tabular}{llrrrrrr}
  \hline \rule{0pt}{3ex} & & \multicolumn{2}{c}{$\mathbf{\boldsymbol{\zeta}=0.1}$} & \multicolumn{2}{c}{$\mathbf{\boldsymbol{\zeta}=0.2}$} & \multicolumn{2}{c}{$\mathbf{\boldsymbol{\zeta}=0.4}$} \\ \cmidrule(lr){1-2} \cmidrule(lr){3-4} \cmidrule(lr){5-6} \cmidrule(lr){7-8}  $\mathbf{\pi_1}$ & \textbf{Method} & \textbf{RCL} & \textbf{PRC} & \textbf{RCL} & \textbf{PRC} & \textbf{RCL} & \textbf{PRC}  \\ \cmidrule(lr){1-2} \cmidrule(lr){3-4} \cmidrule(lr){5-6} \cmidrule(lr){7-8}  \rule{0pt}{3ex} & Split-Lasso-10 & 0.17 & 0.14 & 0.23 & 0.25 & 0.22 & 0.46 \\ 
   & Split-EN-10 & 0.20 & 0.13 & 0.31 & 0.23 & 0.30 & 0.45 \\ 
   & Lasso & 0.02 & 0.12 & 0.02 & 0.21 & 0.02 & 0.41 \\ 
   & Elastic Net & 0.04 & 0.13 & 0.05 & 0.21 & 0.05 & 0.41 \\ 
   & Adaptive & 0.02 & 0.13 & 0.02 & 0.20 & 0.02 & 0.40 \\ 
   & Relaxed & 0.02 & 0.12 & 0.02 & 0.19 & 0.02 & 0.41 \\
\hline
\end{tabular}
}
\end{table}
\begin{table}[H]
\centering
\caption{Mean recall (RCL) and precision (PRC) for Scenario 1, $\rho =$ 0.8, $n =$ 50, $p =$ 1000.}
\resizebox{0.6\textwidth}{!}{
\begin{tabular}{llrrrrrr}
  \hline \rule{0pt}{3ex} & & \multicolumn{2}{c}{$\mathbf{\boldsymbol{\zeta}=0.1}$} & \multicolumn{2}{c}{$\mathbf{\boldsymbol{\zeta}=0.2}$} & \multicolumn{2}{c}{$\mathbf{\boldsymbol{\zeta}=0.4}$} \\ \cmidrule(lr){1-2} \cmidrule(lr){3-4} \cmidrule(lr){5-6} \cmidrule(lr){7-8}  $\mathbf{\pi_1}$ & \textbf{Method} & \textbf{RCL} & \textbf{PRC} & \textbf{RCL} & \textbf{PRC} & \textbf{RCL} & \textbf{PRC}  \\ \cmidrule(lr){1-2} \cmidrule(lr){3-4} \cmidrule(lr){5-6} \cmidrule(lr){7-8}  \rule{0pt}{3ex} & Split-Lasso-10 & 0.13 & 0.13 & 0.20 & 0.27 & 0.18 & 0.50 \\ 
   & Split-EN-10 & 0.19 & 0.13 & 0.32 & 0.24 & 0.30 & 0.46 \\ 
   & Lasso & 0.01 & 0.11 & 0.01 & 0.18 & 0.02 & 0.39 \\ 
   & Elastic Net & 0.03 & 0.11 & 0.04 & 0.21 & 0.04 & 0.40 \\ 
   & Adaptive & 0.01 & 0.11 & 0.02 & 0.19 & 0.02 & 0.38 \\ 
   & Relaxed & 0.01 & 0.11 & 0.01 & 0.17 & 0.01 & 0.40 \\
\hline
\end{tabular}
}
\end{table}
\begin{table}[H]
\centering
\caption{Mean recall (RCL) and precision (PRC) for Scenario 1, $\rho =$ 0.2, $n =$ 100, $p =$ 1000.}
\resizebox{0.6\textwidth}{!}{
\begin{tabular}{llrrrrrr}
  \hline \rule{0pt}{3ex} & & \multicolumn{2}{c}{$\mathbf{\boldsymbol{\zeta}=0.1}$} & \multicolumn{2}{c}{$\mathbf{\boldsymbol{\zeta}=0.2}$} & \multicolumn{2}{c}{$\mathbf{\boldsymbol{\zeta}=0.4}$} \\ \cmidrule(lr){1-2} \cmidrule(lr){3-4} \cmidrule(lr){5-6} \cmidrule(lr){7-8}  $\mathbf{\pi_1}$ & \textbf{Method} & \textbf{RCL} & \textbf{PRC} & \textbf{RCL} & \textbf{PRC} & \textbf{RCL} & \textbf{PRC}  \\ \cmidrule(lr){1-2} \cmidrule(lr){3-4} \cmidrule(lr){5-6} \cmidrule(lr){7-8}  \rule{0pt}{3ex} & Split-Lasso-10 & 0.32 & 0.14 & 0.41 & 0.24 & 0.42 & 0.44 \\ 
   & Split-EN-10 & 0.32 & 0.14 & 0.47 & 0.24 & 0.48 & 0.43 \\ 
   & Lasso & 0.04 & 0.17 & 0.05 & 0.27 & 0.05 & 0.43 \\ 
   & Elastic Net & 0.06 & 0.17 & 0.08 & 0.26 & 0.07 & 0.42 \\ 
   & Adaptive & 0.04 & 0.17 & 0.05 & 0.27 & 0.05 & 0.44 \\ 
   & Relaxed & 0.04 & 0.17 & 0.05 & 0.27 & 0.05 & 0.44 \\
\hline
\end{tabular}
}
\end{table}
\begin{table}[H]
\centering
\caption{Mean recall (RCL) and precision (PRC) for Scenario 1, $\rho =$ 0.5, $n =$ 100, $p =$ 1000.}
\resizebox{0.6\textwidth}{!}{
\begin{tabular}{llrrrrrr}
  \hline \rule{0pt}{3ex} & & \multicolumn{2}{c}{$\mathbf{\boldsymbol{\zeta}=0.1}$} & \multicolumn{2}{c}{$\mathbf{\boldsymbol{\zeta}=0.2}$} & \multicolumn{2}{c}{$\mathbf{\boldsymbol{\zeta}=0.4}$} \\ \cmidrule(lr){1-2} \cmidrule(lr){3-4} \cmidrule(lr){5-6} \cmidrule(lr){7-8}  $\mathbf{\pi_1}$ & \textbf{Method} & \textbf{RCL} & \textbf{PRC} & \textbf{RCL} & \textbf{PRC} & \textbf{RCL} & \textbf{PRC}  \\ \cmidrule(lr){1-2} \cmidrule(lr){3-4} \cmidrule(lr){5-6} \cmidrule(lr){7-8}  \rule{0pt}{3ex} & Split-Lasso-10 & 0.20 & 0.14 & 0.33 & 0.25 & 0.33 & 0.46 \\ 
   & Split-EN-10 & 0.25 & 0.14 & 0.39 & 0.24 & 0.41 & 0.44 \\ 
   & Lasso & 0.04 & 0.15 & 0.04 & 0.23 & 0.04 & 0.41 \\ 
   & Elastic Net & 0.05 & 0.15 & 0.07 & 0.23 & 0.06 & 0.41 \\ 
   & Adaptive & 0.04 & 0.15 & 0.04 & 0.23 & 0.04 & 0.41 \\ 
   & Relaxed & 0.03 & 0.18 & 0.04 & 0.24 & 0.03 & 0.43 \\
\hline
\end{tabular}
}
\end{table}
\begin{table}[H]
\centering
\caption{Mean recall (RCL) and precision (PRC) for Scenario 1, $\rho =$ 0.8, $n =$ 100, $p =$ 1000.} \label{tab:sim_sel_1_100_0.8}
\resizebox{0.6\textwidth}{!}{
\begin{tabular}{llrrrrrr}
  \hline \rule{0pt}{3ex} & & \multicolumn{2}{c}{$\mathbf{\boldsymbol{\zeta}=0.1}$} & \multicolumn{2}{c}{$\mathbf{\boldsymbol{\zeta}=0.2}$} & \multicolumn{2}{c}{$\mathbf{\boldsymbol{\zeta}=0.4}$} \\ \cmidrule(lr){1-2} \cmidrule(lr){3-4} \cmidrule(lr){5-6} \cmidrule(lr){7-8}  $\mathbf{\pi_1}$ & \textbf{Method} & \textbf{RCL} & \textbf{PRC} & \textbf{RCL} & \textbf{PRC} & \textbf{RCL} & \textbf{PRC}  \\ \cmidrule(lr){1-2} \cmidrule(lr){3-4} \cmidrule(lr){5-6} \cmidrule(lr){7-8}  \rule{0pt}{3ex} & Split-Lasso-10 & 0.17 & 0.14 & 0.27 & 0.26 & 0.26 & 0.48 \\ 
   & Split-EN-10 & 0.22 & 0.13 & 0.37 & 0.24 & 0.38 & 0.45 \\ 
   & Lasso & 0.02 & 0.13 & 0.02 & 0.21 & 0.03 & 0.41 \\ 
   & Elastic Net & 0.04 & 0.13 & 0.05 & 0.21 & 0.05 & 0.41 \\ 
   & Adaptive & 0.02 & 0.14 & 0.02 & 0.21 & 0.03 & 0.40 \\ 
   & Relaxed & 0.02 & 0.13 & 0.02 & 0.20 & 0.02 & 0.41 \\
\hline
\end{tabular}
}
\end{table}
\FloatBarrier

\subsection{Scenario 2: Main Effects, Differential Correlation}
\begin{table}[H]
\centering
\caption{Mean prediction accuracy (ACC), sensitivity (SNS) and specificity (SPC), area under ROC curve (AUC) and test-sample loss (TSL) for Scenario 2, $\rho_1 =$ 0.5, $\rho_2 =$ 0.2, $n =$ 50, $p =$ 1000.} \label{tab:sim_pred_2_50_0.5_0.2}
\resizebox{\textwidth}{!}{
\begin{tabular}{llrrrrrrrrrrrrrrr}
  \hline \rule{0pt}{3ex} & & \multicolumn{5}{c}{$\mathbf{\boldsymbol{\zeta}=0.1}$} & \multicolumn{5}{c}{$\mathbf{\boldsymbol{\zeta}=0.2}$} & \multicolumn{5}{c}{$\mathbf{\boldsymbol{\zeta}=0.4}$} \\ \cmidrule(lr){1-2} \cmidrule(lr){3-7} \cmidrule(lr){8-12} \cmidrule(lr){13-17}  $\mathbf{\pi_1}$ & \textbf{Method} & \textbf{ACC} & \textbf{SNS} & \textbf{SPC} & \textbf{AUC} & \textbf{TSL} & \textbf{ACC} & \textbf{SNS} & \textbf{SPC} & \textbf{AUC} & \textbf{TSL} & \textbf{ACC} & \textbf{SNS} & \textbf{SPC} & \textbf{AUC} & \textbf{TSL}  \\ \cmidrule(lr){1-2} \cmidrule(lr){3-7} \cmidrule(lr){8-12} \cmidrule(lr){13-17}  \rule{0pt}{3ex} & Split-Lasso-10 & 0.85 & 0.79 & 0.89 & 0.93 & 0.70 & 0.90 & 0.86 & 0.93 & 0.97 & 0.45 & 0.93 & 0.91 & 0.95 & 0.99 & 0.31 \\ 
   & Split-EN-10 & 0.85 & 0.79 & 0.89 & 0.93 & 0.69 & 0.90 & 0.86 & 0.93 & 0.97 & 0.44 & 0.93 & 0.91 & 0.95 & 0.99 & 0.32 \\ 
   & Lasso & 0.82 & 0.75 & 0.87 & 0.90 & 0.84 & 0.87 & 0.81 & 0.90 & 0.95 & 0.60 & 0.89 & 0.85 & 0.92 & 0.97 & 0.51 \\ 
   & Elastic Net & 0.83 & 0.76 & 0.88 & 0.92 & 0.79 & 0.88 & 0.83 & 0.91 & 0.96 & 0.53 & 0.91 & 0.87 & 0.93 & 0.98 & 0.42 \\ 
   & Adaptive & 0.79 & 0.66 & 0.89 & 0.90 & 0.95 & 0.84 & 0.72 & 0.92 & 0.94 & 0.75 & 0.87 & 0.76 & 0.93 & 0.96 & 0.65 \\ 
   & Relaxed & 0.82 & 0.76 & 0.86 & 0.90 & 1.04 & 0.87 & 0.82 & 0.89 & 0.95 & 0.92 & 0.88 & 0.85 & 0.90 & 0.96 & 0.78 \\
\hline
\end{tabular}
}
\end{table}
\begin{table}[H]
\centering
\caption{Mean prediction accuracy (ACC), sensitivity (SNS) and specificity (SPC), area under ROC curve (AUC) and test-sample loss (TSL) for Scenario 2, $\rho_1 =$ 0.8, $\rho_2 =$ 0.2, $n =$ 50, $p =$ 1000.}
\resizebox{\textwidth}{!}{
\begin{tabular}{llrrrrrrrrrrrrrrr}
  \hline \rule{0pt}{3ex} & & \multicolumn{5}{c}{$\mathbf{\boldsymbol{\zeta}=0.1}$} & \multicolumn{5}{c}{$\mathbf{\boldsymbol{\zeta}=0.2}$} & \multicolumn{5}{c}{$\mathbf{\boldsymbol{\zeta}=0.4}$} \\ \cmidrule(lr){1-2} \cmidrule(lr){3-7} \cmidrule(lr){8-12} \cmidrule(lr){13-17}  $\mathbf{\pi_1}$ & \textbf{Method} & \textbf{ACC} & \textbf{SNS} & \textbf{SPC} & \textbf{AUC} & \textbf{TSL} & \textbf{ACC} & \textbf{SNS} & \textbf{SPC} & \textbf{AUC} & \textbf{TSL} & \textbf{ACC} & \textbf{SNS} & \textbf{SPC} & \textbf{AUC} & \textbf{TSL}  \\ \cmidrule(lr){1-2} \cmidrule(lr){3-7} \cmidrule(lr){8-12} \cmidrule(lr){13-17}  \rule{0pt}{3ex} & Split-Lasso-10 & 0.90 & 0.85 & 0.93 & 0.97 & 0.49 & 0.93 & 0.91 & 0.95 & 0.99 & 0.32 & 0.95 & 0.92 & 0.97 & 0.99 & 0.24 \\ 
   & Split-EN-10 & 0.90 & 0.84 & 0.93 & 0.97 & 0.49 & 0.93 & 0.91 & 0.95 & 0.99 & 0.33 & 0.95 & 0.93 & 0.97 & 1.00 & 0.25 \\ 
   & Lasso & 0.87 & 0.80 & 0.91 & 0.95 & 0.63 & 0.90 & 0.86 & 0.92 & 0.97 & 0.48 & 0.92 & 0.87 & 0.95 & 0.98 & 0.38 \\ 
   & Elastic Net & 0.88 & 0.81 & 0.92 & 0.96 & 0.60 & 0.91 & 0.87 & 0.94 & 0.98 & 0.42 & 0.93 & 0.89 & 0.96 & 0.99 & 0.33 \\ 
   & Adaptive & 0.85 & 0.73 & 0.93 & 0.95 & 0.78 & 0.89 & 0.82 & 0.95 & 0.97 & 0.58 & 0.92 & 0.85 & 0.96 & 0.98 & 0.47 \\ 
   & Relaxed & 0.86 & 0.81 & 0.90 & 0.94 & 1.16 & 0.89 & 0.87 & 0.90 & 0.96 & 0.80 & 0.91 & 0.88 & 0.93 & 0.98 & 0.99 \\
\hline
\end{tabular}
}
\end{table}
\begin{table}[H]
\centering
\caption{Mean prediction accuracy (ACC), sensitivity (SNS) and specificity (SPC), area under ROC curve (AUC) and test-sample loss (TSL) for Scenario 2, $\rho_1 =$ 0.8, $\rho_2 =$ 0.5, $n =$ 50, $p =$ 1000.}
\resizebox{\textwidth}{!}{
\begin{tabular}{llrrrrrrrrrrrrrrr}
  \hline \rule{0pt}{3ex} & & \multicolumn{5}{c}{$\mathbf{\boldsymbol{\zeta}=0.1}$} & \multicolumn{5}{c}{$\mathbf{\boldsymbol{\zeta}=0.2}$} & \multicolumn{5}{c}{$\mathbf{\boldsymbol{\zeta}=0.4}$} \\ \cmidrule(lr){1-2} \cmidrule(lr){3-7} \cmidrule(lr){8-12} \cmidrule(lr){13-17}  $\mathbf{\pi_1}$ & \textbf{Method} & \textbf{ACC} & \textbf{SNS} & \textbf{SPC} & \textbf{AUC} & \textbf{TSL} & \textbf{ACC} & \textbf{SNS} & \textbf{SPC} & \textbf{AUC} & \textbf{TSL} & \textbf{ACC} & \textbf{SNS} & \textbf{SPC} & \textbf{AUC} & \textbf{TSL}  \\ \cmidrule(lr){1-2} \cmidrule(lr){3-7} \cmidrule(lr){8-12} \cmidrule(lr){13-17}  \rule{0pt}{3ex} & Split-Lasso-10 & 0.89 & 0.85 & 0.93 & 0.97 & 0.50 & 0.93 & 0.91 & 0.94 & 0.98 & 0.34 & 0.95 & 0.93 & 0.97 & 0.99 & 0.24 \\ 
   & Split-EN-10 & 0.89 & 0.84 & 0.93 & 0.97 & 0.52 & 0.93 & 0.91 & 0.94 & 0.98 & 0.35 & 0.95 & 0.93 & 0.97 & 0.99 & 0.25 \\ 
   & Lasso & 0.87 & 0.82 & 0.91 & 0.95 & 0.62 & 0.90 & 0.87 & 0.92 & 0.97 & 0.46 & 0.92 & 0.88 & 0.95 & 0.98 & 0.36 \\ 
   & Elastic Net & 0.88 & 0.83 & 0.92 & 0.96 & 0.57 & 0.91 & 0.88 & 0.93 & 0.98 & 0.40 & 0.93 & 0.90 & 0.96 & 0.99 & 0.31 \\ 
   & Adaptive & 0.86 & 0.75 & 0.93 & 0.95 & 0.75 & 0.89 & 0.82 & 0.94 & 0.97 & 0.57 & 0.92 & 0.85 & 0.96 & 0.98 & 0.46 \\ 
   & Relaxed & 0.87 & 0.81 & 0.90 & 0.95 & 1.26 & 0.89 & 0.87 & 0.90 & 0.97 & 0.85 & 0.91 & 0.88 & 0.93 & 0.98 & 0.96 \\
\hline
\end{tabular}
}
\end{table}
\begin{table}[H]
\centering
\caption{Mean prediction accuracy (ACC), sensitivity (SNS) and specificity (SPC), area under ROC curve (AUC) and test-sample loss (TSL) for Scenario 2, $\rho_1 =$ 0.5, $\rho_2 =$ 0.2, $n =$ 100, $p =$ 1000.}
\resizebox{\textwidth}{!}{
\begin{tabular}{llrrrrrrrrrrrrrrr}
  \hline \rule{0pt}{3ex} & & \multicolumn{5}{c}{$\mathbf{\boldsymbol{\zeta}=0.1}$} & \multicolumn{5}{c}{$\mathbf{\boldsymbol{\zeta}=0.2}$} & \multicolumn{5}{c}{$\mathbf{\boldsymbol{\zeta}=0.4}$} \\ \cmidrule(lr){1-2} \cmidrule(lr){3-7} \cmidrule(lr){8-12} \cmidrule(lr){13-17}  $\mathbf{\pi_1}$ & \textbf{Method} & \textbf{ACC} & \textbf{SNS} & \textbf{SPC} & \textbf{AUC} & \textbf{TSL} & \textbf{ACC} & \textbf{SNS} & \textbf{SPC} & \textbf{AUC} & \textbf{TSL} & \textbf{ACC} & \textbf{SNS} & \textbf{SPC} & \textbf{AUC} & \textbf{TSL}  \\ \cmidrule(lr){1-2} \cmidrule(lr){3-7} \cmidrule(lr){8-12} \cmidrule(lr){13-17}  \rule{0pt}{3ex} & Split-Lasso-10 & 0.85 & 0.80 & 0.89 & 0.94 & 0.65 & 0.91 & 0.88 & 0.94 & 0.98 & 0.39 & 0.95 & 0.92 & 0.96 & 0.99 & 0.26 \\ 
   & Split-EN-10 & 0.85 & 0.80 & 0.89 & 0.94 & 0.65 & 0.91 & 0.88 & 0.94 & 0.98 & 0.39 & 0.95 & 0.92 & 0.96 & 0.99 & 0.26 \\ 
   & Lasso & 0.84 & 0.78 & 0.88 & 0.92 & 0.74 & 0.89 & 0.84 & 0.92 & 0.96 & 0.51 & 0.91 & 0.87 & 0.94 & 0.98 & 0.40 \\ 
   & Elastic Net & 0.84 & 0.78 & 0.88 & 0.93 & 0.72 & 0.90 & 0.85 & 0.92 & 0.97 & 0.47 & 0.92 & 0.89 & 0.95 & 0.98 & 0.35 \\ 
   & Adaptive & 0.83 & 0.73 & 0.90 & 0.92 & 0.82 & 0.89 & 0.82 & 0.93 & 0.96 & 0.59 & 0.90 & 0.84 & 0.95 & 0.97 & 0.50 \\ 
   & Relaxed & 0.83 & 0.76 & 0.87 & 0.92 & 0.82 & 0.88 & 0.84 & 0.91 & 0.96 & 0.60 & 0.90 & 0.87 & 0.93 & 0.97 & 0.58 \\
\hline
\end{tabular}
}
\end{table}
\begin{table}[H]
\centering
\caption{Mean prediction accuracy (ACC), sensitivity (SNS) and specificity (SPC), area under ROC curve (AUC) and test-sample loss (TSL) for Scenario 2, $\rho_1 =$ 0.8, $\rho_2 =$ 0.2, $n =$ 100, $p =$ 1000.}
\resizebox{\textwidth}{!}{
\begin{tabular}{llrrrrrrrrrrrrrrr}
  \hline \rule{0pt}{3ex} & & \multicolumn{5}{c}{$\mathbf{\boldsymbol{\zeta}=0.1}$} & \multicolumn{5}{c}{$\mathbf{\boldsymbol{\zeta}=0.2}$} & \multicolumn{5}{c}{$\mathbf{\boldsymbol{\zeta}=0.4}$} \\ \cmidrule(lr){1-2} \cmidrule(lr){3-7} \cmidrule(lr){8-12} \cmidrule(lr){13-17}  $\mathbf{\pi_1}$ & \textbf{Method} & \textbf{ACC} & \textbf{SNS} & \textbf{SPC} & \textbf{AUC} & \textbf{TSL} & \textbf{ACC} & \textbf{SNS} & \textbf{SPC} & \textbf{AUC} & \textbf{TSL} & \textbf{ACC} & \textbf{SNS} & \textbf{SPC} & \textbf{AUC} & \textbf{TSL}  \\ \cmidrule(lr){1-2} \cmidrule(lr){3-7} \cmidrule(lr){8-12} \cmidrule(lr){13-17}  \rule{0pt}{3ex} & Split-Lasso-10 & 0.90 & 0.87 & 0.92 & 0.97 & 0.46 & 0.94 & 0.92 & 0.96 & 0.99 & 0.26 & 0.96 & 0.96 & 0.96 & 1.00 & 0.20 \\ 
   & Split-EN-10 & 0.90 & 0.86 & 0.92 & 0.97 & 0.46 & 0.94 & 0.92 & 0.96 & 0.99 & 0.26 & 0.96 & 0.96 & 0.97 & 1.00 & 0.21 \\ 
   & Lasso & 0.88 & 0.84 & 0.91 & 0.96 & 0.56 & 0.92 & 0.89 & 0.94 & 0.98 & 0.36 & 0.94 & 0.92 & 0.94 & 0.99 & 0.30 \\ 
   & Elastic Net & 0.89 & 0.84 & 0.91 & 0.96 & 0.54 & 0.93 & 0.90 & 0.95 & 0.99 & 0.33 & 0.95 & 0.93 & 0.95 & 0.99 & 0.26 \\ 
   & Adaptive & 0.88 & 0.82 & 0.93 & 0.96 & 0.64 & 0.92 & 0.88 & 0.95 & 0.98 & 0.43 & 0.94 & 0.91 & 0.95 & 0.99 & 0.36 \\ 
   & Relaxed & 0.88 & 0.83 & 0.91 & 0.95 & 0.64 & 0.92 & 0.89 & 0.93 & 0.98 & 0.54 & 0.93 & 0.92 & 0.93 & 0.98 & 0.45 \\
\hline
\end{tabular}
}
\end{table}
\begin{table}[H]
\centering
\caption{Mean prediction accuracy (ACC), sensitivity (SNS) and specificity (SPC), area under ROC curve (AUC) and test-sample loss (TSL) for Scenario 2, $\rho_1 =$ 0.8, $\rho_2 =$ 0.5, $n =$ 100, $p =$ 1000.}  \label{tab:sim_pred_2_100_0.8_0.5}
\resizebox{\textwidth}{!}{
\begin{tabular}{llrrrrrrrrrrrrrrr}
  \hline \rule{0pt}{3ex} & & \multicolumn{5}{c}{$\mathbf{\boldsymbol{\zeta}=0.1}$} & \multicolumn{5}{c}{$\mathbf{\boldsymbol{\zeta}=0.2}$} & \multicolumn{5}{c}{$\mathbf{\boldsymbol{\zeta}=0.4}$} \\ \cmidrule(lr){1-2} \cmidrule(lr){3-7} \cmidrule(lr){8-12} \cmidrule(lr){13-17}  $\mathbf{\pi_1}$ & \textbf{Method} & \textbf{ACC} & \textbf{SNS} & \textbf{SPC} & \textbf{AUC} & \textbf{TSL} & \textbf{ACC} & \textbf{SNS} & \textbf{SPC} & \textbf{AUC} & \textbf{TSL} & \textbf{ACC} & \textbf{SNS} & \textbf{SPC} & \textbf{AUC} & \textbf{TSL}  \\ \cmidrule(lr){1-2} \cmidrule(lr){3-7} \cmidrule(lr){8-12} \cmidrule(lr){13-17}  \rule{0pt}{3ex} & Split-Lasso-10 & 0.90 & 0.87 & 0.92 & 0.97 & 0.48 & 0.94 & 0.92 & 0.96 & 0.99 & 0.26 & 0.96 & 0.96 & 0.97 & 1.00 & 0.19 \\ 
   & Split-EN-10 & 0.90 & 0.86 & 0.92 & 0.97 & 0.48 & 0.94 & 0.92 & 0.96 & 0.99 & 0.26 & 0.96 & 0.96 & 0.97 & 1.00 & 0.19 \\ 
   & Lasso & 0.88 & 0.84 & 0.91 & 0.96 & 0.55 & 0.93 & 0.89 & 0.94 & 0.98 & 0.35 & 0.94 & 0.92 & 0.95 & 0.99 & 0.28 \\ 
   & Elastic Net & 0.89 & 0.85 & 0.91 & 0.96 & 0.53 & 0.93 & 0.90 & 0.95 & 0.99 & 0.32 & 0.95 & 0.94 & 0.96 & 0.99 & 0.24 \\ 
   & Adaptive & 0.88 & 0.81 & 0.93 & 0.96 & 0.63 & 0.92 & 0.88 & 0.95 & 0.98 & 0.43 & 0.94 & 0.92 & 0.96 & 0.99 & 0.35 \\ 
   & Relaxed & 0.88 & 0.84 & 0.91 & 0.96 & 0.61 & 0.92 & 0.90 & 0.94 & 0.98 & 0.52 & 0.93 & 0.92 & 0.94 & 0.99 & 0.53 \\
\hline
\end{tabular}
}
\end{table}
\begin{table}[H]
\centering
\caption{Mean recall (RCL) and precision (PRC) for Scenario 2, $\rho_1 =$ 0.5, $\rho_2 =$ 0.2, $n =$ 50, $p =$ 1000.} \label{tab:sim_sel_2_50_0.5_0.2}
\resizebox{0.6\textwidth}{!}{
\begin{tabular}{llrrrrrr}
  \hline \rule{0pt}{3ex} & & \multicolumn{2}{c}{$\mathbf{\boldsymbol{\zeta}=0.1}$} & \multicolumn{2}{c}{$\mathbf{\boldsymbol{\zeta}=0.2}$} & \multicolumn{2}{c}{$\mathbf{\boldsymbol{\zeta}=0.4}$} \\ \cmidrule(lr){1-2} \cmidrule(lr){3-4} \cmidrule(lr){5-6} \cmidrule(lr){7-8}  $\mathbf{\pi_1}$ & \textbf{Method} & \textbf{RCL} & \textbf{PRC} & \textbf{RCL} & \textbf{PRC} & \textbf{RCL} & \textbf{PRC}  \\ \cmidrule(lr){1-2} \cmidrule(lr){3-4} \cmidrule(lr){5-6} \cmidrule(lr){7-8}  \rule{0pt}{3ex} & Split-Lasso-10 & 0.51 & 0.35 & 0.45 & 0.48 & 0.34 & 0.67 \\ 
   & Split-EN-10 & 0.54 & 0.35 & 0.50 & 0.46 & 0.43 & 0.66 \\ 
   & Lasso & 0.11 & 0.59 & 0.07 & 0.73 & 0.05 & 0.82 \\ 
   & Elastic Net & 0.16 & 0.55 & 0.13 & 0.68 & 0.09 & 0.80 \\ 
   & Adaptive & 0.10 & 0.58 & 0.07 & 0.69 & 0.04 & 0.80 \\ 
   & Relaxed & 0.10 & 0.64 & 0.07 & 0.81 & 0.04 & 0.90 \\
\hline
\end{tabular}
}
\end{table}
\begin{table}[H]
\centering
\caption{Mean recall (RCL) and precision (PRC) for Scenario 2, $\rho_1 =$ 0.8, $\rho_2 =$ 0.2, $n =$ 50, $p =$ 1000.}
\resizebox{0.6\textwidth}{!}{
\begin{tabular}{llrrrrrr}
  \hline \rule{0pt}{3ex} & & \multicolumn{2}{c}{$\mathbf{\boldsymbol{\zeta}=0.1}$} & \multicolumn{2}{c}{$\mathbf{\boldsymbol{\zeta}=0.2}$} & \multicolumn{2}{c}{$\mathbf{\boldsymbol{\zeta}=0.4}$} \\ \cmidrule(lr){1-2} \cmidrule(lr){3-4} \cmidrule(lr){5-6} \cmidrule(lr){7-8}  $\mathbf{\pi_1}$ & \textbf{Method} & \textbf{RCL} & \textbf{PRC} & \textbf{RCL} & \textbf{PRC} & \textbf{RCL} & \textbf{PRC}  \\ \cmidrule(lr){1-2} \cmidrule(lr){3-4} \cmidrule(lr){5-6} \cmidrule(lr){7-8}  \rule{0pt}{3ex} & Split-Lasso-10 & 0.53 & 0.28 & 0.39 & 0.41 & 0.26 & 0.57 \\ 
   & Split-EN-10 & 0.73 & 0.31 & 0.62 & 0.44 & 0.45 & 0.63 \\ 
   & Lasso & 0.08 & 0.51 & 0.05 & 0.55 & 0.03 & 0.66 \\ 
   & Elastic Net & 0.18 & 0.55 & 0.13 & 0.65 & 0.08 & 0.73 \\ 
   & Adaptive & 0.07 & 0.50 & 0.05 & 0.52 & 0.03 & 0.62 \\ 
   & Relaxed & 0.07 & 0.69 & 0.04 & 0.82 & 0.02 & 0.91 \\
\hline
\end{tabular}
}
\end{table}
\begin{table}[H]
\centering
\caption{Mean recall (RCL) and precision (PRC) for Scenario 2, $\rho_1 =$ 0.8, $\rho_2 =$ 0.5, $n =$ 50, $p =$ 1000.}
\resizebox{0.6\textwidth}{!}{
\begin{tabular}{llrrrrrr}
  \hline \rule{0pt}{3ex} & & \multicolumn{2}{c}{$\mathbf{\boldsymbol{\zeta}=0.1}$} & \multicolumn{2}{c}{$\mathbf{\boldsymbol{\zeta}=0.2}$} & \multicolumn{2}{c}{$\mathbf{\boldsymbol{\zeta}=0.4}$} \\ \cmidrule(lr){1-2} \cmidrule(lr){3-4} \cmidrule(lr){5-6} \cmidrule(lr){7-8}  $\mathbf{\pi_1}$ & \textbf{Method} & \textbf{RCL} & \textbf{PRC} & \textbf{RCL} & \textbf{PRC} & \textbf{RCL} & \textbf{PRC}  \\ \cmidrule(lr){1-2} \cmidrule(lr){3-4} \cmidrule(lr){5-6} \cmidrule(lr){7-8}  \rule{0pt}{3ex} & Split-Lasso-10 & 0.45 & 0.35 & 0.35 & 0.47 & 0.26 & 0.63 \\ 
   & Split-EN-10 & 0.61 & 0.36 & 0.53 & 0.48 & 0.42 & 0.66 \\ 
   & Lasso & 0.08 & 0.52 & 0.05 & 0.57 & 0.03 & 0.70 \\ 
   & Elastic Net & 0.15 & 0.55 & 0.11 & 0.62 & 0.08 & 0.74 \\ 
   & Adaptive & 0.07 & 0.49 & 0.04 & 0.56 & 0.03 & 0.68 \\ 
   & Relaxed & 0.07 & 0.62 & 0.04 & 0.78 & 0.02 & 0.84 \\
\hline
\end{tabular}
}
\end{table}
\begin{table}[H]
\centering
\caption{Mean recall (RCL) and precision (PRC) for Scenario 2, $\rho_1 =$ 0.5, $\rho_2 =$ 0.2, $n =$ 100, $p =$ 1000.}
\resizebox{0.6\textwidth}{!}{
\begin{tabular}{llrrrrrr}
  \hline \rule{0pt}{3ex} & & \multicolumn{2}{c}{$\mathbf{\boldsymbol{\zeta}=0.1}$} & \multicolumn{2}{c}{$\mathbf{\boldsymbol{\zeta}=0.2}$} & \multicolumn{2}{c}{$\mathbf{\boldsymbol{\zeta}=0.4}$} \\ \cmidrule(lr){1-2} \cmidrule(lr){3-4} \cmidrule(lr){5-6} \cmidrule(lr){7-8}  $\mathbf{\pi_1}$ & \textbf{Method} & \textbf{RCL} & \textbf{PRC} & \textbf{RCL} & \textbf{PRC} & \textbf{RCL} & \textbf{PRC}  \\ \cmidrule(lr){1-2} \cmidrule(lr){3-4} \cmidrule(lr){5-6} \cmidrule(lr){7-8}  \rule{0pt}{3ex} & Split-Lasso-10 & 0.68 & 0.25 & 0.59 & 0.38 & 0.46 & 0.63 \\ 
   & Split-EN-10 & 0.72 & 0.25 & 0.67 & 0.36 & 0.56 & 0.60 \\ 
   & Lasso & 0.15 & 0.60 & 0.11 & 0.67 & 0.07 & 0.83 \\ 
   & Elastic Net & 0.21 & 0.55 & 0.16 & 0.64 & 0.11 & 0.81 \\ 
   & Adaptive & 0.15 & 0.56 & 0.10 & 0.66 & 0.07 & 0.82 \\ 
   & Relaxed & 0.14 & 0.66 & 0.10 & 0.74 & 0.06 & 0.89 \\
\hline
\end{tabular}
}
\end{table}
\begin{table}[H]
\centering
\caption{Mean recall (RCL) and precision (PRC) for Scenario 2, $\rho_1 =$ 0.8, $\rho_2 =$ 0.2, $n =$ 100, $p =$ 1000.}
\resizebox{0.6\textwidth}{!}{
\begin{tabular}{llrrrrrr}
  \hline \rule{0pt}{3ex} & & \multicolumn{2}{c}{$\mathbf{\boldsymbol{\zeta}=0.1}$} & \multicolumn{2}{c}{$\mathbf{\boldsymbol{\zeta}=0.2}$} & \multicolumn{2}{c}{$\mathbf{\boldsymbol{\zeta}=0.4}$} \\ \cmidrule(lr){1-2} \cmidrule(lr){3-4} \cmidrule(lr){5-6} \cmidrule(lr){7-8}  $\mathbf{\pi_1}$ & \textbf{Method} & \textbf{RCL} & \textbf{PRC} & \textbf{RCL} & \textbf{PRC} & \textbf{RCL} & \textbf{PRC}  \\ \cmidrule(lr){1-2} \cmidrule(lr){3-4} \cmidrule(lr){5-6} \cmidrule(lr){7-8}  \rule{0pt}{3ex} & Split-Lasso-10 & 0.65 & 0.23 & 0.51 & 0.36 & 0.37 & 0.55 \\ 
   & Split-EN-10 & 0.82 & 0.20 & 0.72 & 0.37 & 0.55 & 0.57 \\ 
   & Lasso & 0.12 & 0.48 & 0.08 & 0.58 & 0.05 & 0.65 \\ 
   & Elastic Net & 0.22 & 0.50 & 0.16 & 0.61 & 0.10 & 0.71 \\ 
   & Adaptive & 0.12 & 0.50 & 0.08 & 0.55 & 0.05 & 0.64 \\ 
   & Relaxed & 0.09 & 0.69 & 0.06 & 0.84 & 0.04 & 0.87 \\
\hline
\end{tabular}
}
\end{table}
\begin{table}[H]
\centering
\caption{Mean recall (RCL) and precision (PRC) for Scenario 2, $\rho_1 =$ 0.8, $\rho_2 =$ 0.5, $n =$ 100, $p =$ 1000.} \label{tab:sim_sel_2_100_0.8_0.5}
\resizebox{0.6\textwidth}{!}{
\begin{tabular}{llrrrrrr}
  \hline \rule{0pt}{3ex} & & \multicolumn{2}{c}{$\mathbf{\boldsymbol{\zeta}=0.1}$} & \multicolumn{2}{c}{$\mathbf{\boldsymbol{\zeta}=0.2}$} & \multicolumn{2}{c}{$\mathbf{\boldsymbol{\zeta}=0.4}$} \\ \cmidrule(lr){1-2} \cmidrule(lr){3-4} \cmidrule(lr){5-6} \cmidrule(lr){7-8}  $\mathbf{\pi_1}$ & \textbf{Method} & \textbf{RCL} & \textbf{PRC} & \textbf{RCL} & \textbf{PRC} & \textbf{RCL} & \textbf{PRC}  \\ \cmidrule(lr){1-2} \cmidrule(lr){3-4} \cmidrule(lr){5-6} \cmidrule(lr){7-8}  \rule{0pt}{3ex} & Split-Lasso-10 & 0.59 & 0.28 & 0.50 & 0.42 & 0.36 & 0.62 \\ 
   & Split-EN-10 & 0.70 & 0.29 & 0.67 & 0.40 & 0.53 & 0.61 \\ 
   & Lasso & 0.11 & 0.54 & 0.07 & 0.59 & 0.05 & 0.73 \\ 
   & Elastic Net & 0.19 & 0.53 & 0.15 & 0.61 & 0.10 & 0.76 \\ 
   & Adaptive & 0.11 & 0.51 & 0.07 & 0.60 & 0.05 & 0.72 \\ 
   & Relaxed & 0.09 & 0.67 & 0.07 & 0.73 & 0.04 & 0.83 \\
\hline
\end{tabular}
}
\end{table}
\FloatBarrier

\subsection{Scenario 3: Main Effects, Block Correlation}
\begin{table}[H]
\centering
\caption{Mean prediction accuracy (ACC), sensitivity (SNS) and specificity (SPC), area under ROC curve (AUC) and test-sample loss (TSL) for Scenario 3, $\rho_1 =$ 0.5, $\rho_2 =$ 0.2, $n =$ 50, $p =$ 1000.} \label{tab:sim_pred_3_50_0.5_0.2}
\resizebox{\textwidth}{!}{
\begin{tabular}{llrrrrrrrrrrrrrrr}
  \hline \rule{0pt}{3ex} & & \multicolumn{5}{c}{$\mathbf{\boldsymbol{\zeta}=0.1}$} & \multicolumn{5}{c}{$\mathbf{\boldsymbol{\zeta}=0.2}$} & \multicolumn{5}{c}{$\mathbf{\boldsymbol{\zeta}=0.4}$} \\ \cmidrule(lr){1-2} \cmidrule(lr){3-7} \cmidrule(lr){8-12} \cmidrule(lr){13-17}  $\mathbf{\pi_1}$ & \textbf{Method} & \textbf{ACC} & \textbf{SNS} & \textbf{SPC} & \textbf{AUC} & \textbf{TSL} & \textbf{ACC} & \textbf{SNS} & \textbf{SPC} & \textbf{AUC} & \textbf{TSL} & \textbf{ACC} & \textbf{SNS} & \textbf{SPC} & \textbf{AUC} & \textbf{TSL}  \\ \cmidrule(lr){1-2} \cmidrule(lr){3-7} \cmidrule(lr){8-12} \cmidrule(lr){13-17}  \rule{0pt}{3ex} & Split-Lasso-10 & 0.78 & 0.68 & 0.84 & 0.86 & 0.95 & 0.85 & 0.76 & 0.90 & 0.93 & 0.69 & 0.89 & 0.84 & 0.93 & 0.97 & 0.49 \\ 
   & Split-EN-10 & 0.78 & 0.68 & 0.84 & 0.87 & 0.94 & 0.85 & 0.76 & 0.90 & 0.93 & 0.69 & 0.90 & 0.85 & 0.93 & 0.97 & 0.47 \\ 
   & Lasso & 0.74 & 0.62 & 0.82 & 0.82 & 1.09 & 0.80 & 0.69 & 0.87 & 0.88 & 0.92 & 0.82 & 0.74 & 0.88 & 0.91 & 0.78 \\ 
   & Elastic Net & 0.75 & 0.64 & 0.82 & 0.84 & 1.03 & 0.82 & 0.72 & 0.88 & 0.90 & 0.83 & 0.85 & 0.78 & 0.90 & 0.94 & 0.66 \\ 
   & Adaptive & 0.67 & 0.40 & 0.86 & 0.76 & 1.20 & 0.74 & 0.48 & 0.91 & 0.82 & 1.06 & 0.78 & 0.60 & 0.90 & 0.87 & 0.95 \\ 
   & Relaxed & 0.72 & 0.62 & 0.79 & 0.80 & 1.32 & 0.79 & 0.69 & 0.86 & 0.88 & 1.10 & 0.81 & 0.75 & 0.86 & 0.90 & 1.04 \\
\hline
\end{tabular}
}
\end{table}
\begin{table}[H]
\centering
\caption{Mean prediction accuracy (ACC), sensitivity (SNS) and specificity (SPC), area under ROC curve (AUC) and test-sample loss (TSL) for Scenario 3, $\rho_1 =$ 0.8, $\rho_2 =$ 0.2, $n =$ 50, $p =$ 1000.}
\resizebox{\textwidth}{!}{
\begin{tabular}{llrrrrrrrrrrrrrrr}
  \hline \rule{0pt}{3ex} & & \multicolumn{5}{c}{$\mathbf{\boldsymbol{\zeta}=0.1}$} & \multicolumn{5}{c}{$\mathbf{\boldsymbol{\zeta}=0.2}$} & \multicolumn{5}{c}{$\mathbf{\boldsymbol{\zeta}=0.4}$} \\ \cmidrule(lr){1-2} \cmidrule(lr){3-7} \cmidrule(lr){8-12} \cmidrule(lr){13-17}  $\mathbf{\pi_1}$ & \textbf{Method} & \textbf{ACC} & \textbf{SNS} & \textbf{SPC} & \textbf{AUC} & \textbf{TSL} & \textbf{ACC} & \textbf{SNS} & \textbf{SPC} & \textbf{AUC} & \textbf{TSL} & \textbf{ACC} & \textbf{SNS} & \textbf{SPC} & \textbf{AUC} & \textbf{TSL}  \\ \cmidrule(lr){1-2} \cmidrule(lr){3-7} \cmidrule(lr){8-12} \cmidrule(lr){13-17}  \rule{0pt}{3ex} & Split-Lasso-10 & 0.81 & 0.75 & 0.86 & 0.91 & 0.81 & 0.85 & 0.79 & 0.90 & 0.94 & 0.65 & 0.88 & 0.82 & 0.92 & 0.96 & 0.55 \\ 
   & Split-EN-10 & 0.81 & 0.75 & 0.86 & 0.91 & 0.82 & 0.86 & 0.79 & 0.91 & 0.94 & 0.64 & 0.88 & 0.83 & 0.92 & 0.96 & 0.53 \\ 
   & Lasso & 0.78 & 0.71 & 0.84 & 0.87 & 0.95 & 0.80 & 0.70 & 0.87 & 0.88 & 0.88 & 0.82 & 0.73 & 0.88 & 0.91 & 0.80 \\ 
   & Elastic Net & 0.79 & 0.71 & 0.85 & 0.88 & 0.91 & 0.82 & 0.72 & 0.88 & 0.90 & 0.81 & 0.84 & 0.77 & 0.89 & 0.93 & 0.71 \\ 
   & Adaptive & 0.73 & 0.53 & 0.87 & 0.84 & 1.09 & 0.75 & 0.56 & 0.89 & 0.84 & 1.02 & 0.77 & 0.59 & 0.89 & 0.88 & 0.96 \\ 
   & Relaxed & 0.77 & 0.70 & 0.83 & 0.86 & 1.14 & 0.79 & 0.70 & 0.86 & 0.88 & 1.02 & 0.82 & 0.75 & 0.86 & 0.90 & 1.05 \\
\hline
\end{tabular}
}
\end{table}
\begin{table}[H]
\centering
\caption{Mean prediction accuracy (ACC), sensitivity (SNS) and specificity (SPC), area under ROC curve (AUC) and test-sample loss (TSL) for Scenario 3, $\rho_1 =$ 0.8, $\rho_2 =$ 0.5, $n =$ 50, $p =$ 1000.}
\resizebox{\textwidth}{!}{
\begin{tabular}{llrrrrrrrrrrrrrrr}
  \hline \rule{0pt}{3ex} & & \multicolumn{5}{c}{$\mathbf{\boldsymbol{\zeta}=0.1}$} & \multicolumn{5}{c}{$\mathbf{\boldsymbol{\zeta}=0.2}$} & \multicolumn{5}{c}{$\mathbf{\boldsymbol{\zeta}=0.4}$} \\ \cmidrule(lr){1-2} \cmidrule(lr){3-7} \cmidrule(lr){8-12} \cmidrule(lr){13-17}  $\mathbf{\pi_1}$ & \textbf{Method} & \textbf{ACC} & \textbf{SNS} & \textbf{SPC} & \textbf{AUC} & \textbf{TSL} & \textbf{ACC} & \textbf{SNS} & \textbf{SPC} & \textbf{AUC} & \textbf{TSL} & \textbf{ACC} & \textbf{SNS} & \textbf{SPC} & \textbf{AUC} & \textbf{TSL}  \\ \cmidrule(lr){1-2} \cmidrule(lr){3-7} \cmidrule(lr){8-12} \cmidrule(lr){13-17}  \rule{0pt}{3ex} & Split-Lasso-10 & 0.86 & 0.82 & 0.89 & 0.94 & 0.65 & 0.91 & 0.88 & 0.92 & 0.98 & 0.43 & 0.94 & 0.91 & 0.95 & 0.99 & 0.31 \\ 
   & Split-EN-10 & 0.86 & 0.82 & 0.89 & 0.94 & 0.64 & 0.91 & 0.89 & 0.92 & 0.98 & 0.42 & 0.94 & 0.92 & 0.96 & 0.99 & 0.29 \\ 
   & Lasso & 0.84 & 0.78 & 0.87 & 0.92 & 0.77 & 0.88 & 0.85 & 0.89 & 0.96 & 0.58 & 0.90 & 0.86 & 0.93 & 0.97 & 0.47 \\ 
   & Elastic Net & 0.85 & 0.80 & 0.88 & 0.93 & 0.70 & 0.89 & 0.86 & 0.91 & 0.97 & 0.50 & 0.92 & 0.88 & 0.94 & 0.98 & 0.39 \\ 
   & Adaptive & 0.81 & 0.70 & 0.89 & 0.91 & 0.88 & 0.85 & 0.75 & 0.91 & 0.94 & 0.72 & 0.87 & 0.76 & 0.94 & 0.96 & 0.64 \\ 
   & Relaxed & 0.83 & 0.78 & 0.86 & 0.91 & 1.04 & 0.87 & 0.84 & 0.88 & 0.95 & 0.85 & 0.88 & 0.85 & 0.91 & 0.96 & 0.86 \\
\hline
\end{tabular}
}
\end{table}
\begin{table}[H]
\centering
\caption{Mean prediction accuracy (ACC), sensitivity (SNS) and specificity (SPC), area under ROC curve (AUC) and test-sample loss (TSL) for Scenario 3, $\rho_1 =$ 0.5, $\rho_2 =$ 0.2, $n =$ 100, $p =$ 1000.}
\resizebox{\textwidth}{!}{
\begin{tabular}{llrrrrrrrrrrrrrrr}
  \hline \rule{0pt}{3ex} & & \multicolumn{5}{c}{$\mathbf{\boldsymbol{\zeta}=0.1}$} & \multicolumn{5}{c}{$\mathbf{\boldsymbol{\zeta}=0.2}$} & \multicolumn{5}{c}{$\mathbf{\boldsymbol{\zeta}=0.4}$} \\ \cmidrule(lr){1-2} \cmidrule(lr){3-7} \cmidrule(lr){8-12} \cmidrule(lr){13-17}  $\mathbf{\pi_1}$ & \textbf{Method} & \textbf{ACC} & \textbf{SNS} & \textbf{SPC} & \textbf{AUC} & \textbf{TSL} & \textbf{ACC} & \textbf{SNS} & \textbf{SPC} & \textbf{AUC} & \textbf{TSL} & \textbf{ACC} & \textbf{SNS} & \textbf{SPC} & \textbf{AUC} & \textbf{TSL}  \\ \cmidrule(lr){1-2} \cmidrule(lr){3-7} \cmidrule(lr){8-12} \cmidrule(lr){13-17}  \rule{0pt}{3ex} & Split-Lasso-10 & 0.79 & 0.72 & 0.85 & 0.88 & 0.88 & 0.89 & 0.84 & 0.92 & 0.96 & 0.51 & 0.90 & 0.87 & 0.93 & 0.97 & 0.44 \\ 
   & Split-EN-10 & 0.79 & 0.72 & 0.85 & 0.88 & 0.88 & 0.88 & 0.83 & 0.92 & 0.96 & 0.52 & 0.91 & 0.87 & 0.93 & 0.97 & 0.43 \\ 
   & Lasso & 0.78 & 0.71 & 0.83 & 0.87 & 0.93 & 0.85 & 0.78 & 0.89 & 0.93 & 0.67 & 0.86 & 0.80 & 0.90 & 0.94 & 0.65 \\ 
   & Elastic Net & 0.78 & 0.71 & 0.84 & 0.87 & 0.92 & 0.86 & 0.79 & 0.90 & 0.94 & 0.62 & 0.87 & 0.82 & 0.91 & 0.95 & 0.57 \\ 
   & Adaptive & 0.77 & 0.63 & 0.87 & 0.87 & 1.01 & 0.83 & 0.69 & 0.91 & 0.92 & 0.80 & 0.84 & 0.73 & 0.91 & 0.93 & 0.75 \\ 
   & Relaxed & 0.77 & 0.70 & 0.82 & 0.86 & 0.97 & 0.84 & 0.77 & 0.89 & 0.92 & 0.77 & 0.86 & 0.81 & 0.89 & 0.94 & 0.73 \\
\hline
\end{tabular}
}
\end{table}
\begin{table}[H]
\centering
\caption{Mean prediction accuracy (ACC), sensitivity (SNS) and specificity (SPC), area under ROC curve (AUC) and test-sample loss (TSL) for Scenario 3, $\rho_1 =$ 0.8, $\rho_2 =$ 0.2, $n =$ 100, $p =$ 1000.}
\resizebox{\textwidth}{!}{
\begin{tabular}{llrrrrrrrrrrrrrrr}
  \hline \rule{0pt}{3ex} & & \multicolumn{5}{c}{$\mathbf{\boldsymbol{\zeta}=0.1}$} & \multicolumn{5}{c}{$\mathbf{\boldsymbol{\zeta}=0.2}$} & \multicolumn{5}{c}{$\mathbf{\boldsymbol{\zeta}=0.4}$} \\ \cmidrule(lr){1-2} \cmidrule(lr){3-7} \cmidrule(lr){8-12} \cmidrule(lr){13-17}  $\mathbf{\pi_1}$ & \textbf{Method} & \textbf{ACC} & \textbf{SNS} & \textbf{SPC} & \textbf{AUC} & \textbf{TSL} & \textbf{ACC} & \textbf{SNS} & \textbf{SPC} & \textbf{AUC} & \textbf{TSL} & \textbf{ACC} & \textbf{SNS} & \textbf{SPC} & \textbf{AUC} & \textbf{TSL}  \\ \cmidrule(lr){1-2} \cmidrule(lr){3-7} \cmidrule(lr){8-12} \cmidrule(lr){13-17}  \rule{0pt}{3ex} & Split-Lasso-10 & 0.84 & 0.76 & 0.89 & 0.92 & 0.71 & 0.89 & 0.85 & 0.91 & 0.96 & 0.51 & 0.90 & 0.88 & 0.92 & 0.97 & 0.44 \\ 
   & Split-EN-10 & 0.83 & 0.76 & 0.88 & 0.92 & 0.72 & 0.89 & 0.84 & 0.91 & 0.96 & 0.52 & 0.91 & 0.88 & 0.92 & 0.97 & 0.43 \\ 
   & Lasso & 0.82 & 0.73 & 0.87 & 0.90 & 0.80 & 0.85 & 0.79 & 0.90 & 0.94 & 0.66 & 0.86 & 0.81 & 0.89 & 0.94 & 0.64 \\ 
   & Elastic Net & 0.82 & 0.73 & 0.88 & 0.91 & 0.79 & 0.86 & 0.80 & 0.90 & 0.94 & 0.63 & 0.87 & 0.83 & 0.90 & 0.95 & 0.58 \\ 
   & Adaptive & 0.80 & 0.65 & 0.91 & 0.90 & 0.90 & 0.84 & 0.72 & 0.91 & 0.93 & 0.78 & 0.84 & 0.75 & 0.90 & 0.93 & 0.75 \\ 
   & Relaxed & 0.81 & 0.73 & 0.87 & 0.90 & 0.84 & 0.85 & 0.79 & 0.89 & 0.93 & 0.72 & 0.86 & 0.82 & 0.89 & 0.94 & 0.73 \\
\hline
\end{tabular}
}
\end{table}
\begin{table}[H]
\centering
\caption{Mean prediction accuracy (ACC), sensitivity (SNS) and specificity (SPC), area under ROC curve (AUC) and test-sample loss (TSL) for Scenario 3, $\rho_1 =$ 0.8, $\rho_2 =$ 0.5, $n =$ 100, $p =$ 1000.} \label{tab:sim_pred_3_100_0.8_0.5}
\resizebox{\textwidth}{!}{
\begin{tabular}{llrrrrrrrrrrrrrrr}
  \hline \rule{0pt}{3ex} & & \multicolumn{5}{c}{$\mathbf{\boldsymbol{\zeta}=0.1}$} & \multicolumn{5}{c}{$\mathbf{\boldsymbol{\zeta}=0.2}$} & \multicolumn{5}{c}{$\mathbf{\boldsymbol{\zeta}=0.4}$} \\ \cmidrule(lr){1-2} \cmidrule(lr){3-7} \cmidrule(lr){8-12} \cmidrule(lr){13-17}  $\mathbf{\pi_1}$ & \textbf{Method} & \textbf{ACC} & \textbf{SNS} & \textbf{SPC} & \textbf{AUC} & \textbf{TSL} & \textbf{ACC} & \textbf{SNS} & \textbf{SPC} & \textbf{AUC} & \textbf{TSL} & \textbf{ACC} & \textbf{SNS} & \textbf{SPC} & \textbf{AUC} & \textbf{TSL}  \\ \cmidrule(lr){1-2} \cmidrule(lr){3-7} \cmidrule(lr){8-12} \cmidrule(lr){13-17}  \rule{0pt}{3ex} & Split-Lasso-10 & 0.87 & 0.82 & 0.90 & 0.95 & 0.60 & 0.92 & 0.89 & 0.93 & 0.98 & 0.39 & 0.94 & 0.92 & 0.95 & 0.99 & 0.27 \\ 
   & Split-EN-10 & 0.86 & 0.82 & 0.90 & 0.95 & 0.60 & 0.92 & 0.89 & 0.94 & 0.98 & 0.38 & 0.94 & 0.92 & 0.95 & 0.99 & 0.27 \\ 
   & Lasso & 0.85 & 0.80 & 0.89 & 0.94 & 0.66 & 0.89 & 0.85 & 0.92 & 0.97 & 0.49 & 0.91 & 0.88 & 0.93 & 0.98 & 0.40 \\ 
   & Elastic Net & 0.86 & 0.80 & 0.90 & 0.94 & 0.65 & 0.90 & 0.86 & 0.93 & 0.97 & 0.46 & 0.92 & 0.89 & 0.94 & 0.98 & 0.35 \\ 
   & Adaptive & 0.84 & 0.75 & 0.91 & 0.93 & 0.75 & 0.88 & 0.81 & 0.93 & 0.96 & 0.59 & 0.90 & 0.84 & 0.94 & 0.97 & 0.50 \\ 
   & Relaxed & 0.85 & 0.79 & 0.89 & 0.93 & 0.72 & 0.89 & 0.85 & 0.91 & 0.96 & 0.58 & 0.91 & 0.88 & 0.92 & 0.97 & 0.54 \\
\hline
\end{tabular}
}
\end{table}
\begin{table}[H]
\centering
\caption{Mean recall (RCL) and precision (PRC) for Scenario 3, $\rho_1 =$ 0.5, $\rho_2 =$ 0.2, $n =$ 50, $p =$ 1000.}  \label{tab:sim_sel_3_50_0.5_0.2}
\resizebox{0.6\textwidth}{!}{
\begin{tabular}{llrrrrrr}
  \hline \rule{0pt}{3ex} & & \multicolumn{2}{c}{$\mathbf{\boldsymbol{\zeta}=0.1}$} & \multicolumn{2}{c}{$\mathbf{\boldsymbol{\zeta}=0.2}$} & \multicolumn{2}{c}{$\mathbf{\boldsymbol{\zeta}=0.4}$} \\ \cmidrule(lr){1-2} \cmidrule(lr){3-4} \cmidrule(lr){5-6} \cmidrule(lr){7-8}  $\mathbf{\pi_1}$ & \textbf{Method} & \textbf{RCL} & \textbf{PRC} & \textbf{RCL} & \textbf{PRC} & \textbf{RCL} & \textbf{PRC}  \\ \cmidrule(lr){1-2} \cmidrule(lr){3-4} \cmidrule(lr){5-6} \cmidrule(lr){7-8}  \rule{0pt}{3ex} & Split-Lasso-10 & 0.30 & 0.22 & 0.30 & 0.31 & 0.29 & 0.49 \\ 
   & Split-EN-10 & 0.33 & 0.22 & 0.34 & 0.31 & 0.39 & 0.48 \\ 
   & Lasso & 0.05 & 0.32 & 0.04 & 0.36 & 0.03 & 0.49 \\ 
   & Elastic Net & 0.08 & 0.28 & 0.07 & 0.36 & 0.06 & 0.49 \\ 
   & Adaptive & 0.05 & 0.27 & 0.04 & 0.36 & 0.03 & 0.48 \\ 
   & Relaxed & 0.04 & 0.34 & 0.04 & 0.37 & 0.03 & 0.50 \\
\hline
\end{tabular}
}
\end{table}
\begin{table}[H]
\centering
\caption{Mean recall (RCL) and precision (PRC) for Scenario 3, $\rho_1 =$ 0.8, $\rho_2 =$ 0.2, $n =$ 50, $p =$ 1000.}
\resizebox{0.6\textwidth}{!}{
\begin{tabular}{llrrrrrr}
  \hline \rule{0pt}{3ex} & & \multicolumn{2}{c}{$\mathbf{\boldsymbol{\zeta}=0.1}$} & \multicolumn{2}{c}{$\mathbf{\boldsymbol{\zeta}=0.2}$} & \multicolumn{2}{c}{$\mathbf{\boldsymbol{\zeta}=0.4}$} \\ \cmidrule(lr){1-2} \cmidrule(lr){3-4} \cmidrule(lr){5-6} \cmidrule(lr){7-8}  $\mathbf{\pi_1}$ & \textbf{Method} & \textbf{RCL} & \textbf{PRC} & \textbf{RCL} & \textbf{PRC} & \textbf{RCL} & \textbf{PRC}  \\ \cmidrule(lr){1-2} \cmidrule(lr){3-4} \cmidrule(lr){5-6} \cmidrule(lr){7-8}  \rule{0pt}{3ex} & Split-Lasso-10 & 0.40 & 0.24 & 0.37 & 0.38 & 0.29 & 0.50 \\ 
   & Split-EN-10 & 0.45 & 0.23 & 0.49 & 0.35 & 0.39 & 0.50 \\ 
   & Lasso & 0.05 & 0.31 & 0.04 & 0.37 & 0.02 & 0.40 \\ 
   & Elastic Net & 0.10 & 0.34 & 0.08 & 0.42 & 0.06 & 0.47 \\ 
   & Adaptive & 0.05 & 0.32 & 0.04 & 0.38 & 0.02 & 0.39 \\ 
   & Relaxed & 0.05 & 0.45 & 0.03 & 0.41 & 0.02 & 0.42 \\
\hline
\end{tabular}
}
\end{table}
\input{tabs/tabs_Selection_3b_50_1000_0.8_0.5.txt}
\input{tabs/tabs_Selection_3b_100_1000_0.5_0.2.txt}
\input{tabs/tabs_Selection_3b_100_1000_0.8_0.2.txt}
\input{tabs/tabs_Selection_3b_100_1000_0.8_0.5.txt}
\FloatBarrier

\subsection{Scenario 4: Interactions, Block Correlation}
\input{tabs/tabs_Prediction_3b_50_1000_0.5_0.2_interaction.txt}
\input{tabs/tabs_Prediction_3b_50_1000_0.8_0.2_interaction.txt}
\input{tabs/tabs_Prediction_3b_50_1000_0.8_0.5_interaction.txt}
\input{tabs/tabs_Prediction_3b_100_1000_0.5_0.2_interaction.txt}
\input{tabs/tabs_Prediction_3b_100_1000_0.8_0.2_interaction.txt}
\input{tabs/tabs_Prediction_3b_100_1000_0.8_0.5_interaction.txt}
\input{tabs/tabs_Selection_3b_50_1000_0.5_0.2_interaction.txt}
\input{tabs/tabs_Selection_3b_50_1000_0.8_0.2_interaction.txt}
\input{tabs/tabs_Selection_3b_50_1000_0.8_0.5_interaction.txt}
\input{tabs/tabs_Selection_3b_100_1000_0.5_0.2_interaction.txt}
\input{tabs/tabs_Selection_3b_100_1000_0.8_0.2_interaction.txt}
\input{tabs/tabs_Selection_3b_100_1000_0.8_0.5_interaction.txt}
\FloatBarrier

\subsection{Scenario 5: Non-Linear Effects, Block Correlation}
\input{tabs/tabs_Prediction_3b_50_1000_0.5_0.2_nonlinear.txt}
\input{tabs/tabs_Prediction_3b_50_1000_0.8_0.2_nonlinear.txt}
\input{tabs/tabs_Prediction_3b_50_1000_0.8_0.5_nonlinear.txt}
\input{tabs/tabs_Prediction_3b_100_1000_0.5_0.2_nonlinear.txt}
\input{tabs/tabs_Prediction_3b_100_1000_0.8_0.2_nonlinear.txt}
\input{tabs/tabs_Prediction_3b_100_1000_0.8_0.5_nonlinear.txt}
\input{tabs/tabs_Selection_3b_50_1000_0.5_0.2_nonlinear.txt}
\input{tabs/tabs_Selection_3b_50_1000_0.8_0.2_nonlinear.txt}
\input{tabs/tabs_Selection_3b_50_1000_0.8_0.5_nonlinear.txt}
\input{tabs/tabs_Selection_3b_100_1000_0.5_0.2_nonlinear.txt}
\input{tabs/tabs_Selection_3b_100_1000_0.8_0.2_nonlinear.txt}
\input{tabs/tabs_Selection_3b_100_1000_0.8_0.5_nonlinear.txt}

\FloatBarrier

\section{Full Results for Medical Genomics Data} \label{app:full_results_genomics}

This section contains the detailed performance tables for the benchmark study on ten gene expression datasets, as summarized in Section \ref{app:results_genomics} of this supplement and referenced in Section 6 of the main article.

For each of the ten datasets, we report the performance of the fourteen methods for all considered training set proportions and numbers of selected genes ($p \in \{100, 250, 500, 1000\}$). To facilitate a clear comparison across methods for each specific setting, the tables present \textbf{relative performance metrics}. The values for prediction accuracy (ACC) are scaled relative to the method with the \textit{highest} mean accuracy. The values for test sample loss (TSL) are scaled relative to the method with the \textit{lowest} mean TSL.

Therefore, in the tables below:
\begin{itemize}
	\item For \textbf{ACC}, a value of \textbf{1.00} indicates the best-performing method (highest accuracy), and lower values represent proportionally lower accuracy.
	\item For \textbf{TSL}, a value of \textbf{1.00} indicates the best-performing method (lowest loss), and higher values represent proportionally higher (worse) loss.
\end{itemize}

The results are organized by dataset as follows:
\begin{itemize}
	\item \textbf{GSE20347 (Esophageal Cancer):} Table \ref{tab:results_GSE20347}
	\item \textbf{GSE23400 (Esophageal Cancer, Part 1):} Tables \ref{tab:results_GSE23400_1_35} and \ref{tab:results_GSE23400_1_50}
	\item \textbf{GSE23400 (Esophageal Cancer, Part 2):} Tables \ref{tab:results_GSE23400_2_35} and \ref{tab:results_GSE23400_2_50}
	\item \textbf{GSE5364 (Esophageal Cancer):} Table \ref{tab:results_GSE5364_Esophageal}
	\item \textbf{GSE25869 (Gastric Cancer):} Tables \ref{tab:results_GSE25869_35} and \ref{tab:results_GSE25869_50}
	\item \textbf{GSE10245 (Lung Cancer):} Tables \ref{tab:results_GSE10245_35} and \ref{tab:results_GSE10245_50}
	\item \textbf{GSE5364 (Lung Cancer):} Table \ref{tab:results_GSE5364_Lung}
	\item \textbf{GSE5364 (Thyroid Cancer):} Table \ref{tab:results_GSE5364_Thyroid}
	\item \textbf{GSE21942 (Multiple Sclerosis):} Table \ref{tab:results_GSE21942}
	\item \textbf{GSE14905 (Psoriasis):} Table \ref{tab:results_GSE14905}
\end{itemize}


\input{tabs/gene_results_GSE20347_50.txt}
\input{tabs/gene_results_GSE23400_1_35.txt}
\input{tabs/gene_results_GSE23400_1_50.txt}
\input{tabs/gene_results_GSE23400_2_35.txt}
\input{tabs/gene_results_GSE23400_2_50.txt}
\input{tabs/gene_results_GSE5364_Esophagus_50.txt}
\input{tabs/gene_results_GSE25869_35.txt}
\input{tabs/gene_results_GSE25869_50.txt}
\input{tabs/gene_results_GSE10245_35.txt}
\input{tabs/gene_results_GSE10245_50.txt}
\input{tabs/gene_results_GSE5364_Lung_50.txt}
\input{tabs/gene_results_GSE5364_Thyroid_50.txt}
\input{tabs/gene_results_GSE21942_50.txt}
\input{tabs/gene_results_GSE14905_50.txt}

\FloatBarrier

	\clearpage
	
	\bibliographystyle{Chicago}
	\bibliography{SplitGLMs_Paper}

\begin{thebibliography}{}

\bibitem[\protect\citeauthoryear{Aguila, Morris, Spina, Bar, Schraner, Vinkler,
  Sohn, and Welford}{Aguila et~al.}{2019}]{aguila2019ig}
Aguila, B., A.~B. Morris, R.~Spina, E.~Bar, J.~Schraner, R.~Vinkler, J.~W.
  Sohn, and S.~M. Welford (2019).
\newblock The ig superfamily protein ptgfrn coordinates survival signaling in
  glioblastoma multiforme.
\newblock {\em Cancer letters\/}~{\em 462}, 33--42.

\bibitem[\protect\citeauthoryear{Asghar, Magnusson, Kemppainen, Sukumaran,
  L{\"o}f, Pulli, Kalhori, and T{\"o}rnquist}{Asghar
  et~al.}{2015}]{asghar2015transient}
Asghar, M.~Y., M.~Magnusson, K.~Kemppainen, P.~Sukumaran, C.~L{\"o}f, I.~Pulli,
  V.~Kalhori, and K.~T{\"o}rnquist (2015).
\newblock Transient receptor potential canonical 1 (trpc1) channels as
  regulators of sphingolipid and vegf receptor expression: Implications for
  thyroid cancer cell migration and proliferation.
\newblock {\em Journal of Biological Chemistry\/}~{\em 290\/}(26),
  16116--16131.

\bibitem[\protect\citeauthoryear{Barrett, Wilhite, Ledoux, Evangelista, Kim,
  Tomashevsky, Marshall, Phillippy, Sherman, Holko, et~al.}{Barrett
  et~al.}{2012}]{barrett2012ncbi}
Barrett, T., S.~E. Wilhite, P.~Ledoux, C.~Evangelista, I.~F. Kim,
  M.~Tomashevsky, K.~A. Marshall, K.~H. Phillippy, P.~M. Sherman, M.~Holko,
  et~al. (2012).
\newblock Ncbi geo: archive for functional genomics data sets—update.
\newblock {\em Nucleic Acids Research\/}~{\em 41\/}(D1), D991--D995.

\bibitem[\protect\citeauthoryear{Biau, Devroye, and Lugosi}{Biau
  et~al.}{2008}]{biau2008consistency}
Biau, G., L.~Devroye, and G.~Lugosi (2008).
\newblock Consistency of random forests and other averaging classifiers.
\newblock {\em Journal of Machine Learning Research\/}~{\em 9\/}(9).

\bibitem[\protect\citeauthoryear{Breheny and Huang}{Breheny and
  Huang}{2011}]{ncvreg_package}
Breheny, P. and J.~Huang (2011).
\newblock Coordinate descent algorithms for nonconvex penalized regression,
  with applications to biological feature selection.
\newblock {\em The Annals of Applied Statistics\/}~{\em 5\/}(1), 232--253.

\bibitem[\protect\citeauthoryear{Breiman}{Breiman}{1996a}]{breiman1996bagging}
Breiman, L. (1996a).
\newblock Bagging predictors.
\newblock {\em Machine Learning\/}~{\em 24\/}(2), 123--140.

\bibitem[\protect\citeauthoryear{Breiman}{Breiman}{1996b}]{breiman-stacked}
Breiman, L. (1996b).
\newblock Stacked regressions.
\newblock {\em Machine Learning\/}~{\em 24\/}(1), 49--64.

\bibitem[\protect\citeauthoryear{Breiman}{Breiman}{2001}]{RF}
Breiman, L. (2001, October).
\newblock Random forests.
\newblock {\em Machine Learning\/}~{\em 45\/}(1), 5--32.

\bibitem[\protect\citeauthoryear{Brown, Wyatt, and Ti{\v{n}}o}{Brown
  et~al.}{2005}]{brown2005managing}
Brown, G., J.~L. Wyatt, and P.~Ti{\v{n}}o (2005).
\newblock Managing diversity in regression ensembles.
\newblock {\em Journal of Machine Learning Research\/}~{\em 6\/}(Sep),
  1621--1650.

\bibitem[\protect\citeauthoryear{B{\"u}hlmann and van~de Geer}{B{\"u}hlmann and
  van~de Geer}{2011}]{buhlmann-book}
B{\"u}hlmann, P. and S.~van~de Geer (2011).
\newblock {\em Statistics for High-Dimensional Data: Methods, Theory and
  Applications}.
\newblock Springer Series in Statistics. Springer Berlin Heidelberg.

\bibitem[\protect\citeauthoryear{B{\"u}hlmann and Yu}{B{\"u}hlmann and
  Yu}{2003}]{buhlmann2003boosting}
B{\"u}hlmann, P. and B.~Yu (2003).
\newblock Boosting with the l 2 loss: regression and classification.
\newblock {\em Journal of the American Statistical Association\/}~{\em
  98\/}(462), 324--339.

\bibitem[\protect\citeauthoryear{Chen and Guestrin}{Chen and
  Guestrin}{2016}]{chen2016xgboost}
Chen, T. and C.~Guestrin (2016).
\newblock Xgboost: A scalable tree boosting system.
\newblock In {\em Proceedings of the 22nd acm sigkdd international conference
  on knowledge discovery and data mining}, pp.\  785--794.

\bibitem[\protect\citeauthoryear{Chen, He, Benesty, Khotilovich, Tang, Cho,
  Chen, Mitchell, Cano, Zhou, Li, Xie, Lin, Geng, and Li}{Chen
  et~al.}{2020}]{xgboost_package}
Chen, T., T.~He, M.~Benesty, V.~Khotilovich, Y.~Tang, H.~Cho, K.~Chen,
  R.~Mitchell, I.~Cano, T.~Zhou, M.~Li, J.~Xie, M.~Lin, Y.~Geng, and Y.~Li
  (2020).
\newblock {\em xgboost: Extreme Gradient Boosting}.
\newblock R package version 1.1.1.1.

\bibitem[\protect\citeauthoryear{Christidis, {Van Aelst}, and Zamar}{Christidis
  et~al.}{2021}]{splitglm_package}
Christidis, A., S.~{Van Aelst}, and R.~Zamar (2021).
\newblock {\em SplitGLM: Split Generalized Linear Models}.
\newblock R package version 1.0.2.

\bibitem[\protect\citeauthoryear{Christidis, Lakshmanan, Smucler, and
  Zamar}{Christidis et~al.}{2020}]{christidis2020}
Christidis, A.-A., L.~Lakshmanan, E.~Smucler, and R.~Zamar (2020).
\newblock Split regularized regression.
\newblock {\em Technometrics\/}~{\em 62\/}(3), 330--338.

\bibitem[\protect\citeauthoryear{Czubak-Prowizor, Babinska, and
  Swiatkowska}{Czubak-Prowizor et~al.}{2022}]{czubak2022f11}
Czubak-Prowizor, K., A.~Babinska, and M.~Swiatkowska (2022).
\newblock The f11 receptor (f11r)/junctional adhesion molecule-a
  (jam-a)(f11r/jam-a) in cancer progression.
\newblock {\em Molecular and Cellular Biochemistry\/}~{\em 477\/}(1), 79--98.

\bibitem[\protect\citeauthoryear{Debeer and Strobl}{Debeer and
  Strobl}{2020}]{debeer2020conditional}
Debeer, D. and C.~Strobl (2020).
\newblock Conditional permutation importance revisited.
\newblock {\em BMC bioinformatics\/}~{\em 21}, 1--30.

\bibitem[\protect\citeauthoryear{Donoho and Johnstone}{Donoho and
  Johnstone}{1994}]{donoho1994ideal}
Donoho, D.~L. and J.~M. Johnstone (1994).
\newblock Ideal spatial adaptation by wavelet shrinkage.
\newblock {\em Biometrika\/}~{\em 81\/}(3), 425--455.

\bibitem[\protect\citeauthoryear{Dorani, Hu, Woods, and Zhai}{Dorani
  et~al.}{2018}]{dorani2018ensemble}
Dorani, F., T.~Hu, M.~O. Woods, and G.~Zhai (2018).
\newblock Ensemble learning for detecting gene-gene interactions in colorectal
  cancer.
\newblock {\em PeerJ\/}~{\em 6}, e5854.

\bibitem[\protect\citeauthoryear{Dudoit, Fridlyand, and Speed}{Dudoit
  et~al.}{2002}]{dudoit2002comparison}
Dudoit, S., J.~Fridlyand, and T.~P. Speed (2002).
\newblock Comparison of discrimination methods for the classification of tumors
  using gene expression data.
\newblock {\em Journal of the American statistical association\/}~{\em
  97\/}(457), 77--87.

\bibitem[\protect\citeauthoryear{Edgar, Domrachev, and Lash}{Edgar
  et~al.}{2002}]{edgar2002gene}
Edgar, R., M.~Domrachev, and A.~E. Lash (2002).
\newblock Gene expression omnibus: Ncbi gene expression and hybridization array
  data repository.
\newblock {\em Nucleic Acids Research\/}~{\em 30\/}(1), 207--210.

\bibitem[\protect\citeauthoryear{Fan and Li}{Fan and Li}{2001}]{SCAD}
Fan, J. and R.~Li (2001).
\newblock Variable selection via nonconcave penalized likelihood and its oracle
  properties.
\newblock {\em Journal of the American Statistical Association\/}~{\em
  96\/}(456), 1348--1360.

\bibitem[\protect\citeauthoryear{Fan and Lv}{Fan and Lv}{2008}]{SIS}
Fan, J. and J.~Lv (2008).
\newblock Sure independence screening for ultrahigh dimensional feature space.
\newblock {\em Journal of the Royal Statistical Society: Series B (Statistical
  Methodology)\/}~{\em 70\/}(5), 849--911.

\bibitem[\protect\citeauthoryear{Friedman, Hastie, and Tibshirani}{Friedman
  et~al.}{2010}]{friedman2010regularization}
Friedman, J., T.~Hastie, and R.~Tibshirani (2010).
\newblock Regularization paths for generalized linear models via coordinate
  descent.
\newblock {\em Journal of Statistical Software\/}~{\em 33\/}(1), 1.

\bibitem[\protect\citeauthoryear{Friedman}{Friedman}{2001}]{GBM}
Friedman, J.~H. (2001, 10).
\newblock Greedy function approximation: A gradient boosting machine.
\newblock {\em The Annals of Statistics\/}~{\em 29\/}(5), 1189--1232.

\bibitem[\protect\citeauthoryear{Friedman and Popescu}{Friedman and
  Popescu}{2008}]{friedman2008predictive}
Friedman, J.~H. and B.~E. Popescu (2008).
\newblock Predictive learning via rule ensembles.

\bibitem[\protect\citeauthoryear{Fumera and Roli}{Fumera and
  Roli}{2003}]{fumera2003linear}
Fumera, G. and F.~Roli (2003).
\newblock Linear combiners for classifier fusion: Some theoretical and
  experimental results.
\newblock In {\em International Workshop on Multiple Classifier Systems}, pp.\
  74--83. Springer.

\bibitem[\protect\citeauthoryear{Giacinto and Roli}{Giacinto and
  Roli}{2001}]{giacinto2001design}
Giacinto, G. and F.~Roli (2001).
\newblock Design of effective neural network ensembles for image classification
  purposes.
\newblock {\em Image and Vision Computing\/}~{\em 19\/}(9-10), 699--707.

\bibitem[\protect\citeauthoryear{Hastie, Tibshirani, and Wainwright}{Hastie
  et~al.}{2015}]{hastie2015statistical}
Hastie, T., R.~Tibshirani, and M.~Wainwright (2015).
\newblock {\em Statistical learning with sparsity: the lasso and
  generalizations}.
\newblock CRC press.

\bibitem[\protect\citeauthoryear{Heinze, Boulesteix, Kammer, Morris, White, and
  of~the STRATOS~Initiative}{Heinze et~al.}{2024}]{heinze2024phases}
Heinze, G., A.-L. Boulesteix, M.~Kammer, T.~P. Morris, I.~R. White, and S.~P.
  of~the STRATOS~Initiative (2024).
\newblock Phases of methodological research in biostatistics—building the
  evidence base for new methods.
\newblock {\em Biometrical Journal\/}~{\em 66\/}(1), 2200222.

\bibitem[\protect\citeauthoryear{Ho}{Ho}{1998}]{ho1998random}
Ho, T.~K. (1998).
\newblock The random subspace method for constructing decision forests.
\newblock {\em IEEE Transactions on Pattern Analysis and Machine
  Intelligence\/}~{\em 20\/}(8), 832--844.

\bibitem[\protect\citeauthoryear{Hoerl and Kennard}{Hoerl and
  Kennard}{1970}]{Ridge}
Hoerl, A.~E. and R.~W. Kennard (1970).
\newblock Ridge regression: Biased estimation for nonorthogonal problems.
\newblock {\em Technometrics\/}~{\em 12\/}(1), 55--67.

\bibitem[\protect\citeauthoryear{Holub}{Holub}{2022}]{xrf_package}
Holub, K. (2022).
\newblock {\em xrf: eXtreme RuleFit}.
\newblock R package version 0.2.2.

\bibitem[\protect\citeauthoryear{Huang, Prasad, Lemon, Hampel, Wright,
  Kornacker, LiVolsi, Frankel, Kloos, Eng, et~al.}{Huang
  et~al.}{2001}]{huang2001gene}
Huang, Y., M.~Prasad, W.~J. Lemon, H.~Hampel, F.~A. Wright, K.~Kornacker,
  V.~LiVolsi, W.~Frankel, R.~T. Kloos, C.~Eng, et~al. (2001).
\newblock Gene expression in papillary thyroid carcinoma reveals highly
  consistent profiles.
\newblock {\em Proceedings of the National Academy of Sciences\/}~{\em
  98\/}(26), 15044--15049.

\bibitem[\protect\citeauthoryear{Kohavi, Wolpert, et~al.}{Kohavi
  et~al.}{1996}]{kohavi1996bias}
Kohavi, R., D.~H. Wolpert, et~al. (1996).
\newblock Bias plus variance decomposition for zero-one loss functions.
\newblock In {\em ICML}, Volume~96, pp.\  275--83.

\bibitem[\protect\citeauthoryear{Krogh and Vedelsby}{Krogh and
  Vedelsby}{1995}]{krogh1995neural}
Krogh, A. and J.~Vedelsby (1995).
\newblock Neural network ensembles, cross validation, and active learning.
\newblock In {\em Advances in neural information processing systems}, pp.\
  231--238.

\bibitem[\protect\citeauthoryear{Kuncheva and Whitaker}{Kuncheva and
  Whitaker}{2003}]{kuncheva2003measures}
Kuncheva, L.~I. and C.~J. Whitaker (2003).
\newblock Measures of diversity in classifier ensembles and their relationship
  with the ensemble accuracy.
\newblock {\em Machine Learning\/}~{\em 51\/}(2), 181--207.

\bibitem[\protect\citeauthoryear{Liu, Lu, Guo, Zhang, Ye, Du, Li, Wu, Yu, Zhai,
  et~al.}{Liu et~al.}{2020}]{liu2020lncrna}
Liu, H.~Y., S.~R. Lu, Z.~H. Guo, Z.~S. Zhang, X.~Ye, Q.~Du, H.~Li, Q.~Wu,
  B.~Yu, Q.~Zhai, et~al. (2020).
\newblock lncrna slc16a1-as1 as a novel prognostic biomarker in non-small cell
  lung cancer.
\newblock {\em Journal of Investigative Medicine\/}~{\em 68\/}(1), 52--59.

\bibitem[\protect\citeauthoryear{Ma, Ma, Zhou, Chen, An, and Zhang}{Ma
  et~al.}{2017}]{ma2017protease}
Ma, C., W.~Ma, N.~Zhou, N.~Chen, L.~An, and Y.~Zhang (2017).
\newblock Protease serine s1 family member 8 (prss8) inhibits tumor growth in
  vitro and in vivo in human non-small cell lung cancer.
\newblock {\em Oncology Research\/}~{\em 25\/}(5), 781.

\bibitem[\protect\citeauthoryear{Marquez, Dong, Hayashi, and Serrero}{Marquez
  et~al.}{2024}]{marquez2024prostaglandin}
Marquez, J., J.~Dong, J.~Hayashi, and G.~Serrero (2024).
\newblock Prostaglandin f2 receptor negative regulator (ptgfrn) expression
  correlates with a metastatic-like phenotype in epidermoid carcinoma,
  pediatric medulloblastoma, and mesothelioma.
\newblock {\em Journal of Cellular Biochemistry\/}~{\em 125\/}(8), e30616.

\bibitem[\protect\citeauthoryear{McCullagh and Nelder}{McCullagh and
  Nelder}{1989}]{mccullagh1989monographs}
McCullagh, P. and J.~A. Nelder (1989).
\newblock Monographs on statistics and applied probability.
\newblock {\em Generalized Linear Models\/}~{\em 37}.

\bibitem[\protect\citeauthoryear{Meier, Van De~Geer, and B{\"u}hlmann}{Meier
  et~al.}{2008}]{group}
Meier, L., S.~Van De~Geer, and P.~B{\"u}hlmann (2008).
\newblock The group lasso for logistic regression.
\newblock {\em Journal of the Royal Statistical Society: Series B (Statistical
  Methodology)\/}~{\em 70\/}(1), 53--71.

\bibitem[\protect\citeauthoryear{Meinshausen}{Meinshausen}{2007}]{meinshausen2007relaxed}
Meinshausen, N. (2007).
\newblock Relaxed lasso.
\newblock {\em Computational Statistics \& Data Analysis\/}~{\em 52\/}(1),
  374--393.

\bibitem[\protect\citeauthoryear{Morris, White, and Crowther}{Morris
  et~al.}{2019}]{morris2019using}
Morris, T.~P., I.~R. White, and M.~J. Crowther (2019).
\newblock Using simulation studies to evaluate statistical methods.
\newblock {\em Statistics in medicine\/}~{\em 38\/}(11), 2074--2102.

\bibitem[\protect\citeauthoryear{Murdoch, Singh, Kumbier, Abbasi-Asl, and
  Yu}{Murdoch et~al.}{2019}]{murdoch2019definitions}
Murdoch, W.~J., C.~Singh, K.~Kumbier, R.~Abbasi-Asl, and B.~Yu (2019).
\newblock Definitions, methods, and applications in interpretable machine
  learning.
\newblock {\em Proceedings of the National Academy of Sciences\/}~{\em
  116\/}(44), 22071--22080.

\bibitem[\protect\citeauthoryear{Nembrini, K{\"o}nig, and Wright}{Nembrini
  et~al.}{2018}]{nembrini2018revival}
Nembrini, S., I.~R. K{\"o}nig, and M.~N. Wright (2018).
\newblock The revival of the gini importance?
\newblock {\em Bioinformatics\/}~{\em 34\/}(21), 3711--3718.

\bibitem[\protect\citeauthoryear{Pardalos, {\v{Z}}ilinskas, and
  {\v{Z}}ilinskas}{Pardalos et~al.}{2017}]{pardalos2017non}
Pardalos, P.~M., A.~{\v{Z}}ilinskas, and J.~{\v{Z}}ilinskas (2017).
\newblock {\em Non-convex multi-objective optimization}.
\newblock Springer.

\bibitem[\protect\citeauthoryear{Partridge and Krzanowski}{Partridge and
  Krzanowski}{1997}]{partridge1997software}
Partridge, D. and W.~Krzanowski (1997).
\newblock Software diversity: practical statistics for its measurement and
  exploitation.
\newblock {\em Information and Software Technology\/}~{\em 39\/}(10), 707--717.

\bibitem[\protect\citeauthoryear{{R Core Team}}{{R Core Team}}{2024}]{CRAN}
{R Core Team} (2024).
\newblock {\em R: A Language and Environment for Statistical Computing}.
\newblock Vienna, Austria: R Foundation for Statistical Computing.

\bibitem[\protect\citeauthoryear{Rejchel and Bogdan}{Rejchel and
  Bogdan}{2020}]{rejchel2020rank}
Rejchel, W. and M.~Bogdan (2020).
\newblock Rank-based lasso-efficient methods for high-dimensional robust model
  selection.
\newblock {\em The Journal of Machine Learning Research\/}~{\em 21\/}(1),
  9838--9884.

\bibitem[\protect\citeauthoryear{Rudin}{Rudin}{2019}]{rudin2019stop}
Rudin, C. (2019).
\newblock Stop explaining black box machine learning models for high stakes
  decisions and use interpretable models instead.
\newblock {\em Nature Machine Intelligence\/}~{\em 1\/}(5), 206--215.

\bibitem[\protect\citeauthoryear{Rudin, Chen, Chen, Huang, Semenova, and
  Zhong}{Rudin et~al.}{2022}]{rudin2022interpretable}
Rudin, C., C.~Chen, Z.~Chen, H.~Huang, L.~Semenova, and C.~Zhong (2022).
\newblock Interpretable machine learning: Fundamental principles and 10 grand
  challenges.
\newblock {\em Statistics Surveys\/}~{\em 16}, 1--85.

\bibitem[\protect\citeauthoryear{Saldana and Feng}{Saldana and
  Feng}{2018}]{SIS_package}
Saldana, D.~F. and Y.~Feng (2018).
\newblock {SIS}: An {R} package for sure independence screening in
  ultrahigh-dimensional statistical models.
\newblock {\em Journal of Statistical Software\/}~{\em 83\/}(2), 1--25.

\bibitem[\protect\citeauthoryear{Schapire and Freund}{Schapire and
  Freund}{2012}]{boosting}
Schapire, R.~E. and Y.~Freund (2012).
\newblock {\em Boosting: Foundations and Algorithms}.
\newblock The MIT Press.

\bibitem[\protect\citeauthoryear{Shen, Diamond, Udell, Gu, and Boyd}{Shen
  et~al.}{2017}]{shen2017disciplined}
Shen, X., S.~Diamond, M.~Udell, Y.~Gu, and S.~Boyd (2017).
\newblock Disciplined multi-convex programming.
\newblock In {\em 2017 29th Chinese Control And Decision Conference (CCDC)},
  pp.\  895--900. IEEE.

\bibitem[\protect\citeauthoryear{Skalak et~al.}{Skalak
  et~al.}{1996}]{skalak1996sources}
Skalak, D.~B. et~al. (1996).
\newblock The sources of increased accuracy for two proposed boosting
  algorithms.
\newblock In {\em Proc. American Association for Artificial Intelligence,
  AAAI-96, Integrating Multiple Learned Models Workshop}, Volume 1129, pp.\
  1133. Citeseer.

\bibitem[\protect\citeauthoryear{Song and Langfelder}{Song and
  Langfelder}{2013}]{randomGLM_package}
Song, L. and P.~Langfelder (2013).
\newblock {\em randomGLM: Random General Linear Model Prediction}.
\newblock R package version 1.02-1.

\bibitem[\protect\citeauthoryear{Song, Langfelder, and Horvath}{Song
  et~al.}{2013}]{rglm}
Song, L., P.~Langfelder, and S.~Horvath (2013).
\newblock Random generalized linear model: a highly accurate and interpretable
  ensemble predictor.
\newblock {\em BMC Bioinformatics\/}~{\em 14\/}(1), 5.

\bibitem[\protect\citeauthoryear{Storey}{Storey}{2002}]{storey2002direct}
Storey, J.~D. (2002).
\newblock A direct approach to false discovery rates.
\newblock {\em Journal of the Royal Statistical Society Series B: Statistical
  Methodology\/}~{\em 64\/}(3), 479--498.

\bibitem[\protect\citeauthoryear{Storey, Bass, Dabney, and Robinson}{Storey
  et~al.}{2023}]{qvalue}
Storey, J.~D., A.~J. Bass, A.~Dabney, and D.~Robinson (2023).
\newblock {\em qvalue: Q-value estimation for false discovery rate control}.
\newblock R package version 2.32.0.

\bibitem[\protect\citeauthoryear{Strobl, Boulesteix, Zeileis, and
  Hothorn}{Strobl et~al.}{2007}]{strobl2007bias}
Strobl, C., A.-L. Boulesteix, A.~Zeileis, and T.~Hothorn (2007).
\newblock Bias in random forest variable importance measures: Illustrations,
  sources and a solution.
\newblock {\em BMC bioinformatics\/}~{\em 8}, 1--21.

\bibitem[\protect\citeauthoryear{Tan, Wang, Liu, Xu, Che, Liu, Hu, Hu, Li, and
  Zhou}{Tan et~al.}{2023}]{tan2023c11orf54}
Tan, J., W.~Wang, X.~Liu, J.~Xu, Y.~Che, Y.~Liu, J.~Hu, L.~Hu, J.~Li, and
  Q.~Zhou (2023).
\newblock C11orf54 promotes dna repair via blocking cma-mediated degradation of
  hif1a.
\newblock {\em Communications Biology\/}~{\em 6\/}(1), 606.

\bibitem[\protect\citeauthoryear{Tang, Huang, Li, and Gu}{Tang
  et~al.}{2023}]{tang2023comprehensive}
Tang, J., Q.~Huang, X.~Li, and S.~Gu (2023).
\newblock Comprehensive analysis of the oncogenic and immunological role of
  spon2 in human tumors.
\newblock {\em Medicine\/}~{\em 102\/}(37), e35122.

\bibitem[\protect\citeauthoryear{Tibshirani}{Tibshirani}{1996}]{Lasso}
Tibshirani, R. (1996).
\newblock Regression shrinkage and selection via the lasso.
\newblock {\em Journal of the Royal Statistical Society: Series B (Statistical
  Methodology)\/}~{\em 58\/}(1), 267--288.

\bibitem[\protect\citeauthoryear{Tibshirani, Saunders, Rosset, Zhu, and
  Knight}{Tibshirani et~al.}{2005}]{tibshirani2005sparsity}
Tibshirani, R., M.~Saunders, S.~Rosset, J.~Zhu, and K.~Knight (2005).
\newblock Sparsity and smoothness via the fused lasso.
\newblock {\em Journal of the Royal Statistical Society: Series B (Statistical
  Methodology)\/}~{\em 67\/}(1), 91--108.

\bibitem[\protect\citeauthoryear{Tseng}{Tseng}{2001}]{tseng2001convergence}
Tseng, P. (2001).
\newblock Convergence of a block coordinate descent method for
  nondifferentiable minimization.
\newblock {\em Journal of Optimization Theory and Applications\/}~{\em
  109\/}(3), 475--494.

\bibitem[\protect\citeauthoryear{Tumer and Ghosh}{Tumer and
  Ghosh}{1996}]{tumer1996error}
Tumer, K. and J.~Ghosh (1996).
\newblock Error correlation and error reduction in ensemble classifiers.
\newblock {\em Connection Science\/}~{\em 8\/}(3-4), 385--404.

\bibitem[\protect\citeauthoryear{Ueda and Nakano}{Ueda and
  Nakano}{1996}]{ueda1996generalization}
Ueda, N. and R.~Nakano (1996).
\newblock Generalization error of ensemble estimators.
\newblock In {\em Proceedings of International Conference on Neural Networks
  (ICNN'96)}, Volume~1, pp.\  90--95. IEEE.

\bibitem[\protect\citeauthoryear{Wright and Ziegler}{Wright and
  Ziegler}{2017}]{ranger_package}
Wright, M.~N. and A.~Ziegler (2017).
\newblock {ranger}: A fast implementation of random forests for high
  dimensional data in {C++} and {R}.
\newblock {\em Journal of Statistical Software\/}~{\em 77\/}(1), 1--17.

\bibitem[\protect\citeauthoryear{Xiao, Lu, Chen, Zou, Liu, Li, He, He, and
  Chen}{Xiao et~al.}{2017}]{xiao2017eight}
Xiao, J., X.~Lu, X.~Chen, Y.~Zou, A.~Liu, W.~Li, B.~He, S.~He, and Q.~Chen
  (2017).
\newblock Eight potential biomarkers for distinguishing between lung
  adenocarcinoma and squamous cell carcinoma.
\newblock {\em Oncotarget\/}~{\em 8\/}(42), 71759.

\bibitem[\protect\citeauthoryear{Xu, Shi, Wang, Huang, Xu, Han, Li, and
  Wang}{Xu et~al.}{2022}]{xu2022lactb}
Xu, Y., H.~Shi, M.~Wang, P.~Huang, M.~Xu, S.~Han, H.~Li, and Y.~Wang (2022).
\newblock Lactb suppresses carcinogenesis in lung cancer and regulates the emt
  pathway.
\newblock {\em Experimental and Therapeutic Medicine\/}~{\em 23\/}(3), 247.

\bibitem[\protect\citeauthoryear{Xu and Yin}{Xu and Yin}{2013}]{xu2013block}
Xu, Y. and W.~Yin (2013).
\newblock A block coordinate descent method for regularized multiconvex
  optimization with applications to nonnegative tensor factorization and
  completion.
\newblock {\em SIAM Journal on Imaging Sciences\/}~{\em 6\/}(3), 1758--1789.

\bibitem[\protect\citeauthoryear{Yang, Pesavento, Luo, and Ottersten}{Yang
  et~al.}{2019}]{yang2019inexact}
Yang, Y., M.~Pesavento, Z.-Q. Luo, and B.~Ottersten (2019).
\newblock Inexact block coordinate descent algorithms for nonsmooth nonconvex
  optimization.
\newblock {\em IEEE Transactions on Signal Processing\/}.

\bibitem[\protect\citeauthoryear{Yang and Zou}{Yang and
  Zou}{2017}]{gcdnet_package}
Yang, Y. and H.~Zou (2017).
\newblock {\em gcdnet: LASSO and Elastic Net (Adaptive) Penalized Least
  Squares, Logistic Regression, HHSVM, Squared Hinge SVM and Expectile
  Regression using a Fast GCD Algorithm}.
\newblock R package version 1.0.5.

\bibitem[\protect\citeauthoryear{Yousefi, Hua, and Dougherty}{Yousefi
  et~al.}{2011}]{yousefi2011multiple}
Yousefi, M.~R., J.~Hua, and E.~R. Dougherty (2011).
\newblock Multiple-rule bias in the comparison of classification rules.
\newblock {\em Bioinformatics\/}~{\em 27\/}(12), 1675--1683.

\bibitem[\protect\citeauthoryear{Yu, Qiu, Chen, Ma, Jiang, Zhou, and Ma}{Yu
  et~al.}{2020}]{yu2020submito}
Yu, B., W.~Qiu, C.~Chen, A.~Ma, J.~Jiang, H.~Zhou, and Q.~Ma (2020).
\newblock Submito-xgboost: predicting protein submitochondrial localization by
  fusing multiple feature information and extreme gradient boosting.
\newblock {\em Bioinformatics\/}~{\em 36\/}(4), 1074--1081.

\bibitem[\protect\citeauthoryear{Zahoor and Zafar}{Zahoor and
  Zafar}{2020}]{zahoor2020classification}
Zahoor, J. and K.~Zafar (2020).
\newblock Classification of microarray gene expression data using an
  infiltration tactics optimization (ito) algorithm.
\newblock {\em Genes\/}~{\em 11\/}(7), 819.

\bibitem[\protect\citeauthoryear{Zhang}{Zhang}{2010}]{MCP}
Zhang, C.-H. (2010, 04).
\newblock Nearly unbiased variable selection under minimax concave penalty.
\newblock {\em The Annals of Statistics\/}~{\em 38\/}(2), 894--942.

\bibitem[\protect\citeauthoryear{Zhang, Huang, Xu, Cao, Shen, Liu, and
  Luo}{Zhang et~al.}{2025}]{zhang2025misp}
Zhang, F., B.~Huang, Y.~Xu, G.~Cao, M.~Shen, C.~Liu, and J.~Luo (2025).
\newblock Misp suppresses ferroptosis via mst1/2 kinases to facilitate yap
  activation in non-small cell lung cancer.
\newblock {\em Advanced Science\/}, 2415814.

\bibitem[\protect\citeauthoryear{Zhang and Coombes}{Zhang and
  Coombes}{2012}]{zhang2012sources}
Zhang, J. and K.~R. Coombes (2012).
\newblock Sources of variation in false discovery rate estimation include
  sample size, correlation, and inherent differences between groups.
\newblock {\em BMC Bioinformatics\/}~{\em 13\/}(S13), S1.

\bibitem[\protect\citeauthoryear{Zou}{Zou}{2006}]{zou2006adaptive}
Zou, H. (2006).
\newblock The adaptive lasso and its oracle properties.
\newblock {\em Journal of the American Statistical Association\/}~{\em
  101\/}(476), 1418--1429.

\bibitem[\protect\citeauthoryear{Zou and Hastie}{Zou and Hastie}{2005}]{EN}
Zou, H. and T.~Hastie (2005).
\newblock Regularization and variable selection via the elastic net.
\newblock {\em Journal of the Royal Statistical Society: Series B (Statistical
  Methodology)\/}~{\em 67\/}(2), 301--320.

\bibitem[\protect\citeauthoryear{Zuo, Cui, Yu, Li, and Ressom}{Zuo
  et~al.}{2017}]{zuo2017incorporating}
Zuo, Y., Y.~Cui, G.~Yu, R.~Li, and H.~W. Ressom (2017).
\newblock Incorporating prior biological knowledge for network-based
  differential gene expression analysis using differentially weighted graphical
  lasso.
\newblock {\em BMC Bioinformatics\/}~{\em 18\/}(1), 1--14.

\end{thebibliography}
\end{document}